\documentclass[12pt]{article}

\usepackage{axodraw}

\makeatletter

\def\paragraph{\@startsection{paragraph}{4}{\z@}{+2.00ex plus
 +1ex minus +.2ex}{1.5ex plus .2ex}{\it\normalsize}}

\def\section{\@startsection {section}{1}{\z@}{+3.0ex plus +1ex minus
  +.2ex}{2.3ex plus .2ex}{\normalsize\bf\boldmath}}
\def\subsection{\@startsection{subsection}{2}{\z@}{+2.5ex plus +1ex
minus +.2ex}{1.5ex plus .2ex}{\normalsize\bf\boldmath}}
\def\subsubsection{\@startsection{subsubsection}{3}{\z@}{+3.25ex plus
 +1ex minus +.2ex}{1.5ex plus .2ex}{\normalsize\it}}

\expandafter\ifx\csname mathrm\endcsname\relax\def\mathrm#1{{\rm #1}}\fi

\newcounter{saveeqn}

\@addtoreset{equation}{section}

\newcount\@tempcntc
\def\@citex[#1]#2{\if@filesw\immediate\write\@auxout{\string\citation{#2}}\fi
  \@tempcnta\z@\@tempcntb\m@ne\def\@citea{}\@cite{\@for\@citeb:=#2\do
    {\@ifundefined
       {b@\@citeb}{\@citeo\@tempcntb\m@ne\@citea
        \def\@citea{,\penalty\@m\ }{\bf ?}\@warning
       {Citation `\@citeb' on page \thepage \space undefined}}%
    {\setbox\z@\hbox{\global\@tempcntc0\csname
b@\@citeb\endcsname\relax}%
     \ifnum\@tempcntc=\z@ \@citeo\@tempcntb\m@ne
       \@citea\def\@citea{,\penalty\@m}
       \hbox{\csname b@\@citeb\endcsname}%
     \else
      \advance\@tempcntb\@ne
      \ifnum\@tempcntb=\@tempcntc
      \else\advance\@tempcntb\m@ne\@citeo
      \@tempcnta\@tempcntc\@tempcntb\@tempcntc\fi\fi}}\@citeo}{#1}}

\def\@citeo{\ifnum\@tempcnta>\@tempcntb\else\@citea
  \def\@citea{,\penalty\@m}%
  \ifnum\@tempcnta=\@tempcntb\the\@tempcnta\else
   {\advance\@tempcnta\@ne\ifnum\@tempcnta=\@tempcntb \else
\def\@citea{--}\fi
    \advance\@tempcnta\m@ne\the\@tempcnta\@citea\the\@tempcntb}\fi\fi}

\def\nl{\nonumber\\}

\def\nln{\nl*[-1ex]}

\def\asymp#1%
{\mathrel{\raisebox{-.4em}{$\widetilde{\scriptstyle #1}$}}}
\def\Nlim#1{\mathrel{\raisebox{-.4em}
{$\stackrel{\disp\longrightarrow}{\scriptstyle#1}$}}}
\def\Nequal#1%
{\mathrel{\raisebox{-.5em}{$\stackrel{=}{\scriptstyle\rm#1}$}}}
\newcommand{\dsl}[1]{\not \hspace{-0.7mm}#1}
\def\dsl{\mathpalette\make@slash}
\def\make@slash#1#2{\setbox\z@\hbox{$#1#2$}%
  \hbox to 0pt{\hss$#1/$\hss\kern-\wd0}\box0}

\def\beq{\begin{equation}}
\def\eeq{\end{equation}}
\def\beqar{\begin{eqnarray}}
\def\eeqar{\end{eqnarray}}
\def\barr#1{\begin{array}{#1}}
\def\earr{\end{array}}
\def\bfi{\begin{figure}}
\def\efi{\end{figure}}
\def\btab{\begin{table}}
\def\etab{\end{table}}
\def\bce{\begin{center}}
\def\ece{\end{center}}
\def\nn{\nonumber}
\def\disp{\displaystyle}
\def\text{\textstyle}

\def\al{\alpha}
\def\be{\beta}

\def\ga{\gamma}

\def\eps{\epsilon}
\def\veps{\varepsilon}

\def\la{\lambda}

\def\refeq#1{\mbox{(\ref{#1})}}

\def\reffi#1{\mbox{Figure~\ref{#1}}}

\def\refse#1{\mbox{Section~\ref{#1}}}

\def\citere#1{\mbox{Ref.~\cite{#1}}}
\def\citeres#1{\mbox{Refs.~\cite{#1}}}

\newcommand{\ri}{{\mathrm{i}}}
\newcommand{\rd}{{\mathrm{d}}}

\newcommand{\ord}{\mathswitch{{\cal{O}}}}

\def\mathswitchr#1{\relax\ifmmode{\mathrm{#1}}\else$\mathrm{#1}$\fi}

\newcommand{\PH}{\mathswitchr H}
\newcommand{\Pp}{\mathswitchr p}

\newcommand{\Pt}{\mathswitchr t}

\def\mathswitch#1{\relax\ifmmode#1\else$#1$\fi}

\def\ie{i.e.,\ }

\newcommand{\soft}{{\mathrm{soft}}}
\newcommand{\coll}{{\mathrm{coll}}}

\newcommand{\DOfin}{D_0^{\mathrm{(fin)}}}

\def\Li{\mathop{\mathrm{Li}_2}\nolimits}

\def\cLi{\mathop{{\cal L}i_2}\nolimits}
\def\Re{\mathop{\mathrm{Re}}\nolimits}
\def\Im{\mathop{\mathrm{Im}}\nolimits}
\def\sgn{\mathop{\mathrm{sgn}}\nolimits}

\newcommand{\D}{{\cal{D}}}

\newcommand{\DOp}[6]{\rlap{\raisebox{-4.5em}{ 
{ \unitlength 1pt
\begin{picture}(140,100)
\Line(20,30)( 50,40)
\Line(20,90)( 50,80)
\Line(50,80)( 50,40)
\Line(50,80)( 90,80)
\Line(90,80)(120,90)
\Line(50,40)( 90,40)
\Line(90,40)(120,30)
\Line(90,40)( 90,80)
\Vertex( 50, 40){3}
\Vertex( 50, 80){3}
\Vertex( 90, 40){3}
\Vertex( 90, 80){3}
\CArc(70,60)(50,160,200)
\CArc(70,70)(50,240,300)
\Text( 15, 90)[r]{$#1$}
\Text( 15, 30)[r]{$#2$}
\Text(125, 30)[l]{$#3$}
\Text(125, 90)[l]{$#4$}
\Text( 13, 60)[r]{$#5$}
\Text( 70,  7)[c]{$#6$}
\end{picture}
} }}}
\newcommand{\DOm}[5]{\raisebox{-4.5em}{ 
{ \unitlength 1pt
\begin{picture}(140,90)
\Text( 70, 89)[c]{$#1$}
\Text( 45, 60)[r]{$#2$}
\Text( 70, 30)[c]{$#3$}
\Text( 95, 60)[l]{$#4$}
\Text(160, 90)[l]{$#5$}
\Text( 70, 70)[c]{${}^0$}
\Text( 55, 57)[l]{${}^1$}
\Text( 70, 45)[c]{${}^2$}
\Text( 85, 57)[r]{${}^3$}
\end{picture}
} }}

\def\draftdate{\relax}
\def\mda{\relax}
\def\mua{\relax}
\def\mla{\relax}
\def\draft{
\def\thtystars{******************************}
\def\sixtystars{\thtystars\thtystars}
\typeout{}
\typeout{\sixtystars**}
\typeout{* Draft mode!
         For final version remove \protect\draft\space in source file *}
\typeout{\sixtystars**}
\typeout{}
\def\draftdate{\today}
\def\mua{\marginpar[\boldmath\hfil$\uparrow$]%
                   {\boldmath$\uparrow$\hfil}%
                    \typeout{marginpar: $\uparrow$}\ignorespaces}
\def\mda{\marginpar[\boldmath\hfil$\downarrow$]%
                   {\boldmath$\downarrow$\hfil}%
                    \typeout{marginpar: $\downarrow$}\ignorespaces}
\def\mla{\marginpar[\boldmath\hfil$\rightarrow$]%
                   {\boldmath$\leftarrow $\hfil}%
                    \typeout{marginpar: $\leftrightarrow$}\ignorespaces}
\def\Mua{\marginpar[\boldmath\hfil$\Uparrow$]%
                   {\boldmath$\Uparrow$\hfil}%
                    \typeout{marginpar: $\uparrow$}\ignorespaces}
\def\Mda{\marginpar[\boldmath\hfil$\Downarrow$]%
                   {\boldmath$\Downarrow$\hfil}%
                    \typeout{marginpar: $\downarrow$}\ignorespaces}
\def\Mla{\marginpar[\boldmath\hfil$\Rightarrow$]%
                   {\boldmath$\Leftarrow $\hfil}%
                    \typeout{marginpar: $\leftrightarrow$}\ignorespaces}
\overfullrule 5pt
\oddsidemargin -15mm
\marginparwidth 29mm
}

\def\stars{\strut\leaders\hbox{*}\hfill\strut}
\def\starline{\hfil\strut\hfil\hbox to \textwidth {\stars}\hfil}

\begin{document}
\thispagestyle{empty}
\def\thefootnote{\fnsymbol{footnote}}
\setcounter{footnote}{1}
\null
\draftdate
\strut\hfill FR-PHENO-2010-020\\
\strut\hfill PSI-PR-10-10
\vfill
\begin{center}
{\large \bf\boldmath
Scalar one-loop 4-point integrals 
\par} \vskip 1em
\vspace{1cm}
{\large
{\sc A.\ Denner$^1$ and S.\ Dittmaier$^2$} }
\\[.8cm]
$^1$ {\it Paul Scherrer Institut, W\"urenlingen und Villigen\\
CH-5232 Villigen PSI, Switzerland} \\[0.5cm]
$^2$ {\it Albert-Ludwigs-Universit\"at Freiburg, Physikalisches Institut,
\\
D-79104 Freiburg, Germany}\\[0.5cm]
%$^2$ {\it Max-Planck-Institut f\"ur Physik 
%(Werner-Heisenberg-Institut) \\
%F\"ohringer Ring 6, 
%D-80805 M\"unchen, Germany} 
%\\[0.5cm]
\par 
\end{center}\par
\vfill \vskip 2.0cm {\bf Abstract:} \par 
We provide a complete set of results for the scalar 4-point function
appearing in one-loop calculations in QCD, QED, the electroweak
Standard Model and popular extensions thereof.  Complex internal
masses, which are needed for calculations involving unstable
particles, are supported throughout, whereas complex momenta are not
supported. In particular, for the most general, regular case we
present two independent results in terms of 72 and 32 dilogarithms.
In addition we list explicit results for all soft- and/or
collinear-singular cases in dimensional regularization, mass
regularization, and in regularizations of mixed type.  The exceptional
case with a vanishing modified Cayley determinant, which hardly
appears in applications, is not considered.
\par
\vskip 1cm
\noindent
May 2010 
\null
\setcounter{page}{0}
\clearpage
\def\thefootnote{\arabic{footnote}}
\setcounter{footnote}{0}

\section{Introduction}
\label{se:intro}

Experimental tests of the Standard Model of particle physics or its
extensions require precise theoretical predictions. These are to a
large extent obtained in perturbation theory. For an adequate
investigation of most processes at the LHC or future particle
colliders typically at least the next-to-leading order (NLO) in QCD
must be known, but for many processes also electroweak one-loop
corrections are required. Various interesting processes involve three
or more particles in the final state and internal resonances.

In recent years the particle theory community made a big effort to
calculate processes for the LHC at NLO. In this context, the
traditional approach of evaluating Feynman diagrams has been further
developed (see, e.g.,
\citere{delAguila:2004nf,Binoth:2005ff,Ellis:2005zh,Denner:2005nn} and
references therein),
and new methods, which abandon the use of individual diagrams, have
been proposed and applied
\cite{Ossola:2006us,Bern:2007dw,Ellis:2007br,Berger:2008sj,vanHameren:2009vq}.
Details on these developments and more references to methods and
results can, e.g., be found in \citere{Buttar:2006zd}.  Apart from
purely numerical approaches
\cite{Ferroglia:2002mz,Nagy:2003qn,Kurihara:2005ja,Catani:2008xa,Kilian:2009wy},
independent of the actual method, finally at the one-loop order all
results are expressed as linear combinations of
a set of master integrals, typically the standard scalar one-loop integrals.%
\footnote{Other methods that reduce to a different set of basic integrals also
  rely in general on numerical evaluation of the basic integrals
  \cite{Binoth:2002xh}.} In
four space--time dimensions~\cite{Me65,'tHooft:1978xw}, and in
dimensional regularization near four dimensions~\cite{Bern:1993kr},
only the scalar 1-point, 2-point, 3-point, and 4-point functions are
needed, since higher functions can be reduced to those (see references
in \citere{Buttar:2006zd} for different variants of these
decompositions).  For the calculation of electroweak corrections to
multiparticle processes these are required for the general case, \ie
with arbitrary internal masses and external momenta.  For processes
involving unstable particles these functions are also needed for
complex parameters. To this end, we support 
complex internal masses while keeping momentum variables real,
i.e.\ we consider the generalisation required by the
so-called ``complex-mass scheme''~\cite{Denner:2005fg},
which is frequently employed to describe one-loop corrections to
processes with particle resonances.

In \citere{'tHooft:1978xw} pioneering work in the calculation of
scalar one-loop integrals has been carried out by 't~Hooft and
Veltman.  For the 1-point, 2-point, and 3-point functions compact
explicit expressions are provided that are valid for complex internal
masses and arbitrary physical momenta. For the scalar 4-point function
with real masses a result in terms of 24 dilogarithms has been derived
in \citere{'tHooft:1978xw}, but for complex masses only a sketch for a
result in terms of 108 dilogarithms has been 
given there. This result has
been implemented into a {\sc FORTRAN} code recently
\cite{Nhung:2009pm}.  In \citere{Denner:1991qq} a result for the
general scalar 4-point function has been derived involving 16
dilogarithms, but this compact result is only valid for real masses.

Besides the regular scalar functions, in most calculations involving
massless or nearly massless particles special soft- or
collinear-singular cases appear. This is, in particular, the case for
NLO calculations in QCD, where usually dimensional regularization is
applied, or in QED and the electroweak Standard Model, where the masses
of the light leptons are often taken to be infinitesimally small.  For
the 2-point function, the singular limits are easy to extract, and
for the 3-point function the singular cases have been summarised in
\citere{Dittmaier:2003bc}. However, for the 4-point functions a complete set
exists only in dimensional regularization so far. The scalar 4-point function
with vanishing internal masses and no or one non-zero external mass
(in addition to the two non-vanishing Mandelstam variables) is provided in
\citere{Fabricius:1979tb,Papadopoulos:1981ju}. All cases with vanishing
internal masses are given in
\citeres{Bern:1993kr,Duplancic:2000sk,Duplancic:2002dh}. 
Finally, using also various other special cases spread in the literature,
all singular
cases relevant in dimensional regularization have been collected in 
\citere{Ellis:2007qk}. 

The purpose of this paper is to provide a complete set of formulas for
the scalar 4-point function valid for complex masses and in all
regular and soft- and/or collinear-singular cases both in dimensional
and in mass regularization.
In detail, we provide two independent results for the scalar 4-point
function with arbitrary complex squared masses (with negative
imaginary parts), which are both valid in all kinematical regions
relevant for physical processes.  One of these results is obtained
similar to the method of 't~Hooft and Veltman~\cite{'tHooft:1978xw}
and involves 72 dilogarithms, the other result follows the strategy of
\citere{Denner:1991qq}, leading to only 32 dilogarithms.  For several
specific choices of complex masses and kinematical invariants we have
numerically compared our results to the ones obtained with the code
{\D0C}~\cite{Nhung:2009pm}, which is based on 108 dilogarithms, and
have found agreement.  Moreover, we provide complete analytic
continuations of the results of \citere{Denner:1991qq} for the cases
involving one or more zero masses.  Additionally we summarise results
for the scalar 4-point function for soft- and collinear-singular cases
in dimensional regularization, in mass regularization, and in
regularizations of mixed type. As only restriction we require that
particles with infinitesimally small masses do not change their mass
when emitting a massless particle. This is the case for all
applications in QCD, QED, the electroweak Standard Model and popular
extensions thereof.

We have derived the singular cases in two different ways.  The first
method is based on the approach of \citere{Dittmaier:2003bc}. Here, a
singular integral is explicitly calculated in a suitable
regularization scheme and then translated to all other regularization
schemes by adding and subtracting appropriate 3-point functions. As an
independent check, many integrals have been directly
evaluated in addition, others have been taken from the literature. 
Specifically, those that do not require dimensional
regularization have been derived as limiting cases using the methods
and results of \citere{Denner:1991qq}, those that require dimensional
regularization have been taken from \citere{Ellis:2007qk} or from other
papers, or derived
as special cases therefrom. We have taken care to provide concise
results that are valid in all relevant regions of phase space.

However, we do not provide results for exceptional phase-space points,
where the denominator in front of the dilogarithms and logarithms
vanishes. These cases do only rarely appear in practical calculations,
and can often be avoided by introducing a finite width for unstable
particles (for examples see \citeres{Denner:1996ug,Boudjema:2008zn}).
Either they correspond to Landau singularities, where
the scalar 4-point function is singular, or to exceptional non-singular
cases that practically do not show up in calculations.  If needed,
they can be derived as limiting cases of our results.
We also do not provide results for complex momentum variables that are
required for unstable external particles \cite{Passarino:2010qk}.

The paper is organised as follows: In \refse{se:conventions} we summarise
our conventions and notation. Results for the general non-singular
case with complex masses and for the corresponding massless cases are
derived in \refse{se:regints}. Results for all soft- and/or
collinear-singular cases are listed in \refse{se:singints}, and
Section~\ref{se:sum} contains a short summary. 

\section{Conventions and notation}
\label{se:conventions}

The one-loop scalar 4-point integral is defined in $D$ space-time
dimensions as 
\beqar\label{D0def}
\DOp{p_{10}^2}{p_{21}^2}{p_{32}^2}{p_{30}^2}%
{p_{20}^2}{p_{31}^2}% 
\DOm{m_0^2}{m_1^2}{m_2^2}{m_3^2}{} \qquad
&\equiv& \; D_0(p_1,p_2,p_3,m^2_0,m^2_1,m^2_2,m^2_3)
\nn\\[-3.5em]
&\equiv& \; D_0(p_{10}^2,p_{21}^2,p_{32}^2,p_{30}^2,p_{20}^2,p_{31}^2,
m^2_0,m^2_1,m^2_2,m^2_3)
\nn\\[6ex]
&& \quad {}
\hspace{-15em}
= \frac{(2\pi\mu)^{4-D}}{\ri\pi^2} \int \rd^Dq \,
\frac{1}{(q^2-m_0^2)[(q+p_1)^2-m_1^2][(q+p_2)^2-m_2^2][(q+p_3)^2-m_3^2]}
\nl
\eeqar
with $p_0=0$, $p_{ij}=p_i-p_j$, and $\mu$ is the mass scale of
dimensional regularization. The internal squared masses $m_i^2$ are
allowed to be complex with phases between $0$ and $-\pi/2$.  If not
stated otherwise, real squared masses are assumed to have an
infinitesimally small negative imaginary part. If $m_i$ appears in the
following expressions it is
defined as the square root of $m_i^2$ with positive real part. The
external momenta are real, but over-lined squared momenta are
understood to receive an infinitesimally small positive imaginary
part, \ie $\bar p_{ij}^2=p_{ij}^2+\ri0$.  Note that for $D\ne4$ our
normalisation deviates from the one used in
\citeres{Bern:1993kr,Ellis:2007qk}.

After introducing Feynman parameters and performing the momentum
integration \refeq{D0def} can be written as
\beqar\label{D0fpint}
D_0 &=& \Gamma(1+\eps)(4\pi\mu^2)^\eps (1+\eps)
\nl&&\times
\int_0^1\rd x_1\,\int_0^{1-x_1}\rd x_2\,\int_0^{1-x_1-x_2}\rd x_3\,
\frac{1}{[P(1-x_1-x_2-x_3,x_1,x_2,x_3)]^{2+\eps}},
\eeqar
where $D=4-2\eps$ and
the integral extends over the 3-dimensional unit simplex.
The polynomial $P$, which is homogeneous of degree 2 in the four variables 
$x_0$, \dots, $x_3$, is given by
\beq
P(x_0,x_1,x_2,x_3) = \sum_{i=0}^3 m_i^2 x_i^2
+ \sum_{i,j=0 \atop i<j}^3 Y_{ij} x_i x_j
\label{eq:P}
\eeq
with the modified Cayley matrix $Y$ defined by
\beq\label{defY}
Y_{ij} = Y_{ji} = m_i^2 + m_j^2 - p_{ij}^2.
%, \quad p_{ij}=p_i-p_j, \quad p_0=0.
\eeq
The representation \refeq{D0fpint} shows the symmetry of the
4-point function resulting from permutations of the four internal
lines. This symmetry is directly realized in permutations of the
indices $\{0,1,2,3\}$ of the squared masses and momenta.

In the results, the roots of quadratic equations
\beq\label{rijdef}
0=m_i^2+Y_{ij} x+m_j^2 x^2            
=m_j^2\left(x+\frac{1}{r_{ij,1}}\right)\left(x+\frac{1}{r_{ij,2}}\right)
=m_i^2\left(1+r_{ij,1}x\right)\left(1+r_{ij,2}x\right)
\eeq
appear which we denote as $-1/r_{ij,1}$ and $-1/r_{ij,2}$.
It is easy to show that for real momenta and squared masses with phases
between $0$ and $-\pi/2$ the variables $r_{ij,k}$ vary only
  on the first Riemann sheet, \ie they cannot become real and negative.
For real masses this is ensured by the infinitesimal imaginary parts.
The sign of the infinitesimal imaginary part of $r_{ij,k}$ resulting
from the infinitesimal imaginary parts of the masses can be
determined as
\beq
\sgn(\Im r_{ij,k}) = \sgn\left( m_i^2 r_{ij,k}-\frac{m_j^2}{ r_{ij,k}} \right)
\quad \mbox{for $\Re(r_{ij,k})<0$ and $\Im(r_{ij,k})={}$infinitesimal}.
\eeq
For negative $r_{ij,k}$, \ie where the infinitesimal imaginary part matters,
an infinitesimal imaginary part of $\bar p_{ij}^2$ yields the same
result for $\sgn(\Im r_{ij,k})$.

Instead of using the variables $r_{ij,k}$, we sometimes employ
the shorthand
\beq\label{xijdef}
x_{ij} = \frac{\sqrt{1-4m_i m_j/\Big[\bar p_{ij}^2-(m_i-m_j)^2\Big]}-1}
{\sqrt{1-4m_i m_j/\Big[\bar p_{ij}^2-(m_i-m_j)^2\Big]}+1},
\label{eq:xij}
\eeq
which is---ignoring the infinitesimal imaginary part for a 
moment---related to the roots of \refeq{rijdef} via
$r_{ij,1}=({m_j}/{m_i})x_{ij}$ and 
$r_{ij,2}=({m_j}/{m_i})x_{ij}^{-1}$.

The 4-point function is evaluated in terms of logarithms and
dilogarithms.  As usual we choose the principal value of the logarithm
such that the cut lies along the negative real axis. Hence one finds
for $a$, $b$, and $ab$ not on the negative real axis
\begin{equation}
\ln(ab) -\ln(a) -\ln(b) = \eta (a,b) \label{eta}
\end{equation}
with
\beq
\eta (a,b) = 2 \pi \ri \Bigl\{ \theta (-\Im a)\, \theta (- \Im b) \,\theta 
                ( \Im (ab))    
-  \theta ( \Im  a) \,\theta ( \Im b) \,\theta (- \Im (ab) ) \Bigr\} .  
\eeq
The principal value of the dilogarithms for $(1-z)$ not real and
negative is given by
\begin{equation}
\Li(z) = - \int_{0}^{1}\frac{\rd t}{t}\, {\ln (1-zt)} \, . 
\end{equation}
More details and various useful relations for the dilogarithm can be found
in \citere{Lewin:1989}.

Most of our results involve only functions of the type 
\beq
\ln\left(\prod_{i=1}^n x_i\right), \qquad 
\Li\left(1-\prod_{i=1}^n x_i\right),
\eeq
where the variables $x_i$ vary only on the first Riemann sheet, \ie
$-\pi<\arg(x_i)<\pi$. Once these results have been derived in the 
Euclidean region and for real masses they can be easily analytically
continued by the substitutions \cite{Beenakker:1990jr}
\beqar
\ln\left(\prod_{i=1}^n x_i\right)&\to& \sum_{i=1}^n
\ln\left(x_i\right),\\
\label{eq:licont}
\Li\left(1-\prod_{i=1}^n x_i\right) &\to& \cLi(x_1,\ldots,x_n) \\
&\equiv& 
\Li\left(1-\prod_{i=1}^n x_i\right) 
+\left[\ln\left(\prod_{i=1}^n x_i\right)
-\sum_{i=1}^n \ln\left(x_i\right)\right]
\Biggr[\ln\left(1-\prod_{i=1}^n x_i\right)\nl
&&{}-\theta\left(\left|\prod_{i=1}^n x_i\right|-1\right)
\left(\ln\left(-\prod_{i=1}^n x_i\right)
-\frac{1}{2}\ln\left(\prod_{i=1}^n x_i\right)
-\frac{1}{2}\sum_{i=1}^n \ln\left(x_i\right)\right)\Biggr],
\nn
\eeqar
where $\theta$ is the Heaviside step function.  In most of our results
we need only the analytically continued dilogarithm for  two variables
$x_1$, $x_2$. This simplifies to
\beq
\cLi(x_1,x_2) = \Li\left(1-x_1 x_2\right) 
+\eta(x_1,x_2) \ln\left(1-x_1 x_2\right),
\eeq
because the term proportional to the $\theta$-function in
\refeq{eq:licont} does not contribute for $n=2$ if the variables $x_i$
stay on the first Riemann sheet.

\section{Regular 4-point integrals}
\label{se:regints}

\subsection{4-point integral for general complex masses---method
similar to 't~Hooft's and Veltman's}

In this section we evaluate the general scalar 4-point integral $D_0$
with complex internal masses $m_i$ widely following the method of
\citere{'tHooft:1978xw}. As stated above, we implicitly assume that
the mass squares $m_i^2$ possess either a non-zero negative imaginary
part or an infinitesimally small imaginary part that is negative.
Moreover, in this section we use the fact the the real part of $m_i^2$
is non-negative (a restriction that is relaxed in the next section).
The integral is both UV and IR finite, so that we set $D=4$ from the
beginning, and \refeq{D0fpint} simplifies to
\beq\label{D0genint}
D_0 = \int_0^1\rd x_1\,\int_0^{1-x_1}\rd x_2\,\int_0^{1-x_1-x_2}\rd x_3\,
\frac{1}{[P(1-x_1-x_2-x_3,x_1,x_2,x_3)]^2}
\eeq
with the polynomial $P$ as defined in \refeq{eq:P}.
For later convenience we define the polynomials
\beqar
\label{defPk}
P_k(x_0,x_1,x_2,x_3) &\equiv&
\frac{\partial}{\partial x_k}P(x_0,x_1,x_2,x_3)
= 2m_k^2 x_k + \sum_{i=0 \atop i\ne k}^3 Y_{ik} x_i
= \sum_{i=0}^3 Y_{ik} x_i, 
\nn\\
&&\qquad k=0,\dots,3,\nl
P_a(x_0,x_1,x_2,x_3;\al) &\equiv&
\Bigl[-(1+\al)P_0+\al P_2+P_3\Bigr](x_0,x_1,x_2,x_3),
\nn\\
P_b(x_0,x_1,x_2,x_3;\be) &\equiv&
\Bigl[-(1+\be)P_0+P_1+\be P_2\Bigr](x_0,x_1,x_2,x_3),
\eeqar
which are linear and homogeneous in the $x_i$,
and the real-valued function
\beqar
P_{ab}(\al,\be) &=& 2(1+\al)(1+\be)m_0^2 -(1+\al)Y_{01}
-(\al+\be+2\al\be)Y_{02} -(1+\be)Y_{03}
\nn\\
&& {}
+\al Y_{12} +Y_{13} +2\al\be m_2^2 +\be Y_{23}
\nn\\
&=& (1+\al)p^2_{10} +(\al+\be+2\al\be)p^2_{20} +(1+\be)p^2_{30}
-\al p^2_{21} -p^2_{31} -\be p^2_{32},
\eeqar
with $\al$ and $\be$ being real parameters.

Before performing the first integration, we rescale the integration parameters
according to
\beq
x_2 = (1-x_1)\xi_2, \quad
x_3 = (1-x_1)\xi_3,
\eeq
so that the non-trivial part of the
integration boundary simplifies to the 2-dimensional unit simplex,
\beq
D_0 = \int_0^1\rd x_1\,\int_0^1\rd \xi_2\,\int_0^{1-\xi_2}\rd \xi_3\,
\frac{(1-x_1)^2}{
\Bigl[P\Bigl((1-x_1)(1-\xi_2-\xi_3),x_1,(1-x_1)\xi_2,(1-x_1)\xi_3\Bigr)\Bigr]^2}\,.
\eeq
Now we perform the ``Euler shift'' 
\beq
\xi_2 = y_2 + \al \xi_3
\eeq
in order to substitute variable $\xi_2$ by $y_2$. The parameter $\al$
is chosen to obey the condition
\beq
0 = p_{30}^2  + \al(p_{30}^2-p_{32}^2+p_{20}^2) + \al^2 p_{20}^2,
\eeq
which renders to coefficient of $\xi_3^2$ in $P$ vanishing.  At this
point it is important to realize that the parameter $\al$ is a real
quantity for all possible momentum configurations occurring in any
scattering or decay (sub)process. For the next intermediate steps we
assume $\al>0$ and analytically continue the final result to arbitrary
real $\al$ later. For $\al>0$ the integration over $\xi_2$ and $\xi_3$
can be written as
\beqar
\int_0^1\rd \xi_2\,\int_0^{1-\xi_2}\rd \xi_3 &=& 
\int_0^1\rd \xi_3\,\int_{-\al \xi_3}^{1-(1+\al)\xi_3}\rd y_2 
\nn\\
&=&
\int_0^{-\al}\rd y_2\,\int_0^{-y_2/\al}\rd \xi_3
+\int_{-\al}^1\rd y_2\,\int_0^{(1-y_2)/(1+\al)}\rd \xi_3.
\label{eq:intx2x3}
\eeqar
Exploiting the linearity of $P$ in $\xi_3$, the $\xi_3$ dependence
can be isolated,
\beq
P\Bigl((1-x_1)(1-y_2-(1+\al)\xi_3),x_1,(1-x_1)(y_2+\al \xi_3),(1-x_1)\xi_3\Bigr) 
= (1-x_1)\xi_3A+B,
\eeq
with
\beqar
A &\equiv& A(x_1,y_2) = P_a\Bigl((1-x_1)(1-y_2),x_1,(1-x_1)y_2,0;\al\Bigr),
\nn\\
B &\equiv& B(x_1,y_2) = P\Bigl((1-x_1)(1-y_2),x_1,(1-x_1)y_2,0\Bigr).
\eeqar
The $\xi_3$-integration is carried out as follows,
\beqar
D_0 &=&
\int_0^1\rd x_1\,\left\{
\int_0^{-\al}\rd y_2\,\int_0^{-y_2/\al}\rd \xi_3
+\int_{-\al}^1\rd y_2\,\int_0^{(1-y_2)/(1+\al)}\rd \xi_3
\right\} \frac{(1-x_1)^2}{[(1-x_1)\xi_3A+B]^2} 
%\nn\\
%&=&
%\int_0^1\rd x_1\,
%\int_0^{-\al}\rd y_2\, (1-x_1)
%\left(\frac{1}{AB}-\frac{\al}{A(-A(1-x_1)y_2+B\al)}\right) 
%\nn\\
%&& {}
%+ \int_0^1\rd x_1\,\int_{-\al}^1\rd y_2\, (1-x_1)
%\left( \frac{1}{AB}-\frac{1+\al}{A[A(1-x_1)(1-y_2)+B(1+\al)]} \right)
\nn\\
&=&
-\int_0^1\rd x_1\,\int_0^{-\al}\rd y_2\, 
\frac{\al(1-x_1)}{A[-(1-x_1)y_2A+\al B]}
\nn\\
&& {}
-\int_0^1\rd x_1\,\int_{-\al}^1\rd y_2\, 
\frac{(1+\al)(1-x_1)}{A[(1-x_1)(1-y_2)A+(1+\al)B]} 
+\int_0^1\rd x_1\,\int_0^1\rd y_2\, \frac{1-x_1}{AB}.
\nn\\
\label{eq:ABints}
\eeqar
In the last integral we combined the two contributions resulting from
the boundary $\xi_3=0$ to a single integral using
$\int_0^{-\al}\rd y_2+\int_{-\al}^1\rd y_2=\int_0^1\rd y_2$.
We perform the following three different substitutions in the three
integrals of \refeq{eq:ABints}, respectively,
\beq
y_2 = -\al+\frac{\al \eta_2}{1-x_1}, \quad
y_2 = 1-\frac{(1+\al)\eta_2}{1-x_1}, \quad
y_2 = \frac{\eta_2}{1-x_1}
\label{eq:y2trafo}
\eeq
and introduce the shorthands $A_i$ and $B_i$ for the functions
of $x_1$ and $\eta_2$ that result from $A(x_1,y_2)$ and $B(x_1,y_2)$
upon making the three substitutions \refeq{eq:y2trafo} that are
successively labelled by $i=1,2,3$.
The functions $A_i$ and $B_i$ are polynomials in $x_1$ and $\eta_2$,
in particular, the $A_i$ are of linear and the $B_i$ of quadratic degree.
This is explicit after expressing them in terms of $P$,
\beqar
A_1 &\equiv& A_1(x_1,\eta_2) = 
P_a\Bigl( (1+\al)(1-x_1)-\al \eta_2, x_1, -\al(1-x_1-\eta_2), 0; \al \Bigr),
\nn\\
A_2 &\equiv& A_2(x_1,\eta_2) = 
P_a\Bigl( (1+\al)\eta_2, x_1, 1-x_1-(1+\al)\eta_2, 0; \al \Bigr),
\nn\\
A_3 &\equiv& A_3(x_1,\eta_2) = 
P_a\Bigl( 1-x_1-\eta_2, x_1, \eta_2, 0; \al \Bigr),
\nn\\
B_1 &\equiv& B_1(x_1,\eta_2) = 
P\Bigl( (1+\al)(1-x_1)-\al \eta_2, x_1, -\al(1-x_1-\eta_2), 0 \Bigr),
\nn\\
B_2 &\equiv& B_2(x_1,\eta_2) = 
P\Bigl( (1+\al)\eta_2, x_1, 1-x_1-(1+\al)\eta_2, 0 \Bigr),
\nn\\
B_3 &\equiv& B_3(x_1,\eta_2) = 
P\Bigl( 1-x_1-\eta_2, x_1, \eta_2, 0 \Bigr).
\eeqar
The $D_0$ integral becomes
\beq
D_0 =
\int_0^1\rd x_1\,
\int_0^{1-x_1}\rd \eta_2\, 
\left\{
\frac{\al}{A_1[(1-x_1-\eta_2)A_1+B_1]}
-\frac{1+\al}{A_2[\eta_2A_2+B_2]} 
+\frac{1}{A_3B_3}
\right\}.
\label{eq:intx1y2}
\eeq
In this representation it is obvious that the contributions
involving $A_1$, $B_1$ or $A_2$, $B_2$ are zero in the special
cases $\al=0$ or $\al=-1$, respectively.
In order to eliminate the $x_1^2$ terms in the factors containing $B_i$
in \refeq{eq:intx1y2}, we again introduce Euler shifts
\beq
\eta_2 = z+\be_i x_1
\eeq
with three different variables $\be_i$ which correspond to the parts
containing $A_i$ and $B_i$.  The following conditions render the
above-mentioned coefficients of $x_1^2$ zero, but at the same time
lead to real-valued values of $\be_i$ (for physical decay and
scattering processes),
\beqar\label{defbetai}
0 &=&
p_{31}^2 + \be_1 (p_{31}^2 - p_{10}^2 + p_{30}^2) + \be_1^2 p_{30}^2,
\nn\\
0 &=&
p_{21}^2 + \be_2 (p_{21}^2 - p_{31}^2 + p_{32}^2) + \be_2^2 p_{32}^2,
\nn\\
0 &=&
p_{10}^2 + \be_3 (p_{10}^2 - p_{21}^2 + p_{20}^2) + \be_3^2 p_{20}^2.
\eeqar
For intermediate steps we again impose the condition $\be_i>0$ and
perform analytical continuations to arbitrary real $\be_i$ at the end. 
Having eliminated the $x_1^2$ terms in the $B_i$ parts of
\refeq{eq:intx1y2}, each of the three terms in curly brackets can
be decomposed via partial fractioning in $x_1$, so that only 
terms of the form $1/(ax_1+b)$ with $x_1$-independent terms $a$ and $b$
result. To make this step explicit, we isolate the $x_1$ terms in
$A_i$ and $B_i$,
\beqar
A_i(x_1,z+\be_ix_1) &=& A_i^{(0)}(z) + x_1 A_i^{(1)}(z), \nl
B_i(x_1,z+\be_ix_1) &=& B_i^{(0)}(z) + x_1 B_i^{(1)}(z)+ x_1^2 B_i^{(2)}(z).
\eeqar
The arguments of $ A_i^{(k)}$ and $B_i^{(k)}$ are suppressed in the
following.
The coefficients $B_i^{(2)}$ are fixed by the above elimination of $x_1^2$ 
terms,
\beq
B_1^{(2)} = (1+\be_1) A_1^{(1)}, \quad
B_2^{(2)} = -\be_2 A_2^{(1)}, \quad
B_3^{(2)} = 0,
\eeq
and drop out in the subsequent manipulations.
The other coefficients read 
\beqar
A_1^{(0)} &=& 
P_a\Bigl( 1+\al(1-z), 0, -\al(1-z), 0; \al \Bigr),
\qquad
A_1^{(1)} =
P_{ab}\Bigl(\al,\al(1+\be_1)\Bigr),
\nn\\
B_1^{(0)} &=& 
P\Bigl( 1+\al(1-z), 0, -\al(1-z), 0 \Bigr),
\nn\\
B_1^{(1)} &=& 
P_b\Bigl( 1+\al(1-z), 0, -\al(1-z), 0; \al(1+\be_1) \Bigr),
\nn\\[1em]
A_2^{(0)} &=& 
P_a\Bigl( (1+\al)z, 0, 1-(1+\al)z, 0; \al \Bigr),
\qquad
A_2^{(1)} = 
P_{ab}\Bigl(\al,-1-(1+\al)\be_2\Bigr),
\nn\\
B_2^{(0)} &=& 
P\Bigl( (1+\al)z, 0, 1-(1+\al)z, 0 \Bigr),
\nn\\
B_2^{(1)} &=& 
P_b\Bigl( (1+\al)z, 0, 1-(1+\al)z, 0; -1-(1+\al)\be_2 \Bigr),
\nn\\[1em]
A_3^{(0)} &=& 
P_a( 1-z, 0, z, 0; \al ),
\qquad
A_3^{(1)} = 
P_{ab}(\al,\be_3),
\nn\\
B_3^{(0)} &=& 
P( 1-z, 0, z, 0 ),
\qquad
B_3^{(1)} = 
P_b( 1-z, 0, z, 0; \be_3 ).
\eeqar
The partial fractioning of the integrand in \refeq{eq:intx1y2} 
with respect to its $x_1$ dependence yields
\beqar
&&
\frac{1}{C_1(z)} \Biggl[
\Biggl(x_1+\frac{A_1^{(0)}}{A_1^{(1)}}\Biggr)^{-1}
-\Biggl(x_1+\frac{(1-z)A_1^{(0)}+B_1^{(0)}}
{-(1+\be_1)A_1^{(0)}+(1-z)A_1^{(1)}+B_1^{(1)}}\Biggr)^{-1}
\Biggr]
\nn\\
&& {}
+\frac{1}{C_2(z)} \Biggl[
\Biggl(x_1+\frac{A_2^{(0)}}{A_2^{(1)}}\Biggr)^{-1}
-\Biggl(x_1+\frac{B_2^{(0)}+zA_2^{(0)}}
{\be_2 A_2^{(0)}+B_2^{(1)}+zA_2^{(1)}}\Biggr)^{-1}
\Biggr]
\nn\\
&& {}
+\frac{1}{C_3(z)} \Biggl[
\Biggl(x_1+\frac{A_3^{(0)}}{A_3^{(1)}}\Biggr)^{-1}
-\Biggl(x_1+\frac{B_3^{(0)}}{B_3^{(1)}}\Biggr)^{-1}
\Biggr]
\eeqar
with
\beqar
C_1(z) &=& \frac{1}{\al} \Bigl[
A_1^{(1)}B_1^{(0)} -A_1^{(0)}B_1^{(1)} +(A_1^{(0)})^2(1+\be_1)
\Bigr],
\nn\\
C_2(z) &=& \frac{1}{1+\al} \Bigl[
-A_2^{(1)} B_2^{(0)} + A_2^{(0)} B_2^{(1)} +(A_2^{(0)})^2\be_2 
\Bigr],
\nn\\
C_3(z) &=& A_3^{(1)} B_3^{(0)}-A_3^{(0)} B_3^{(1)},
\eeqar
which are quadratic polynomials in $z$.
In order to carry out the $x_1$-integration we manipulate
to double integral analogously to \refeq{eq:intx2x3},
\beqar
\int_0^1\rd x_1\,\int_0^{1-x_1}\rd \eta_2 &=& 
\int_0^1\rd x_1\,\int_{-\be_i x_1}^{1-(1+\be_i)x_1}\rd z
\nn\\
&=&
\int_0^{-\be_i}\rd z\,\int_0^{-z/\be_i}\rd x_1
+\int_{-\be_i}^1\rd z\,\int_0^{(1-z)/(1+\be_i)}\rd x_1.
\eeqar
The result of the integration is
\beqar
D_0 &=& \int_0^{-\be_1}\rd z\, \frac{1}{C_1(z)} \Biggl\{
\ln A_1\biggl(-\frac{z}{\be_1},0\biggr)
-\ln\biggl[\biggl(1+\frac{z}{\be_1}\biggr)
A_1\biggl(-\frac{z}{\be_1},0\biggr)+B_1\biggl(-\frac{z}{\be_1},0\biggr) \biggr]
\Biggr\}
\nn\\
&& {}
+ \int_0^{-\be_2}\rd z\, \frac{1}{C_2(z)} \Biggl\{
\ln A_2\biggl(-\frac{z}{\be_2},0\biggr)
-\ln B_2\biggl(-\frac{z}{\be_2},0\biggr) \Biggr\}
\nn\\
&& {}
+ \int_0^{-\be_3}\rd z\, \frac{1}{C_3(z)} \Biggl\{
\ln A_3\biggl(-\frac{z}{\be_3},0\biggr)
-\ln B_3\biggl(-\frac{z}{\be_3},0\biggr) \Biggr\}
\nn\\
&& {}
+\int_{-\be_1}^1\rd z\, \frac{1}{C_1(z)} \Biggl\{
\ln A_1\biggl(\frac{1-z}{1+\be_1},\frac{z+\be_1}{1+\be_1}\biggr)
-\ln B_1\biggl(\frac{1-z}{1+\be_1},\frac{z+\be_1}{1+\be_1}\biggr) \Biggr\}
\nn\\
&& {}
+\int_{-\be_2}^1\rd z\, \frac{1}{C_2(z)} \Biggl\{
\ln A_2\biggl(\frac{1-z}{1+\be_2},\frac{z+\be_2}{1+\be_2}\biggr)
\nn\\
&& \quad {}
-\ln\biggl[\frac{z+\be_2}{1+\be_2}
A_2\biggl(\frac{1-z}{1+\be_2},\frac{z+\be_2}{1+\be_2}\biggr)
+B_2\biggl(\frac{1-z}{1+\be_2},\frac{z+\be_2}{1+\be_2}\biggr) \biggr] \Biggr\}
\nn\\
&& {}
+\int_{-\be_3}^1\rd z\, \frac{1}{C_3(z)} \Biggl\{
\ln A_3\biggl(\frac{1-z}{1+\be_3},\frac{z+\be_3}{1+\be_3}\biggr)
-\ln B_3\biggl(\frac{1-z}{1+\be_3},\frac{z+\be_3}{1+\be_3}\biggr) \Biggr\}
\nn\\
&& {}
-\int_0^1\rd z\, \frac{1}{C_1(z)} \Bigl\{
\ln A_1(0,z) -\ln\Bigl[(1-z)A_1(0,z)+B_1(0,z)\Bigr]  \Bigr\}
\nn\\
&& {}
-\int_0^1\rd z\, \frac{1}{C_2(z)} \Bigl\{
\ln A_2(0,z) -\ln\Bigl[z A_2(0,z)+B_2(0,z)\Bigr] \Bigr\}
\nn\\
&& {}
-\int_0^1\rd z\, \frac{1}{C_3(z)} \Bigl\{
\ln A_3(0,z) -\ln B_3(0,z) \Bigr\}.
\label{eq:D0zint}
\eeqar
Organising the logarithms in this way, of course, requires some care
concerning their cut structures as functions of $z$.
For instance, we have evaluated the integrals containing only $A_i$ functions
as follows,
\beqar
\int_{r_0}^{r_1}\rd x_1\,
\Biggl(x_1+\frac{A_i^{(0)}}{A_i^{(1)}}\Biggr)^{-1}
&=&
\ln\Biggl(r_1+\frac{A_i^{(0)}}{A_i^{(1)}}\Biggr)
-\ln\Biggl(r_0+\frac{A_i^{(0)}}{A_i^{(1)}}\Biggr)
\nn\\
&=&
\ln\Bigl(A_i^{(1)} r_1+A_i^{(0)}\Bigr)
-\ln\Bigl(A_i^{(1)} r_0+A_i^{(0)}\Bigr)
\nn\\
&=&
\ln A_i(r_1,z+\be_i r_1) - \ln A_i(r_0,z+\be_i r_0),
\eeqar
justified by the fact that all $r_i$ and $A_i^{(1)}$ are real valued,
so that the arguments of the two logarithms in the differences have
the same imaginary part.  Similar manipulations are possible for the
integrals containing $B_i$ functions after observing that the
imaginary parts of $A_i^{(0)}$ and $B_i^{(1)}$ are $z$-independent and
that the signs of the imaginary parts in the resulting logarithms do
not change with varying $z$ (i.e.\ no branch cuts are crossed).
Moreover, it can be easily checked that in \refeq{eq:D0zint} the
residues of the zeros of the functions $C_i(z)$ add up to zero.  This
fact would be useful if we proceeded with the final integration as
suggested in \citere{'tHooft:1978xw}; however, we go a somewhat
different way and exploit this fact at the very end in the analytical
continuation in $\al$ and $\be_i$.

For later convenience,
we factorize the quadratic polynomials $C_i(z)$ as follows,
\beq
C_i(z) = a_i z^2+b_i z+c_i = a_i(z-z_{i,1})(z-z_{i,2}),
\eeq
i.e.\ $z_{i,1}$ and $z_{i,2}$ are the solutions of the quadratic
equations $C_i(z)=0$. The constants $a_i$, $b_i$, and $c_i$ are
explicitly given by
\beqar
a_1 &=& \al\Bigl[ p_{20}^2 P_{ab}\Bigl(\al,\al(1+\beta_1)\Bigr)
            -P_a(-1,0,1,0;\al) P_b\Bigl(-1,0,1,0;\al(1+\beta_1)\Bigr)
\nn\\
&& \quad {}
         +(1+\beta_1) P_a(-1,0,1,0;\al)^2 \Bigr],
\nn\\
b_1 &=& (Y_{20}-2m_0^2-2\al p_{20}^2) P_{ab}\Bigl(\al,\al(1+\beta_1)\Bigr) 
       -P_a(1+\al,0,-\al,0;\al) 
\nn\\
&& \quad {} 
       \times P_b\Bigl(-1,0,1,0;\al(1+\beta_1)\Bigr)
       -P_a(-1,0,1,0;\al) P_b\Bigl(1+\al,0,-\al,0;\al(1+\beta_1)\Bigr)
\nn\\
&& \quad {}
       +2(1+\beta_1) P_a(-1,0,1,0;\al) P_a(1+\al,0,-\al,0;\al),
\nn\\
c_1 &=& \frac{1}{\al} \Bigl[ 
        \Bigl( (1+\al)Y_{30}-\al Y_{23}-m_3^2 \Bigr)
        P_{ab}\Bigl(\al,\al(1+\beta_1)\Bigr) 
         -P_a(1+\al,0,-\al,0;\al) 
\nn\\
&& \quad {} 
       \times P_b\Bigl(1+\al,0,-\al,0;\al(1+\beta_1)\Bigr)
       +(1+\beta_1) P_a(1+\al,0,-\al,0;\al)^2 \Bigr],
\nn\\
a_2 &=& (1+\al)\Bigl[ -p_{20}^2 P_{ab}\Bigl(\al,-1-(1+\al)\beta_2\Bigr)
       + P_a(1,0,-1,0;\al) 
\nn\\
&& \quad {}
        \times P_b\Bigl(1,0,-1,0;-1-(1+\al)\beta_2\Bigr)
       + P_a(1,0,-1,0;\al)^2 \beta_2 \Bigr],
\nn\\
b_2 &=& -(Y_{20}-2m_2^2) P_{ab}\Bigl(\al,-1-(1+\al)\beta_2\Bigr)
       +P_a(1,0,-1,0;\al) 
\nn\\
&& \quad {}
        \times P_b\Bigl(0,0,1,0;-1-(1+\al)\beta_2\Bigr)
       +P_a(0,0,1,0;\al) P_b\Bigl(1,0,-1,0;-1-(1+\al)\beta_2\Bigr)
\nn\\
&& \quad {}
       +2\beta_2 P_a(1,0,-1,0;\al) P_a(0,0,1,0;\al),
\nn\\
c_2 &=& \frac{1}{1+\al}\Bigl[
        -m_2^2 P_{ab}\Bigl(\al,-1-(1+\al)\beta_2\Bigr) 
        +P_a(0,0,1,0;\al) P_b\Bigl(0,0,1,0;-1-(1+\al)\beta_2\Bigr)
\nn\\
&& \quad {}
        + \beta_2 P_a(0,0,1,0,\al)^2 \Bigr],
\nn\\
a_3 &=& p_{20}^2 P_{ab}(\al,\beta_3)-P_a(-1,0,1,0;\al) P_b(-1,0,1,0;\beta_3),
\nn\\
b_3 &=& (Y_{20}-2m_0^2) P_{ab}(\al,\beta_3)
        -P_a(1,0,0,0;\al) P_b(-1,0,1,0;\beta_3)
\nn\\
&& \quad {}
        -P_a(-1,0,1,0;\al) P_b(1,0,0,0;\beta_3),
\nn\\
c_3 &=& m_0^2 P_{ab}(\al,\beta_3) -P_a(1,0,0,0;\al) P_b(1,0,0,0;\beta_3),
\eeqar
and obey the non-trivial relations
\beq
\det Y = b_i^2-4a_i c_i, \quad  i=1,2,3,
\eeq
so that we can define
\beq
z_{i,1} = \frac{-b_i+\sqrt{\det Y}}{2a_i}, \quad
z_{i,2} = \frac{-b_i-\sqrt{\det Y}}{2a_i}.
\eeq
In order to simplify the integrand of \refeq{eq:D0zint}, we perform
the substitutions 
\beq
\zeta=-\frac{z}{\be_i}, \quad
\zeta=\frac{1-z}{1+\be_i}, \quad
\zeta=z
\eeq
in the three different blocks of integrals
$\int_0^{-\be_i}\rd z$, $\int_{-\be_i}^1\rd z$, $\int_0^1\rd z$,
respectively, leading to a common range of integration over
$\zeta$, viz.\ $\int_0^1\rd \zeta$.
This transforms the $z_{i,j}$ into
\beq
\zeta'_{i,j}=-\frac{z_{i,j}}{\be_i}, \quad
\zeta''_{i,j}=\frac{1-z_{i,j}}{1+\be_i}, \quad
\zeta_{i,j}=z_{i,j},
\eeq
and integral \refeq{eq:D0zint} becomes
\beqar
\lefteqn{
D_0 = \frac{1}{\sqrt{\det Y}} \int_0^1\rd \zeta\, 
} &&
\nn\\
&& {} \times
\Biggl\{
\Bigl[(\zeta-\zeta'_{1,1})^{-1}-(\zeta-\zeta'_{1,2})^{-1}\Bigr]
\Bigl\{
\ln A_1(\zeta,0)-\ln\Bigl[(1-\zeta) A_1(\zeta,0)+B_1(\zeta,0)\Bigr]
\Bigr\}
\nn\\
&& \qquad {}
+ \Bigl[(\zeta-\zeta'_{2,1})^{-1}-(\zeta-\zeta'_{2,2})^{-1}\Bigr]
\Bigl\{
\ln A_2(\zeta,0)-\ln B_2(\zeta,0) \Bigr\}
\nn\\
&& \qquad {}
+ \Bigl[(\zeta-\zeta'_{3,1})^{-1}-(\zeta-\zeta'_{3,2})^{-1}\Bigr]
\Bigl\{
\ln A_3(\zeta,0)-\ln B_3(\zeta,0) \Bigr\}
\nn\\
&& \qquad {}
- \Bigl[(\zeta-\zeta''_{1,1})^{-1}-(\zeta-\zeta''_{1,2})^{-1}\Bigr]
\Bigl\{
\ln A_1(\zeta,1-\zeta)
-\ln B_1(\zeta,1-\zeta) \Bigr\}
\nn\\
&& \qquad {}
- \Bigl[(\zeta-\zeta''_{2,1})^{-1}-(\zeta-\zeta''_{2,2})^{-1}\Bigr]
\nn\\
&& \qquad\quad {} \times
\Bigl\{
\ln A_2(\zeta,1-\zeta)
-\ln\Bigl[(1-\zeta) A_2(\zeta,1-\zeta)+B_2(\zeta,1-\zeta) \Bigr] \Bigr\}
\nn\\
&& \qquad {}
- \Bigl[(\zeta-\zeta''_{3,1})^{-1}-(\zeta-\zeta''_{3,2})^{-1}\Bigr]
\Bigl\{
\ln A_3(\zeta,1-\zeta)-\ln B_3(\zeta,1-\zeta) \Bigr\}
\nn\\
&& \qquad {}
- \Bigl[(\zeta-\zeta_{1,1})^{-1}-(\zeta-\zeta_{1,2})^{-1}\Bigr]
\Bigl\{
\ln A_1(0,\zeta) -\ln\Bigl[(1-\zeta)A_1(0,\zeta)+B_1(0,\zeta)\Bigr]  \Bigr\}
\nn\\
&& \qquad {}
- \Bigl[(\zeta-\zeta_{2,1})^{-1}-(\zeta-\zeta_{2,2})^{-1}\Bigr]
\Bigl\{
\ln A_2(0,\zeta) -\ln\Bigl[\zeta A_2(0,\zeta)+B_2(0,\zeta)\Bigr] \Bigr\}
\nn\\
&& \qquad {}
- \Bigl[(\zeta-\zeta_{3,1})^{-1}-(\zeta-\zeta_{3,2})^{-1}\Bigr]
\Bigl\{
\ln A_3(0,\zeta) -\ln B_3(0,\zeta) \Bigr\}
\Biggr\}.
\eeqar
Finally, we perform the projective transformation,
\beq
x = \frac{\zeta}{1-\zeta},
\eeq
with
\beq
x'_{i,j} = \frac{\zeta'_{i,j}}{1-\zeta'_{i,j}} 
= -\frac{z_{i,j}}{z_{i,j}+\be_i},
\quad
x''_{i,j} = \frac{\zeta''_{i,j}}{1-\zeta''_{i,j}} 
= \frac{1-z_{i,j}}{\be_i+z_{i,j}},
\quad
x_{i,j} = \frac{\zeta_{i,j}}{1-\zeta_{i,j}} = \frac{z_{i,j}}{1-z_{i,j}},
\eeq
leading to
\beqar
D_0 &=& \frac{1}{\sqrt{\det Y}} \int_0^\infty\rd x\, \biggl\{
\Bigl[(x-x'_{1,1})^{-1}-(x-x'_{1,2})^{-1}\Bigr] l_{31}(x)
\nn\\
&& \qquad {}
+\Bigl[(x-x'_{2,1})^{-1}-(x-x'_{2,2})^{-1}\Bigr] l_{21}(x)
\nn\\
&& \qquad {}
+\Bigl[(x-x'_{3,1})^{-1}-(x-x'_{3,2})^{-1}\Bigr] l_{01}(x)
\nn\\
&& \qquad {}
-\Bigl[(x-x''_{1,1})^{-1}-(x-x''_{1,2})^{-1}\Bigr] l_{01}(x)
\nn\\
&& \qquad {}
-\Bigl[(x-x''_{2,1})^{-1}-(x-x''_{2,2})^{-1}\Bigr] l_{31}(x)
\nn\\
&& \qquad {}
-\Bigl[(x-x''_{3,1})^{-1}-(x-x''_{3,2})^{-1}\Bigr] l_{21}(x)
\nn\\
&& \qquad {}
-\Bigl[(x-x_{1,1})^{-1}-(x-x_{1,2})^{-1}\Bigr] l_{30}(x)
\nn\\
&& \qquad {}
-\Bigl[(x-x_{2,1})^{-1}-(x-x_{2,2})^{-1}\Bigr] l_{23}(x)
\nn\\
&& \qquad {}
-\Bigl[(x-x_{3,1})^{-1}-(x-x_{3,2})^{-1}\Bigr] l_{02}(x)
\biggr\},
\label{eq:D0x}
\eeqar
where we made use of the generic functions
\beqar
l_{ij}(x) &=& 
\ln(1+x)
+\ln\biggl(M^2_i + M^2_j x\biggr)
-\ln\Bigl(m_i^2+Y_{ij} x+m_j^2 x^2\Bigr)
\label{eq:lij}
\\
&=&
\ln M^2_i-\ln m_i^2+\ln(1+x)
+\ln\biggl(1 + \frac{M^2_j}{M^2_i} x\biggr)
-\ln(1+r_{ij,1}x)
-\ln(1+r_{ij,2}x),
\nn
\eeqar
which depend on the constants
\beqar
M^2 &=& -(1+\al)m_0^2 + \al m_2^2 + m_3^2,
\nn\\
M^2_0 &=& M^2 - p_{30}^2 - \al p_{20}^2,
\nn\\
M^2_1 &=& M^2 + (1+\al)p_{10}^2 - \al p_{21}^2 - p_{31}^2,
\nn\\
M^2_2 &=& M^2 + (1+\al)p_{20}^2 - p_{32}^2, 
\nn\\
M^2_3 &=& M^2 + (1+\al)p_{30}^2-\al p_{32}^2.
\eeqar
In the second relation of \refeq{eq:lij} we use the variables
$r_{ij,k}$ as defined in \refeq{rijdef}.
The validity of this relation
for all real positive $x$ becomes obvious after realizing that it is 
trivially fulfilled for $x=0$ and that none of the arguments of the
logarithms crosses a branch cut.
Note that the freedom to add an infinitesimally small negative imaginary
part to each mass $m_i^2$ should be exploited to render the imaginary
part of $M^2$ (and thus of all $M_i^2$) non-zero. This point is
important in the cases when some masses $m_i$ are equal, so that the
imaginary parts entering $M^2$ might compensate, which should be
avoided.  It should also be realized that the contributions involving
the variables $x'_{i,j}$ or $x''_{i,j}$ drop out in the special cases
$\be_i=0$ or $\be_i=-1$, respectively.

The integral representation \refeq{eq:D0x}, together with \refeq{eq:lij},
is well suited to perform the analytical continuation from 
$\al>0$ and $\be_i>0$ to any real values of $\al$ and $\be_i$.
During this continuation the masses $m_i$ are kept fixed.
We first vary the squared momenta $p_{10}^2$, $p_{21}^2$, and $p_{31}^2$ 
to reach the domain where one or more $\be_i\le0$,
while keeping  $p_{30}^2$, $p_{20}^2$, and $p_{32}^2$
(and thus also $\al$, $M^2$, $M_0^2$, $M_2^2$, and $M_3^2$) fixed.
In this procedure, none of the arguments in the logarithms of the
relevant $l_{ij}(x)$ crosses a branch cut.
Moreover, it does not matter whether one of the zeroes
$x_{i,j}$, $x'_{i,j}$, $x''_{i,j}$ runs over the path of the
$x$-integration, which is the positive real axes, or not,
because the residues at the poles at $x_{i,j}$, $x'_{i,j}$, $x''_{i,j}$ 
vanish as explained above. This shows that the continuation to
$\be_i\le0$ is trivial, i.e.\ that \refeq{eq:D0x} is correct
without modification.
The only complication in the continuation to $\al\le0$ stems from the
fact that the imaginary part of $M^2$ (and thus of all $M_i^2$) can
change sign. This sign change can lead to a discontinuity of
$\pm 2\pi\ri$ in the $\ln M_i^2$ term in $l_{ij}(x)$. However,
these possible constant terms in $l_{ij}(x)$ compensate each
other within the sum in \refeq{eq:D0x}, because after the
$x$-integration they are proportional to
\beq
\ln(-x'_{i,j}) - \ln(-x''_{i,j}) - \ln(-x_{i,j}) = 0.
\eeq
This relation holds for all real values of $\be_i$.
Again no extra terms arise if any of the $x_{i,j}$, $x'_{i,j}$, $x''_{i,j}$ 
runs over the integration axis because of the vanishing residues at these
poles.
In summary, \refeq{eq:D0x} is valid without modification for all
real values of $\al$ and $\be_i$.

The $x$-integration in \refeq{eq:D0x} can be easily carried out.
Using the auxiliary integral (B.1) of \citere{Denner:1991qq}, the final
result reads
\beqar
D_0 &=& \frac{1}{\sqrt{\det Y}} \sum_{k=1}^2 (-1)^{k+1} \Bigl\{
L_{31}(x'_{1,k})
+L_{21}(x'_{2,k})
+L_{01}(x'_{3,k})
-L_{01}(x''_{1,k})
-L_{31}(x''_{2,k})
\nn\\
&& \qquad {}
-L_{21}(x''_{3,k})
-L_{30}(x_{1,k})
-L_{23}(x_{2,k})
-L_{02}(x_{3,k})
\Bigr\},
\label{eq:D0result}
\eeqar
with the generic function
\beqar
L_{ij}(x) &=& 
-[\ln M^2_i-\ln m_i^2]\ln(-x)
+\Li(1+x)
\nn\\
&& {}
+\cLi\biggl(\frac{M^2_j}{M^2_i},-x\biggr)
-\cLi(r_{ij,1},-x)
-\cLi(r_{ij,2},-x).
\eeqar
If $M^2_i\to0$ the function $L_{ij}(x)$ reduces to
\beqar
L_{ij}(x) &\Nlim{M^2_i\to0}&
-[\ln M^2_j-\ln m_i^2]\ln(-x)
-\frac{1}{2}\ln^2(-x)
+\Li(1+x)
\nn\\
&& {}
-\cLi(r_{ij,1},-x)
-\cLi(r_{ij,2},-x) 
+\mbox{const.},
\eeqar
\begin{sloppypar}
\noindent
where the $x$-independent term indicated as {``const.''}\ drops out in
the difference $\sum_{k=1}^2 (-1)^{k+1} L_{ij}(x_k)$.  The final
result \refeq{eq:D0result} for the $D_0$ function with non-vanishing
complex masses $m_i$ contains 72 dilogarithms of complex arguments.
\end{sloppypar}

\subsection{4-point integral for general complex masses---method
of \citere{Denner:1991qq}}

Following the method of \citere{Denner:1991qq}, we first derive a
result for the 4-point function involving 16 dilogarithms that is
valid for complex masses if one of the external momenta squared
vanishes. Then we use an identity to express the general 4-point
function in terms of two 4-point functions of this kind.

\subsubsection{General massive 4-point integral for one vanishing external
  momentum squared}
\label{se:d0regrp}

We start by deriving a result for the regular scalar 4-point function
with four internal real masses and two space-like or light-like momenta.
Upon analytic continuation in the masses and one of the momentum squares
we generalise it to the case of complex masses and five arbitrary real
invariants, while the sixth invariant has to fulfil certain
conditions, which hold in particular if this invariant vanishes. We
follow the methods of \citere{Denner:1991qq}.

In the following we need the polynomials
\begin{eqnarray}
Q(x_{0},x_{1},x_{2},x_{3})&=&
\left[-\frac{1}{r_{20}}\frac{\partial}{\partial x_{0}}
+\frac{\partial}{\partial x_{2}}\right] P(x_{0},x_{1},x_{2},x_{3}) 
\label{Q}
\end{eqnarray}
and 
\begin{eqnarray}
\overline{Q}(x_{0},x_{1},x_{2},x_{3})&=&
\left[\,\frac{\partial}{\partial x_{1}}
-\frac{1}{r_{13}}\frac{\partial}{\partial x_{3}}\right] 
    P(x_{0},x_{1},x_{2},x_{3})  , \label{Qq}
\end{eqnarray}
where $r_{20}$ and $r_{13}$ are each one of the two variables
defined in the corresponding equation \refeq{rijdef}.

We start  from the representation \refeq{D0genint} and assume that all
masses are real and that the quantities $r_{20}$ and $r_{13}$ are real and
positive, which is the case if the momenta $p_{20}$ and $p_{31}$
are space-like or light-like.
We apply the projective transformation 
\begin{equation}
\begin{array}{cccc}
\left( \begin{array}{c}
       x_{1}\\
       x_{2}\\
       x_{3}
       \end{array}\right) &
         =&\disp
           \frac{1}{1+x+(1-1/r_{13})y+(1-1/r_{20})z}&
             \left( \begin{array}{c}
                    y\\
                    z\\
                    x-y/r_{13}
                    \end{array}\right) 
\label{proj2} 
\end{array}
\end{equation}
with the Jacobian
\begin{eqnarray}
\frac{\partial(x_{1},x_{2},x_{3})}{\partial(x,y,z)}
&=&\frac{1}{[1+x+(1-1/r_{13})y+(1-1/r_{20})z]^{4}} 
\end{eqnarray}
in order to linearise $P(1-x_1-x_2-x_3,x_1,x_2,x_3)$ 
simultaneously in two variables of integration.
The transformation of the integration region is determined by the inverse
\begin{equation}
\begin{array}{cccc}
\left( \begin{array}{c}
       x\\
       y\\
       z\\
       \end{array}\right) &
         =&\disp
           \frac{1}{1-x_{1}-(1-1/r_{20})x_{2}-x_{3}}&
             \left( \begin{array}{c}
                    x_{3}+x_{1}/r_{13}\\
                    x_{1}\\
                    x_{2} 
                    \end{array}\right)
\end{array} \;. \label{proj2invers}
\end{equation} 
This yields  
\begin{eqnarray}
D_{0}
&=&
\int_{0}^{\infty} \rd x\int_{0}^{x r_{13}}\rd y \int_{0}^{r_{20}}\rd z \, 
\frac{1}{[P(1-z/r_{20},y,z,x-y/r_{13})-\ri\veps]^{2}}\,, 
\end{eqnarray}
where $P$ is linear in both $y$ and $z$. Note that we have included an
explicit infinitesimal imaginary part $-\ri\veps$ consistent with the
infinitesimal imaginary part of the masses. 
However, for later use the masses are still
assumed to have an infinitesimal imaginary part which is
negligible compared to $\ri\veps$.
Performing the first integration and subsequently 
combining the resulting terms over a common denominator results in  
\begin{eqnarray}
D_{0}
&=&
\int_{0}^{\infty} \rd x\int_{0}^{x r_{13}}\rd y \;
\frac{1} {Q(1,y,0,x-y/r_{13})} 
\nl                                   
&& 
\hspace{4em}\times 
\left( 
\frac{1}{P(1,y,0,x-y/r_{13})-\ri\veps } 
- 
\frac{1}{P(0,y,r_{20},x-y/r_{13})-\ri\veps}
\right) \nl
&=&
r_{20}\int_{0}^{\infty}\rd x
\int_{0}^{x r_{13}}\rd y  \,
\frac{1}{P(1,y,0,x-y/r_{13})-\ri\veps}
\frac{1}{P(0,y,r_{20},x-y/r_{13})-\ri\veps} .
\hspace{2em}
\end{eqnarray}
By means of a partial-fraction decomposition
the second integration can be done as well, 
\begin{eqnarray}  
D_{0}
&=&
\int_{0}^{\infty}\rd x \,\frac{1}{R(x)} 
\int_{0}^{x r_{13}} \rd y  
\left(
\frac{\overline{Q}(1,0,0,x)}{P(1,y,0,x-y/r_{13})-\ri\veps }
-
\frac{\overline{Q}(0,0,r_{20},x)}{P(0,y,r_{20},x-y/r_{13})-\ri\veps}
\right)\nl
&=&
\int_{0}^{\infty}\rd x \, \frac{1}{R(x)} 
\Bigl( 
\ln [P(0,0,r_{20},x)-\ri\veps] 
- \ln [P(1,0,0,x) -\ri\veps]  \label{IR}
\nl
&& \hspace{4em}
- \ln [P(0,x r_{13},r_{20},0 )-\ri\veps]
+ \ln [P(1,x r_{13},0,0) -\ri\veps] 
\frac{}{} \Bigr) ,   
\end{eqnarray}
where the $y$-independent function $R(x)$ is defined as
\begin{eqnarray}\label{Rdef}
R(x)
&=&
\frac{1}{r_{20}}\biggl[\overline{Q}(1,0,0,x)[P(0,y,r_{20},x-y/r_{13})-\ri\veps]
\nln &&{}
-\overline{Q}(0,0,r_{20},x)[P(1,y,0,x-y/r_{13})-\ri\veps]\biggr]
\nl&=&
a x^{2} + b x +c + \ri\veps d
= a (x-x_{1}) (x-x_{2}) \label{R5} 
\end{eqnarray}
with
\begin{eqnarray}
a &=& m_1^2 r_{13}(Y_{23}-Y_{03}/r_{20})-m_3^2(Y_{12}-Y_{01}/r_{20}),  \nl
b &=& (m_2^2 r_{20}-m_0^2/r_{20})(m_1^2 r_{13}-m_3^2/r_{13})+Y_{01}Y_{23}-
  Y_{03}Y_{12} , \nl
c &=& m_2^2 r_{20}(Y_{01}-Y_{03}/r_{13})-m_0^2(Y_{12}-Y_{23}/r_{13}),
\nl
d &=& Y_{12}-Y_{01}/r_{20} -Y_{23}/r_{13} +Y_{03}/(r_{20}r_{13})   .
\label{eq:abcd}
\end{eqnarray}
The discriminant of the quadratic form reads
\beq
b^2-4ac = \det Y,
\eeq
where $Y$ is the modified Cayley matrix defined in \refeq{defY},
and the roots $x_{1,2}$ are given by
\beq\label{xdef}
x_{1,2}=\frac{1}{2a}(-b \pm \sqrt{\det Y})\mp \frac{\ri\veps d}{\sqrt{\det Y}}.
\eeq
Note that the infinitesimal imaginary part of the variables
$x_{1,2}$ results only form the $\ri\veps d$ term in \refeq{R5}, i.e.\ real 
mass parameters enter the calculation of $x_{1,2}$ via \refeq{eq:abcd}
as truly real quantities without infinitesimal imaginary parts.

A further partial-fraction decomposition yields an integral
representation valid for real and positive $r_{20},r_{13}$
and real masses
\begin{eqnarray}\label{D0regbasicres}
\lefteqn{a(x_{1}-x_{2})D_{0}=
\int_{0}^{\infty }\rd x \left( \frac{1}{x-x_{1}}-
    \frac{1}{x-x_{2}}\right)}                  
\nl&&{}\times
\left\{  
   \ln \left( \frac{P(0,0,1,x/r_{20})-\ri\veps/r_{20}^2}
                   {P(1,0,0,x)-\ri\veps} \right)
-  \ln \left( \frac{P(0,x r_{13}/r_{20},1,0)-\ri\veps/r_{20}^2}
                   {P(1,x r_{13},0,0)-\ri\veps}\right)
\right\}.
\end{eqnarray}

The following algebraic relations with arbitrary $y$ and $z$ 
hold for $R(x)$ and prove useful for the analytic continuation
\begin{eqnarray}
R(x)
&=&
\left\{ \frac{}{}
[P(1-z/r_{20},0,z,x)-\ri\veps]\,[P(1,y,0,x-y/r_{13})-\ri\veps]
\right. \label{R1} \nl
&&
\hspace{1em} -
\left.
[P(1,0,0,x)-\ri\veps]\,
[P(1-z/r_{20},y,z,x-y/r_{13})-\ri\veps]
\frac{}{} \right\}/(yz) \\
&=&
-
\left\{ \frac{}{}
\overline{Q}(1-z/r_{20},0,z,x) \,
[P(1,y,0,x-y/r_{13}) -\ri\veps] 
\right. \label{R2} \nl
&&
\hspace{1em} -
\left.
\overline{Q}(1,0,0,x) \,
[P(1-z/r_{20},y,z,x-y/r_{13})-\ri\veps]
\frac{}{} \right\}/z  \\
&=&
\biggl\{
{Q}(1,0,0,x) \,
[P(1,y,0,x-y/r_{13})-\ri\veps]
 \label{R3} \nl
&&
\hspace{1em} -
\left.
{Q}(1,y,0,x-y/r_{13}) \,
[P(1,0,0,x) -\ri\veps] 
\frac{}{} \right\}/y
\\ &=&
\left\{ \frac{}{}
\overline{Q}(1-z/r_{20},0,z,x) \, Q(1,y,0,x-y/r_{13})
\right. \label{R4} \nl
&&
\hspace{1em} -
\left.
d \, [P(1-z/r_{20},y,z,x-y/r_{13})-\ri\veps]
\frac{}{} \right\}.
\end{eqnarray}

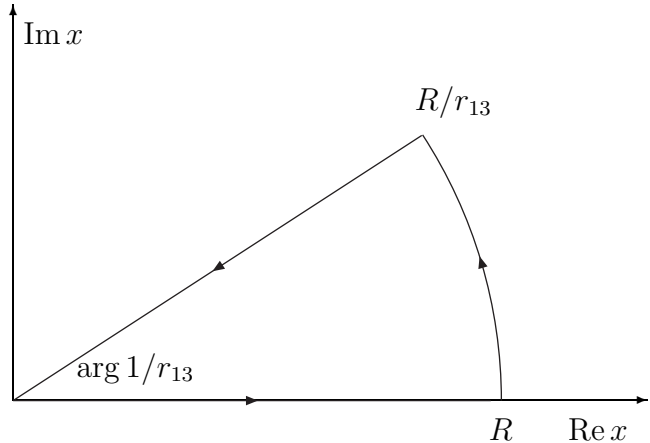
\begin{figure}
\centerline{
\setlength{\unitlength }{1pt}%
\begin{picture}(240,150)(-30,-15)
\put(0,0){\vector(1,0){240}}
\put(210,-15){$\Re x$}
\put(0,0){\vector(0,1){150}}
\put(4,135){$\Im x$}
\put(152,110){$R/r_{13}$}
\put(24,9){$\arg 1/r_{13}$}
%\put(105,45){\circle*{3}}
%\put(108,48){\mbox{\large $z_{0}$}}
\put(180,-15){$R$}
\ArrowArc(0,0)(184,0,33)        
\ArrowLine(0,0)(180,0)
\ArrowLine(154,100)(0,0)
%\Vertex(0,0){3}\Vertex(240,150){3}\Vertex(-30,-15){3}
\end{picture}%
}
\caption{The contour $(0,R ,R/r_{13})$ in the complex $x$-plane}
\label{fi:contour}
\end{figure}

In order to perform the analytic continuation, we rewrite
\refeq{D0regbasicres} as
\begin{eqnarray}
a ( x_{1}-x_{2} ) D_{0}&=& 
  \int_{0}^{\infty } \rd x\left(\frac{1}{x-x_{1}}-\frac{1}{x-x_{2}}\right) 
            \ln \left( \frac{P(0,0,1,x/r_{20})-\ri\veps/r_{20}^2}
                            {P(1,0,0,x)-\ri\veps} \right)
            \nonumber\\   
&&-\int_{0}^{\infty/r_{13} } \rd x
        \left(\frac{1}{x-x_{1}}-\frac{1}{x-x_{2}}\right) 
        \ln \left( \frac{P(0,x r_{13}/r_{20},1,0)-\ri\veps/r_{20}^2}
                        {P(1,x r_{13},0,0)-\ri\veps}\right)
            \nonumber\\
&&  + \sum_{k=1}^{2} (-1)^{k}
   \ln \left( \frac{P(0,0,1,x_{k}/r_{20})-\ri\veps/r_{20}^2}
                   {P(1,0,0,x_{k})-\ri\veps}  \right) 
   \oint_{(0,\infty ,\infty/r_{13})} \frac{\rd x}{x-x_{k}}  
  , \label{D0contour}
\label{D0regbasicrescont}
\end{eqnarray}
where the contour of the integrals in the last line results from the
one shown  
in \reffi{fi:contour} in the limit $R \rightarrow \infty $. 
Equation~\refeq{D0regbasicrescont} obviously coincides with
\refeq{D0regbasicres} for real masses and $r_{13},r_{20}>0$.
Now we allow for complex masses and arbitrary $p_{31}^2$, leading to
complex $r_{13}$, while $r_{20}$ is still required to be real and
positive (which entails some restrictions on $p_{20}^2$ and/or the
masses $m_0^2$ and $m_2^2$).  We note that, in particular, for
$p_{20}^2=0$ we can choose $r_{20}=1$, so that our result holds. We
here allow also squared masses with negative real parts, but always
with negative imaginary parts, a case for which \refeq{D0def} can be
directly analytically continued.
Equation \refeq{D0regbasicrescont} yields the correct analytical
continuation, since:
\begin{itemize}
\item No cut is crossed in the logarithms of the first two integrals
  of \refeq{D0regbasicrescont}, since all polynomials $P$ have
  negative imaginary parts.  Note that $x,x/r_{20}$ in the first
  integral and $xr_{13},xr_{13}/r_{20}$ in the second integral are
  real.
\item If one of the poles $x_{1,2}$ crosses the integration contour in
  one of the first two integrals, it also moves into the closed
  contour $(0,\infty ,\infty/r_{13})$.  The residues add up to zero
  owing to \refeq{R1} where one has to set $(y,z)=(x r_{13},r_{20})$.
  For real and positive $x_{1,2}$ the explicit imaginary parts in
  \refeq{xdef} and \refeq{D0regbasicrescont} proportional to
  $\ri\veps$ are crucial for this cancellation, i.e.\ both have to be
  taken into account consistently (with the same numerical size of
  $\veps$).
\item Finally, we have to show that there are no further
  discontinuities in the last term of \refeq{D0regbasicrescont}, apart
  from the required ones resulting from the possible movement of
  $x_{1,2}$ inside the integration contour.  If $x_k$ is outside the
  contour, nothing is to show, because the term is zero.  If $x_k$ is
  within the closed contour of the integral in the last line of
  \refeq{D0regbasicrescont}, it can be written as $x_k=u+v/r_{13}$
  with real and positive $u$ and $v$. Using \refeq{R1} for $x=x_k$,
  $y=v$ and $z=r_{20}$, we find
\beq\label{ratiowithuandv}
 \frac{P(0,0,1,x_{k}/r_{20})-\ri\veps/r_{20}^2}{P(1,0,0,x_{k})-\ri\veps} 
= \frac{P(0,v/r_{20},1,u/r_{20})-\ri\veps/r_{20}^2}{P(1,v,0,u)-\ri\veps}.
\eeq 
For real and positive $r_{20}$, both $P(0,v/r_{20},1,u/r_{20})-\ri\veps$ and
$P(1,v,0,u)-\ri\veps$ have a negative imaginary part and the ratio does never
cross the cut.

If the ratio approaches the cut of the logarithm, its imaginary part
can be determined from \refeq{ratiowithuandv} as long as $r_{13}$ has
a finite imaginary part (otherwise $v$ is not defined).
If $r_{13}$ is real, the last line of
\refeq{D0regbasicrescont} contributes only if $r_{13}$ is 
negative. 
In this case the imaginary part of the ratio can be
determined using \refeq{R2} via
\beqar
 \frac{P(0,0,1,x_{k}/r_{20})-\ri\veps/r_{20}^2} {P(1,0,0,x_{k})-\ri\veps}
&=& \frac{\overline Q(0,0,r_{20},x_{k})} {r_{20}^2\overline Q(1,0,0,x_{k})}
\nn\\
&=&\frac{r_{20}(Y_{12}-Y_{23}/r_{13})+(m_1^2r_{13}-m_3^2/r_{13})x_k}
{r_{20}^2[Y_{01}-Y_{03}/r_{13}+(m_1^2r_{13}-m_3^2/r_{13})x_k]}
\eeqar
taking into account only the infinitesimal imaginary part of $x_k$
resulting from $\ri\veps$.
Obviously, the imaginary parts of the numerator and denominator of the
last expression have the same imaginary part (from $x_k$) near the
real axis, so that the ratio cannot cross the cut.
\end{itemize}
Consequently, \refeq{D0contour} provides a correct analytical
continuation.  
We finally reemphasize the treatment of infinitesimal negative imaginary 
parts attributed to real mass squares. While these infinitesimal
imaginary parts are made explicit in the calculation of $x_k$ by
the $\ri\veps$ terms (and whenever $\ri\veps$ appears), the 
infinitesimal imaginary parts of the variables $r_{ij,k}$ of \refeq{rijdef} are
implicitly determined from the ones in the squared masses (or equivalently
from $\bar p_{ij}^2$) and are negligible compared to $\ri\veps$, \ie
the infinitesimal imaginary part of ${x_k}r_{13}$ is solely
  determined by the infinitesimal imaginary part of ${x_k}$.

Using the auxiliary integral (B.1) of \citere{Denner:1991qq},
\refeq{D0contour} can be evaluated to
\beqar\label{D0p0res}
\lefteqn{a ( x_{1}-x_{2} ) D_{0}(p_1,p_2,p_3,m_0^2,m_1^2,m_2^2,m_3^2)=}\qquad \nn\\
&=& \sum_{k=1}^{2} (-1)^{k+1}\Biggl\{
 \sum_{l=1}^{2}\biggl[
\cLi\biggl(-x_k,\frac{r_{23,l}}{r_{20}}\biggr)
-\cLi\biggl(-x_k,r_{03,l}\biggr)
\nl&&{}
-\cLi\biggl(-{x_k}{r_{13}},\frac{r_{21,l}}{r_{20}}\biggr)
+\cLi\biggl(-{x_k}{r_{13}},r_{01,l}\biggr)\biggr]
\nl&&{}
-\eta\biggl(-x_k,{r_{13}}\biggr)\biggl[
 \ln \left( \frac{P(0,0,1,x_{k}/r_{20})-\ri\veps/r_{20}^2}{P(1,0,0,x_{k})-\ri\veps} \right)
-\ln\biggr(\frac{m_2^2}{m_0^2}\biggr)\biggr]
\Biggr\}.
\eeqar
Note that also for negative squared masses, $r_{ij}$ do never cross
the negative real axis as long as the squared masses have negative
(infinitesimal) imaginary parts. The result \refeq{D0p0res} is in
particular valid for $p_{20}^2=0$, where $r_{20}=1$, and arbitrary
real values for the other invariants and complex squared masses with
negative imaginary parts (and positive or negative real parts).  Owing
to the permutation symmetry of the arguments, it can be used if any
external squared momentum vanishes, $p_{ij}^2=0$.

\subsubsection{Result for general massive 4-point function via
  propagator identity}

The result of the previous subsection can be used to derive
a second result for the general scalar 4-point integral $D_0$ with
complex internal masses. To this end, we employ a propagator identity of
\citere{'tHooft:1978xw} to express the general $D_0$ in terms of two
4-point functions with one vanishing external momentum squared.  For
the latter functions the result \refeq{D0p0res} is applicable.  Thus,
we end up with a result in terms of 32 dilogarithms for the
general $D_0$ function with complex masses.

The propagator identity (2.2) of \citere{'tHooft:1978xw} allows to
express the product of two propagators in terms of a sum of two
products of two propagators,
\beq\label{propid}
\frac{1}{(q+p_1)^2-m_1^2}\frac{1}{(q+p_3)^2-m_3^2}
=
\frac{1}{(q+p)^2-M^2}
\biggl[\frac{1-\ga}{(q+p_1)^2-m_1^2} +
\frac{\ga}{(q+p_3)^2-m_3^2}\biggr],
\eeq
where $\ga$ is arbitrary and 
\beqar
p&=& \ga p_1 + (1-\ga)p_3,\nl
M^2&=& p^2+\ga(m_1^2-p_1^2)+(1-\ga)(m_3^2-p_3^2).
\eeqar
Fixing $\ga$ in such a way that (recall $p^2_{1}=p^2_{10}$ and
$p^2_{3}=p^2_{30}$)
\beq 
p^2=\ga^2 p^2_{31} + \ga(p^2_{10}-p^2_{31}-p^2_{30}) + p^2_{30}=0
\eeq
 and inserting the identity \refeq{propid}
into \refeq{D0def}, we can express the general scalar 4-point function
as
\beqar\label{D0propid}
\lefteqn {D_0(p_{10}^2,p_{21}^2,p_{32}^2,p_{30}^2,p_{20}^2,p_{31}^2,
m^2_0,m^2_1,m^2_2,m^2_3)}\qquad \nl
&=&\ga  D_0(0,(p_2-p)^2,p_{32}^2,p_{30}^2,p_{20}^2,\ga^2p_{31}^2,
m^2_0,M^2,m^2_2,m^2_3)
\nl&&{}
+ (1-\ga)D_0(p_{10}^2,p_{21}^2,(p_2-p)^2,0,p_{20}^2,(1-\ga)^2p_{31}^2,
m^2_0,m^2_1,m^2_2,M^2)\nl
&=&\ga  D_0(p_{30}^2,\ga^2p_{31}^2,(p_2-p)^2,p_{20}^2,0,p_{32}^2,
m^2_0,m^2_3,M^2,m^2_2)
\nl&&{}
+ (1-\ga)D_0(p_{10}^2,(1-\ga)^2p_{31}^2,(p_2-p)^2,p_{20}^2,0,p_{21}^2,
m^2_0,m^2_1,M^2,m^2_2),\qquad
\eeqar
where %(for $p^2=0$)
\beq
(p_2-p)^2=\ga(p_{21}^2-p_{10}^2)+(1-\ga)(p_{32}^2-p_{30}^2),
\eeq
and we used the symmetry of $D_0$ mentioned after \refeq{defY} to arrange
the arguments such that the 5th momentum squared vanishes.

As mentioned above,  $\ga$ [which corresponds to $-1/\be_1$ of
\refeq{defbetai}] is real for all momentum configurations
appearing in $1\to n$ decays or $2\to n$ scattering processes. 
For such processes, at least one of the momenta $p_{31}$, $p_{30}$,
and $p_{10}$ is time-like. Since the  interchange
$p_{30}\leftrightarrow p_{10}$ replaces $\ga$ by $1-\ga$ and 
$p_{31}\leftrightarrow p_{30}$ transforms $\ga$ into $1/\ga$,
%and the  interchange $p_{31}\leftrightarrow p_{10}$   $\ga$ into $\ga/(\ga-1)$
the momenta can always be permuted such that 
\beq\label{alphacond}
0\le\ga\le1.
\eeq
Then in \refeq{D0propid} the squared masses have negative imaginary parts
(while the real part of $M^2$ may become negative), and the result 
\refeq{D0p0res} holds.

To be specific we select a suitable permutation of the arguments of
$D_0$ as follows:
\begin{itemize}
\item If there is no $t$-channel-like squared momentum among the
  arguments of $D_0$, the centre-of-mass energy squared $s$ must be
  one of the arguments, and we can choose $p_{31}^2=s$ and
\beq\label{alpha1}
\ga=\frac{1}{2p^2_{31}}\Bigl[p^2_{31}+p^2_{30}-p^2_{10}
+\sqrt{(p^2_{31}+p^2_{30}-p^2_{10})^2-4p^2_{31}p^2_{30}}\,\Bigr],
\eeq
where $p_{10}$ can be any sum of outgoing momenta and
$p_{10}=p_{30}-p_{31}$. The choice \refeq{alpha1} fulfils
\refeq{alphacond} since $p_{31}^2\ge\Bigl(\sqrt{p_{30}^2}+\sqrt{p_{10}^2}\Bigr)^2$.
\item For one $t$-channel-like squared momentum $t$ we choose
  $p_{30}^2=t$. Since $p_{31}^2\ge0$, $p_{10}^2\ge0$ and
  $p_{30}^2\le\Bigl(\sqrt{p_{31}^2}-\sqrt{p_{10}^2}\Bigr)^2$ for any choice of
  line~1 of the diagram, $\ga$ defined by \refeq{alpha1} fulfils \refeq{alphacond}
  again.
\item If there are two $t$-channel-like momentum invariants $t_1$ and
  $t_2$, we choose  $p_{31}^2=t_1$ and  $p_{10}^2=t_2$. Since
  $p_{30}^2\ge0$, 
\beq\label{alpha2}
\ga=\frac{1}{2p^2_{31}}\Bigl[p^2_{31}+p^2_{30}-p^2_{10}
-\sqrt{(p^2_{31}+p^2_{30}-p^2_{10})^2-4p^2_{31}p^2_{30}}\Bigr],
\eeq
fulfils \refeq{alphacond}.
\end{itemize}

In conclusion, \refeq{D0propid} together with \refeq{D0p0res} and the
appropriate permutation of momenta and choice of $\ga$ according to
\refeq{alpha1} or \refeq{alpha2} provides a result for the general
scalar 4-point function with non-vanishing complex masses, valid for
all scattering and decay processes.

\subsection{Regular 4-point function with one or more masses vanishing}

For vanishing masses the general results of the last two sections are
not directly applicable. Instead of extracting the corresponding
limits from the final results, it is more convenient to use the result
\refeq{D0regbasicres} which is based on \citere{Denner:1991qq}.
If certain variables diverge in the following formulas (e.g.\ for $a=0$),
a well-defined result is obtained by permuting propagator lines of the diagram,
as long as the integral is non-singular.

We first consider the case of one vanishing mass and derive a result 
for  $m_2^2=0$ and $m^2_0$, $m^2_1$, $m^2_3$ non-zero and possibly complex
with negative imaginary parts. 
It is appropriate to start by considering real masses and space-like
or light-like momenta $p_{20}$ and $p_{31}$, so that $r_{20}$ and $r_{13}$
are real and positive. In order to make use of
the result \refeq{D0regbasicres}, we choose the solution for $r_{20}$
that satisfies
\beq
\frac{1}{r_{20}} \, \mathop{\longrightarrow}\limits_{m_2^2\to0} \, 0,\qquad
m_2^2 r_{20} \, \mathop{\longrightarrow}_{m_2^2\to0} \, Y_{02}.
\eeq
Then, we obtain 
\beqar
\frac{P(0,0,1,x/r_{20})-\ri\veps/r_{20}^2}{P(0,x r_{13}/r_{20},1,0)-\ri\veps/r_{20}^2}
&\,\asymp{m_2^2\to0}\,&\frac{Q(0,0,1,x/r_{20})}{Q(0,x r_{13}/r_{20},1,0)}
\,\asymp{m_2^2\to0}\,\frac{Q(1,0,0,x)}{Q(1,x r_{13},0,0)}
\nn\\
&\,\mathop{\longrightarrow}\limits_{m_2^2\to0} \,&
\frac{Y_{02}+Y_{23}x}{Y_{02}+Y_{12}x r_{13}},
\eeqar
where the explicit $\ri\veps$ terms on the very l.h.s.\ vanish for
$m_2^2\to0$ so that the infinitesimal imaginary parts 
are implicitly fixed by those of the internal masses contained in the
$Q$ functions and $Y_{ij}$ variables.  With these replacements,
\refeq{D0regbasicres} can be written as
\begin{eqnarray}
\lefteqn{a ( x_{1}-x_{2} ) D_{0}(p_1,p_2,p_3,m_0^2,m_1^2,0,m_3^2)} 
\nonumber\qquad\\*
&=&   \int_{0}^{\infty } \rd x\left(\frac{1}{x-x_{1}}-\frac{1}{x-x_{2}}\right) 
            \ln \left( \frac{Q(1,0,0,x)}{P(1,0,0,x)-\ri\veps}    \right)
            \nonumber\\   
&&{}-\int_{0}^{\infty/r_{13}} \rd x\left(\frac{1}{x-x_{1}}-\frac{1}{x-x_{2}}\right) 
        \ln \left( \frac{Q(1,x r_{13},0,0)}{P(1,xr_{13},0,0)-\ri\veps}    \right)
            \nonumber\\
&&  + \sum_{k=1}^{2} (-1)^{k}
   \ln \left( \frac{Q(1,0,0,x_k)}{P(1,0,0,x_k)-\ri\veps}  \right) 
   \oint_{(0,\infty ,\infty/{r}_{13})} \frac{\rd x}{x-x_{k}} \nl 
&=&   \int_{0}^{\infty } \rd x\left(\frac{1}{x-x_{1}}-\frac{1}{x-x_{2}}\right) 
            \ln \left( \frac{Y_{02}+Y_{23}x}{m_0^2+Y_{03}x+m_3^2x^2-\ri\veps}    \right)
            \nonumber\\   
&&{}-\int_{0}^{\infty/r_{13}} \rd x\left(\frac{1}{x-x_{1}}-\frac{1}{x-x_{2}}\right) 
        \ln \left( \frac{Y_{02}+Y_{12}{x}{r_{13}}}{m_0^2+Y_{01}{x}{r_{13}}+m_1^2({x}{r_{13}})^2-\ri\veps}    \right)
            \nonumber\\
&&  + \sum_{k=1}^{2} (-1)^{k}
   \ln \left( \frac{Y_{02}+Y_{23}x_k}{m_0^2+Y_{03}x_k+m_3^2x_k^2-\ri\veps}  \right) 
   \oint_{(0,\infty ,\infty/{r}_{13})} \frac{\rd x}{x-x_{k}}  
, \label{D0m0contour}
\end{eqnarray}           
where again the contour of the last integral goes from zero to infinity on
the positive real axis, along an arc at infinity to the direction of
$1/{r}_{13}$ and back to the origin 
(see \reffi{fi:contour} for $R\to\infty$).
The quantities $x_{1,2}$ are determined from \refeq{Rdef} with 
\begin{eqnarray}
a &=&m_1^2 r_{13} Y_{23} - m_3^2 Y_{12},  \nl
b &=& Y_{02} (m_1^2 r_{13}-m_3^2/ r_{13}) + Y_{01} Y_{23} - Y_{03} Y_{12} ,\nonumber\\
c &=& Y_{02}( Y_{01} - Y_{03}/ r_{13}) 
    - m_0^2(Y_{12}-Y_{23}/r_{13}),\nl
d &=& Y_{12} - Y_{23}/ r_{13} .
\end{eqnarray}

The result \refeq{D0m0contour} is valid for real masses and real and
positive $r_{13}$. The analytic continuation proceeds as for
\refeq{D0contour}. We move the squared masses into the complex
plane (with negative imaginary part) and $p_{31}^2$ to arbitrary real
values. Thereby $r_{13}$ becomes complex and moves on the first
Riemann sheet. 
%It can be shown that it cannot become negative real as long as
%$m^2_1$ and $m^2_3$ have (infinitesimal) negative imaginary  parts. 
Equation \refeq{D0m0contour} yields the correct analytical
continuation, since:
\begin{itemize}
\item The arguments of the logarithms in the first two integrals do
  not cross the cut, because numerator and denominator of the
  fractions both have negative imaginary parts (which in the numerator
  results from the masses).
\item If one of the poles $x_{1,2}$ crosses the integration contour in
  one of the first two integrals, it also moves into the closed
  contour $(0,\infty ,\infty/r_{13})$.  The residues add up to zero
  owing to \refeq{R3} where one has to set $y=x r_{13}$.  
  As stressed also above, for $x_{1,2}$ that are close to the positive real 
  axis the explicit $\ri\veps$ parts 
  in \refeq{xdef} and \refeq{D0m0contour} are
  crucial for this cancellation.
\item If $x_k$ is within the closed contour of the integral in the
  last line, it can be written as $x_k=u+v/r_{13}$ with real and
  positive $u$ and $v$. Using  \refeq{R3} for $x=x_k$ and $y=v$,  we
  find
\beqar\label{argloguv}
 \frac{P(1,0,0,x_{k})-\ri\veps} {Q(1,0,0,x_{k})}
&=& \frac{P(1,v,0,u)-\ri\veps}{Q(1,v,0,u)}\nl
&=& \frac
{m_0^2+m_1^2v^2+m_3^2u^2+Y_{01}v+Y_{03}u+Y_{13}uv-\ri\veps}
{Y_{02}+Y_{12}v+Y_{23}u}.\qquad
\eeqar
For squared masses with negative imaginary parts both $Q(1,v,0,u)$ and
$P(1,v,0,u)-\ri\veps$ have a negative imaginary part and the ratio does not
cross the cut.
If the argument of the logarithm approaches the cut, its imaginary
part can be determined from \refeq{argloguv} as long as $\Im
r_{13}\ne0$ (otherwise $v$ is not defined). 
If the imaginary part of $r_{13}$ vanishes, the imaginary
part of the ratio can be determined via [using \refeq{R4}]
\beq
 \frac{P(1,0,0,x_{k})-\ri\veps} {Q(1,0,0,x_{k})}
=\frac{\overline Q(1,0,0,x_{k})}{d} 
=\frac
{Y_{01}-Y_{03}/r_{13}+(m_1^2r_{13}-m_3^2/r_{13})x_k}{ Y_{12} - 
Y_{23}/r_{13} },
\eeq
taking into account only the infinitesimal imaginary part of $x_k$
resulting from $\ri\veps$.
\end{itemize}

Employing the auxiliary integral (B.1) of \citere{Denner:1991qq},
\refeq{D0m0contour} can be evaluated to 
\beqar
\lefteqn{a ( x_{1}-x_{2} ) D_{0}(p_1,p_2,p_3,m_0^2,m_1^2,0,m_3^2)} \nonumber\qquad\nl
&=& \sum_{k=1}^{2} (-1)^{k+1}\biggl\{
\cLi\biggl(-x_k,\frac{Y_{23}}{Y_{02}}\biggr)
-\cLi\biggl(-x_k,r_{03,1}\biggr)
-\cLi\biggl(-x_k,r_{03,2}\biggr)
\nl&&{}
-\cLi\biggl(-{x_k}{r_{13}},\frac{Y_{12}}{Y_{02}}\biggr)
+\cLi\biggl(-{x_k}{r_{13}},r_{01,1}\biggr)
+\cLi\biggl(-{x_k}{r_{13}},r_{01,2}\biggr)
\nl&&{}
+\eta\biggl(-x_k,{r_{13}}\biggr)\biggl[
\ln \left( \frac{P(1,0,0,x_k)-\ri\veps}{Q(1,0,0,x_k)}  \right) 
+\ln\biggr(\frac{Y_{02}}{m_0^2}\biggr)\biggr]
\biggr\}.
\eeqar
For $Y_{02}=m_0^2-p_{20}^2=0$ and $Y_{23}\ne0$ this result can be used
after exchanging indices 0 and 3 (in the sense of interchanging lines
0 and 3 in the diagram and performing the whole calculation with this
assignment).  For $Y_{02}=Y_{23}=0$, $Y_{02}=Y_{12}=0$, or
$Y_{12}=Y_{23}=0$ the integral becomes soft singular; these cases are
covered in the next section.

Next we consider the case of two vanishing masses and assume 
$m_1^2=m_2^2=0$, but $m_0^2\ne0$ and $m_3^2\ne0$. The other cases with two
zero-masses can be obtained by obvious substitutions.%
\footnote{This includes also the case with vanishing masses on
  opposite lines of the diagram because of the symmetry w.r.t.\ 
  interchange of neighbouring lines,
  $D_0(p_{10}^2,p_{21}^2,p_{32}^2,p_{30}^2,p_{20}^2,p_{31}^2,
  m^2_0,m^2_1,m^2_2,m^2_3) =
  D_0(p_{10}^2,p_{20}^2,p_{32}^2,p_{31}^2,p_{21}^2,p_{30}^2,
  m^2_1,m^2_0,m^2_2,m^2_3)$.}  We choose the roots such that
\beqar
\frac{1}{r_{13}} \,\mathop{\longrightarrow}\limits_{m_1^2\to0} \,0,\qquad
m_1^2 r_{13} \,\mathop{\longrightarrow}_{m_1^2\to0} \,Y_{13},\qquad
{r_{01,1}} \,\mathop{\longrightarrow}\limits_{m_1^2\to0}
\,\frac{Y_{01}}{m_0^2},
\qquad
r_{13}  r_{01,2} \,\mathop{\longrightarrow}_{m_1^2\to0} \,\frac{Y_{13}}{Y_{01}},
\eeqar
and obtain
\beqar\label{D0m00}
\lefteqn{a ( x_{1}-x_{2} ) D_{0}(p_1,p_2,p_3,m_0^2,0,0,m_3^2)} 
\nonumber\qquad\nl
&=& \sum_{k=1}^{2} (-1)^{k+1}\biggl\{
\cLi\biggl(-x_k,\frac{Y_{23}}{Y_{02}}\biggr)
-\cLi\biggl(-x_k,r_{03,1}\biggr)
-\cLi\biggl(-x_k,r_{03,2}\biggr)
\nl&&{}
+\cLi\biggl(-x_k,\frac{Y_{13}}{Y_{01}}\biggr)
-\ln(-x_k)
\biggl[\ln\biggl(\frac{Y_{01}}{Y_{12}}\biggr)
+\ln\biggl(\frac{Y_{02}}{m_0^2}\biggr)\biggr]
\biggr\},
\eeqar
for real positive $Y_{13}$, where the $x_k$ are determined from
\refeq{Rdef} with
\beqar\label{D0m00abc}
a &=& Y_{13}Y_{23} - m_3^2 Y_{12},  \qquad
b = Y_{02}Y_{13} + Y_{01} Y_{23} - Y_{03} Y_{12} ,\nl
c &=& Y_{01} Y_{02} - m_0^2 Y_{12}, \qquad
d = Y_{12}.
\eeqar
By going back to the corresponding integral representation
\refeq{D0m0contour} one can convince oneself that \refeq{D0m00} is
valid for general momenta and complex masses without modification. 
If $Y_{01}=0$ or
$Y_{02}=0$ the result can be used after interchanging indices 0 and 3,
as long as $Y_{23}$ and $Y_{13}$ do not vanish.
For $Y_{01}=Y_{23}=0$ it can be written as
\beqar\label{D0m00alt}
\lefteqn{a ( x_{1}-x_{2} ) D_{0}(p_1,p_2,p_3,m_0^2,0,0,m_3^2)} \nonumber\qquad\nl
&=& \sum_{k=1}^{2} (-1)^{k+1}\biggl\{
-\cLi\biggl(-x_k,r_{03,1}\biggr)
-\cLi\biggl(-x_k,r_{03,2}\biggr)
-\frac{1}{2}\ln^2(-x_k)
\nl&&{}
-\ln(-x_k)
\biggl[\ln\biggl(\frac{Y_{13}}{Y_{12}}\biggr)
+\ln\biggl(\frac{Y_{02}}{m_0^2}\biggr)\biggr]
\biggr\},
\eeqar
which can also be used for $Y_{02}=0$ and $Y_{13}=0$ after
interchanging indices 0 and 3. Other degenerate cases are soft or
collinear singular (see \refse{se:singints}).

For the case of three vanishing masses, consider 
$m_1^2=m_2^2=m_3^2=0$, but $m_0^2\ne0$. The other cases can be 
obtained by obvious substitutions.
Using
\beqar
{r_{03,1}} \,\mathop{\longrightarrow}\limits_{m_3^2\to0}\,
\frac{Y_{03}}{m_0^2},
\qquad
r_{03,2} %\mathop{\longrightarrow}_{m_3\to0} \frac{m_3^2}{Y_{03}}
\,\mathop{\longrightarrow}_{m_3^2\to0} 0\,,
\eeqar
we obtain directly from \refeq{D0m00} and \refeq{D0m00abc}
\beqar\label{D0m000}
\lefteqn{a ( x_{1}-x_{2} ) D_{0}(p_1,p_2,p_3,m_0^2,0,0,0)} \nonumber\qquad\nl
&=& \sum_{k=1}^{2} (-1)^{k+1}\biggl\{
\cLi\biggl(-x_k,\frac{Y_{23}}{Y_{02}}\biggr)
-\cLi\biggl(-x_k,\frac{Y_{03}}{m_{0}^2}\biggr)
\nl&&{}
+\cLi\biggl(-x_k,\frac{Y_{13}}{Y_{01}}\biggr)
-\ln(-x_k)
\biggl[\ln\biggl(\frac{Y_{01}}{Y_{12}}\biggr)
+\ln\biggl(\frac{Y_{02}}{m_0^2}\biggr)\biggr]
\biggr\}
\eeqar
and
\beqar\label{D0m000abc}
a &=& Y_{13} Y_{23}, \phantom{{} - m_0^2 Y_{12}}   \qquad
b = Y_{02}Y_{13} + Y_{01} Y_{23} - Y_{03} Y_{12} ,\nl
c &=& Y_{01} Y_{02} - m_0^2 Y_{12}, \qquad
d = Y_{12}.
\eeqar
Degenerate cases are obtained as for \refeq{D0m00}.

Finally for four vanishing masses, \refeq{D0m000} and \refeq{D0m000abc}
simplify to 
\beqar
\lefteqn{a ( x_{1}-x_{2} ) D_{0}(p_1,p_2,p_3,0,0,0,0)} \nonumber\qquad\nl
&=& \sum_{k=1}^{2} (-1)^{k+1}\biggl\{
\cLi\biggl(-x_k,\frac{Y_{23}}{Y_{02}}\biggr)
-\cLi\biggl(-\frac{1}{x_k},\frac{Y_{01}}{Y_{13}}\biggr)
\nl&&{}
-\ln(-x_k)
\biggl[\ln\biggl(\frac{Y_{13}}{Y_{12}}\biggr)
+\ln\biggl(\frac{Y_{02}}{Y_{03}}\biggr)\biggr]
\biggr\}
\eeqar
and
\beq
a = Y_{13}Y_{23} ,  \qquad
b = Y_{02}Y_{13} + Y_{01} Y_{23} - Y_{03} Y_{12} ,\qquad
c = Y_{01} Y_{02}, \qquad
d = Y_{12}.
\eeq
Degenerate cases are mass singular. Different results for this case 
have been given in \citeres{Bern:1993kr,Duplancic:2002dh}.

\section{Singular 4-point integrals}
\label{se:singints}

Whenever mass parameters
$\lambda$ or $\lambda_i$ appear, they are understood as infinitesimal.
If not all singularities are cured by mass parameters, dimensional
regularization is applied.
Thus, in the following all formulas are valid up to orders
${\cal O}(\lambda)$, ${\cal O}(\lambda_i)$, and ${\cal O}(\eps)$.
For convenience, we define the shorthand
\beq
c_\eps = \Gamma(1+\eps)(4\pi)^\eps
\eeq
and recall our graphical notation,
\beqar
\DOp{p_{10}^2}{p_{21}^2}{p_{32}^2}{p_{30}^2}%
{p_{20}^2}{p_{31}^2}% 
\DOm{m_0^2}{m_1^2}{m_2^2}{m_3^2}{} \qquad
&\equiv& \; D_0(p_1,p_2,p_3,m_0^2,m_1^2,m_2^2,m_3^2)
\nn\\[-3.5em]
&\equiv& \; D_0(p_{10}^2,p_{21}^2,p_{32}^2,p_{30}^2,p_{20}^2,p_{31}^2,
m_0^2,m_1^2,m_2^2,m_3^2)
\nn\\
\eeqar
with $p_{ij}=p_i-p_j$.  Over-lined variables are understood to receive
an infinitesimally small imaginary part, 
$\bar p^2_{ij}=p^2_{ij}+\ri0$, etc., and internal squared masses other than
$\lambda^2$ or $\lambda_i^2$ are allowed to be complex with phases between
$0$ and $-\pi/2$.  We categorise the singular $D_0$ functions according to their
numbers of collinear and soft singularities, $n_{\coll}$ and
$n_{\soft}$, respectively, following the classification of
Kinoshita~\cite{Kinoshita:ur}:
\begin{itemize}
\item
Each of the six squared momenta $p_{ij}^2$ gives rise to a collinear
singularity if the three variables $p_{ij}^2$, $m_i^2$, and $m_j^2$ are
zero or small, i.e.\ of ${\cal O}(\la_{(i)}^2)$. Pictorially this means
that the squared momentum difference between two propagators of small
masses is small, e.g.\ if a particle with a light-like vector
couples to two light internal particles.
In non-exceptional phase-space points of scattering reactions
either $\{p_{10}^2,p_{32}^2\}$ or $\{p_{21}^2,p_{30}^2\}$ or 
$\{p_{20}^2,p_{31}^2\}$ are not small or zero, so that
at most four collinear singularities can be present at a time. 
By convention we take $\{p_{20}^2,p_{31}^2\}$ not small in the following.
\item A soft singularity arises if a light particle is exchanged
  between two on-shell particles. Formally this situation appears if
  $m_i^2$ is zero and $p_{ij}^2=m_j^2$ and $p_{ik}^2=m_k^2$ for two
  different indices $j,k$ not equal to $i$. For non-exceptional
  phase-space points, at most four soft singularities can exist at a
  time.
\end{itemize}
In the following we also define the regularization-scheme-independent
finite part $\DOfin$ of a singular $D_0$ function,
as described in \citere{Dittmaier:2003bc}, upon subtracting an appropriate
linear combination of scalar 3-point integrals $C_0(i)$, which
result from $D_0$ upon shrinking propagator $i$ to a point.

\paragraph{Purely soft-singular cases $(n_{\coll}=0)$}

There are two cases with only one soft, but no collinear
singularity.
Taking $m_0$ as the ``soft mass'',
the common finite part of these soft-singular integrals is given by
\beq
\DOfin = D_0 - \frac{C_0(2)}{p_{20}^2-m_2^2}.
\eeq
Therefore, knowing $D_0$ of \refeq{eq:DOc0s1dimreg} one can obtain 
$D_0$ of \refeq{eq:DOc0s1massreg} from the difference of the
corresponding integrals $C_0(2)$ for $m_0=0$ and $m_0=\la$,
which can be found in Appendix~B of \citere{Dittmaier:2003bc}.
The mass-regularized integral \refeq{eq:DOc0s1massreg} was taken over
from (2.9) [case (i)] of \citere{Beenakker:1990jr}. 
The dimensionally regularized integral can be obtained by the simple
replacement
$$
\ln(\la)\to\frac{c_\eps}{2\eps}+\ln(\mu)+ \ord{(\eps)},
$$
which applies if only soft (but no collinear) singularities are regulated dimensionally,
and the results are given by
\beqar
\DOp{m_1^2}{p_{21}^2}{p_{32}^2}{m_3^2}%
{p_{20}^2}{p_{31}^2}% 
\DOm{0}{m_1^2}{m_2^2}{m_3^2}{(n_{\soft}=1)} 
&=&
\frac{x_{31}}{m_1 m_3(p_{20}^2-m_2^2)(1-x_{31}^2)}\Biggl\{
2\ln(x_{31}) \biggl[
-\frac{c_\eps}{2\eps}
\nn\\[-3em]
&& \quad {}
-\ln\biggl(\frac{\mu m_2}{m_2^2-\bar p_{20}^2}\biggr)
+\ln(1-x_{31}^2)
\biggr]
+\ln^2(x_{21})
+\ln^2(x_{32})
\nn\\[.3em]
&& \quad {}
\hspace{-15em}
+\Li(x_{31}^2)
+\sum_{k,l=\pm1} \cLi(x_{31},x_{21}^k,x_{32}^l)
-\frac{\pi^2}{6}
\Biggr\},
\label{eq:DOc0s1dimreg}
\\[1em]
\DOp{m_1^2}{p_{21}^2}{p_{32}^2}{m_3^2}%
{p_{20}^2}{p_{31}^2}% 
\DOm{\la^2}{m_1^2}{m_2^2}{m_3^2}{(n_{\soft}=1)} 
&=&
\frac{x_{31}}{m_1 m_3(p_{20}^2-m_2^2)(1-x_{31}^2)}\Biggl\{
2\ln(x_{31}) \biggl[
-\ln\biggl(\frac{m_2\la}{m_2^2-\bar p_{20}^2}\biggr)
\nn\\[-3em]
&& \quad {}
+\ln(1-x_{31}^2)
\biggr]
+\ln^2(x_{21})
+\ln^2(x_{32})
+\Li(x_{31}^2)
\nn\\[.3em]
&& \quad {}
\hspace{-15em}
+\sum_{k,l=\pm1} \cLi(x_{31},x_{21}^k,x_{32}^l)
-\frac{\pi^2}{6}
\Biggr\},
\label{eq:DOc0s1massreg}
\eeqar
where the shorthands $x_{ij}$ of \refeq{eq:xij} are used.
Note that the precise definition of $x_{ij}$, 
which implies $|x_{ij}|\le1$, is important for the
analytical continuation.
Owing to the on-shell conditions the masses $m_1$ and $m_3$ are
real here, but $m_2$ can be complex.

The results for $m_2=0$ are obtained by the substitution
\beqar
&& \hspace{-2em}
-2\ln(x_{31})
\ln\biggl(\frac{m_2\nu}{m_2^2-\bar p_{20}^2}\biggr)
+\ln^2(x_{21})
+\ln^2(x_{32})
+\sum_{k,l=\pm1} \cLi(x_{31},x_{21}^k,x_{32}^l)
\nl
&\to\;&
\ln(x_{31})\biggl[
-\frac{1}{2}\ln(x_{31})-\ln\biggl(\frac{\nu^2}{m_1m_3}\biggr)
-\ln\biggl(\frac{m_1^2-\bar p_{21}^2}{-\bar p_{20}^2}\biggr)
-\ln\biggl(\frac{m_3^2-\bar p_{32}^2}{-\bar p_{20}^2}\biggr)
\biggr]
\nn\\[.3em]
&& \quad {}
+\frac{1}{2}\ln^2(y)
+\sum_{l=\pm1} \cLi(x_{31},y^l)
\eeqar
with $\nu=\mu$ or $\nu=\la$,
where
\beq
y=\frac{x_{21}}{x_{32}}
=\frac{m_1(m_3^2-\bar p_{32}^2)}{m_3(m_1^2-\bar p_{21}^2)}.
\eeq

The dimensionally regularized integral \refeq{eq:DOc0s1dimreg}
corresponds to ``Box 16'' of \citere{Ellis:2007qk}.

The following three cases contain two soft singularities.  Their
finite part is given by
\beq
\DOfin = D_0 - \frac{C_0(0)+C_0(2)}{p_{20}^2},
\eeq
i.e.\ the full expressions for the $D_0$ functions can be deduced from
just one out of the three cases upon changing the regularization by
subtracting and re-adding 3-point integrals (see Appendix~B of
\citere{Dittmaier:2003bc}).  The purely mass-regularized case
\refeq{eq:DOc0s2massreg} can be read off from (2.13) [case (iii)] of
\citere{Beenakker:1990jr} (see also \citere{VanNieuwenhuizen:1971yn}),
so that the full results read
\beqar
\DOp{m_1^2}{m_1^2}{m_3^2}{m_3^2}%
{p_{20}^2}{p_{31}^2}% 
\DOm{0}{m_1^2}{0}{m_3^2}{(n_{\soft}=2)}
&=&
\frac{-2x_{31}\ln(x_{31})}{m_1 m_3p_{20}^2(1-x_{31}^2)}
\Biggl\{
\frac{c_\eps}{\eps}+\ln\biggl(\frac{\mu^2}{-\bar p_{20}^2}\biggr)
\Biggr\},
\label{eq:DOc0s2dimreg}
\\[1em]
\DOp{m_1^2}{m_1^2}{m_3^2}{m_3^2}%
{p_{20}^2}{p_{31}^2}% 
\DOm{0}{m_1^2}{\la^2}{m_3^2}{(n_{\soft}=2)} 
&=&
\frac{-2x_{31}\ln(x_{31})}{m_1 m_3p_{20}^2(1-x_{31}^2)}
\Biggl\{
\frac{c_\eps}{2\eps}
+\ln\biggl(\frac{\mu\la}{-\bar p_{20}^2}\biggr)
\Biggr\},
\hspace{2em}
\label{eq:DOc0s2dimmassreg}
\\[1em]
\DOp{m_1^2}{m_1^2}{m_3^2}{m_3^2}%
{p_{20}^2}{p_{31}^2}% 
\DOm{\la_1^2}{m_1^2}{\la_2^2}{m_3^2}{(n_{\soft}=2)} 
&=&
\frac{-2x_{31}\ln(x_{31})}{m_1 m_3p_{20}^2(1-x_{31}^2)}
\ln\biggl(\frac{\la_1\la_2}{-\bar p_{20}^2}\biggr),
\label{eq:DOc0s2massreg}
\eeqar
where $x_{31}$ is again defined as in \refeq{eq:xij}.
Owing to the on-shell conditions the masses $m_1$ and $m_3$ are
real.

The dimensionally regularized integral \refeq{eq:DOc0s2dimreg}
corresponds to  ``Box 14'' of
\citere{Ellis:2007qk}.

\paragraph{Cases with one collinear singularity $(n_{\coll}=1)$}

The following three cases do not contain soft singularities and have
the common finite part
\beq
\DOfin = D_0 - \frac{(p_{31}^2-p_{30}^2)C_0(2)+(p_{20}^2-p_{21}^2)C_0(3)}
{(p_{20}^2-m_2^2)(p_{31}^2-m_3^2)-(p_{21}^2-m_2^2)(p_{30}^2-m_3^2)}.
\eeq
The following results are valid for complex masses $m_2$ and $m_3$ and
$m_3\ne0$. For \refeq{eq:DOc1s0dimreg} and \refeq{eq:DOc1s0massreg2}
results for $m_3=0$ and $m_2\ne0$ result from a permutation of the
arguments, for  \refeq{eq:DOc1s0massreg1} we also give a result that
is valid for $m_3=0$ and $m_2\ne0$:
\beqar
%\DOp{0}{p_{21}^2}{p_{32}^2}{p_{30}^2}%
%{p_{20}^2}{p_{31}^2}% 
%\DOm{0}{0}{m_2^2}{m_3^2}{(n_{\soft}=0)} 
%&=&
%\frac{1}{(p_{20}^2-m_2^2)(p_{31}^2-m_3^2)-(p_{21}^2-m_2^2)(p_{30}^2-m_3^2)}
%\nn\\[-3em]
%&& \quad {}
%\times\Biggl\{ 
%\biggl[ 
%\frac{c_\eps}{\eps}
%+\ln\biggl(\frac{\mu^2}{m_3^2-\bar p_{31}^2}\biggr)
%+\ln\biggl(\frac{m_3^2}{m_3^2-\bar p_{30}^2}\biggr) \biggr]
%\nn\\[.3em]
%&& \quad {}
%\hspace{-15em}
%\times[\ln(x_1)-\ln(x_2)]
%-\ln\biggl(\frac{m_3^2-\bar p_{31}^2}{m_3^2-\bar p_{30}^2}\biggr)
%\ln\biggl(\frac{m_2^2-\bar p_{20}^2}{m_2^2-\bar p_{21}^2}\biggr)
%+2\Li\biggl(\frac{\bar p_{20}^2-\bar p_{21}^2}{m_2^2-\bar p_{21}^2}\biggr)
%\nn\\[.3em]
%&& \quad {}
%\hspace{-15em}
%+2\Li\biggl(\frac{\bar p_{31}^2-\bar p_{30}^2}{m_3^2-\bar p_{30}^2}\biggr)
%-2\cLi(x_2,x_1^{-1})
%+\sum_{k=1}^2 \sum_{l=3}^4 (-1)^k \cLi(x_k,x_l) \Biggr\},
%\label{eq:DOc1s0dimreg}
%%
%\\[1em]\alt
\DOp{0}{p_{21}^2}{p_{32}^2}{p_{30}^2}%
{p_{20}^2}{p_{31}^2}% 
\DOm{0}{0}{m_2^2}{m_3^2}{(n_{\soft}=0)} 
&=&
\frac{1}{(p_{20}^2-m_2^2)(p_{31}^2-m_3^2)-(p_{21}^2-m_2^2)(p_{30}^2-m_3^2)}
\nn\\[-3em]
&& \quad {}
\times\Biggl\{ 
\biggl[ 
\frac{c_\eps}{\eps}
+2\ln\biggl(\frac{\mu m_3}{m_3^2-\bar p_{30}^2}\biggr) \biggr]
%\times
[\ln(x_1)-\ln(x_2)]
\nn\\[.3em]
&& \quad {}
\hspace{-15em}
+2\Li\biggl(\frac{\bar p_{20}^2-\bar p_{21}^2}{m_2^2-\bar p_{21}^2}\biggr)
-2\Li\biggl(\frac{\bar p_{30}^2-\bar p_{31}^2}{m_3^2-\bar p_{31}^2}\biggr)
-2\cLi(x_2,x_1^{-1})
\nn\\[.3em]
&& \quad {}
\hspace{-15em}
+\sum_{k=1}^2 \sum_{l=3}^4 (-1)^k \cLi(x_k,x_l) \Biggr\},
\label{eq:DOc1s0dimreg}
\\[1em]
%
%\DOp{\la^2}{p_{21}^2}{p_{32}^2}{p_{30}^2}%
%{p_{20}^2}{p_{31}^2}% 
%\DOm{0}{\la^2}{m_2^2}{m_3^2}{(n_{\soft}=0)} 
%&=&
%\frac{1}{(p_{20}^2-m_2^2)(p_{31}^2-m_3^2)-(p_{21}^2-m_2^2)(p_{30}^2-m_3^2)}
%\nn\\[-3em]
%&& \quad {}
%\times\Biggl\{ 
%\biggl[ \ln\biggl(\frac{\la^2}{m_3^2-\bar p_{31}^2}\biggr)
%+\ln\biggl(\frac{m_3^2}{m_3^2-\bar p_{30}^2}\biggr) \biggr]
%\nn\\[.3em]
%&& \quad {}
%\hspace{-15em}
%\times[\ln(x_1)-\ln(x_2)]
%-\ln\biggl(\frac{m_3^2-\bar p_{31}^2}{m_3^2-\bar p_{30}^2}\biggr)
%\ln\biggl(\frac{m_2^2-\bar p_{20}^2}{m_2^2-\bar p_{21}^2}\biggr)
%\nn\\[.3em]
%&& \quad {}
%\hspace{-15em}
%-2\cLi(x_2,x_1^{-1})
%+\sum_{k=1}^2 \sum_{l=3}^4 (-1)^k \cLi(x_k,x_l) \Biggr\},
%\label{eq:DOc1s0massreg1}
%%
%\\[1em]\alt
%
\DOp{\la^2}{p_{21}^2}{p_{32}^2}{p_{30}^2}%
{p_{20}^2}{p_{31}^2}% 
\DOm{0}{\la^2}{m_2^2}{m_3^2}{(n_{\soft}=0)} 
&=&
\frac{1}{(p_{20}^2-m_2^2)(p_{31}^2-m_3^2)-(p_{21}^2-m_2^2)(p_{30}^2-m_3^2)}
\nn\\[-3em]
&& \quad {}
\times\Biggl\{ 
2\ln\biggl(\frac{\la m_3}{m_3^2-\bar p_{30}^2}\biggr)[\ln(x_1)-\ln(x_2)]
\nn\\[.3em]
&& \quad {}
\hspace{-15em}
-\ln^2\biggl(\frac{m_2^2-\bar p_{20}^2}{m_2^2-\bar p_{21}^2}\biggr)
+\ln^2\biggl(\frac{m_3^2-\bar p_{30}^2}{m_3^2-\bar p_{31}^2}\biggr)
-2\cLi(x_2,x_1^{-1})
+\sum_{k=1}^2 \sum_{l=3}^4 (-1)^k \cLi(x_k,x_l) \Biggr\}
\nn\\[.3em]
&&{}
\hspace{-15em}
=\frac{1}{(p_{20}^2-m_2^2)(p_{31}^2-m_3^2)-(p_{21}^2-m_2^2)(p_{30}^2-m_3^2)}
\Biggl\{ 
2\ln\biggl(\frac{\la m_2}{m_2^2-{\bar p_{21}^2}}\biggr)[\ln(x_1)-\ln(x_2)]
\nn\\[.3em]
&& \quad {}
\hspace{-15em}
-2\cLi(x_2,x_1^{-1})
-\sum_{k=1}^2 \sum_{l=3}^4 (-1)^k \cLi(x_k^{-1},x_l^{-1}) \Biggr\},
\label{eq:DOc1s0massreg1}
\\[1em]
%
%\DOp{0}{p_{21}^2}{p_{32}^2}{p_{30}^2}%
%{p_{20}^2}{p_{31}^2}% 
%\DOm{\la^2}{\la^2}{m_2^2}{m_3^2}{(n_{\soft}=0)} 
%&=&
%\frac{1}{(p_{20}^2-m_2^2)(p_{31}^2-m_3^2)-(p_{21}^2-m_2^2)(p_{30}^2-m_3^2)}
%\nn\\[-3em]
%&& \quad {}
%\times\Biggl\{ 
%\biggl[ \ln\biggl(\frac{\la^2}{m_3^2-\bar p_{31}^2}\biggr)
%+\ln\biggl(\frac{m_3^2}{m_3^2-\bar p_{30}^2}\biggr) \biggr]
%\nn\\[.3em]
%&& \quad {}
%\hspace{-15em}
%\times[\ln(x_1)-\ln(x_2)]
%-\ln\biggl(\frac{m_3^2-\bar p_{31}^2}{m_3^2-\bar p_{30}^2}\biggr)
%\ln\biggl(\frac{m_2^2-\bar p_{20}^2}{m_2^2-\bar p_{21}^2}\biggr)
%+2\Li\biggl(\frac{\bar p_{20}^2-\bar p_{21}^2}{m_2^2-\bar p_{21}^2}\biggr)
%\nn\\[.3em]
%&& \quad {}
%\hspace{-15em}
%+2\Li\biggl(\frac{\bar p_{31}^2-\bar p_{30}^2}{m_3^2-\bar p_{30}^2}\biggr)
%-2\cLi(x_2,x_1^{-1})
%+\sum_{k=1}^2 \sum_{l=3}^4 (-1)^k \cLi(x_k,x_l) \Biggr\},
%\label{eq:DOc1s0massreg2}
%\\[1em]\alt
%
\DOp{0}{p_{21}^2}{p_{32}^2}{p_{30}^2}%
{p_{20}^2}{p_{31}^2}% 
\DOm{\la^2}{\la^2}{m_2^2}{m_3^2}{(n_{\soft}=0)} 
&=&
\frac{1}{(p_{20}^2-m_2^2)(p_{31}^2-m_3^2)-(p_{21}^2-m_2^2)(p_{30}^2-m_3^2)}
\nn\\[-3em]
&& \quad {}
\times\Biggl\{ 
2\ln\biggl(\frac{\la m_3}{m_3^2-\bar p_{30}^2}\biggr)[\ln(x_1)-\ln(x_2)]
\nn\\[.3em]
&& \quad {}
\hspace{-15em}
+2\Li\biggl(\frac{\bar p_{20}^2-\bar p_{21}^2}{m_2^2-\bar p_{21}^2}\biggr)
-2\Li\biggl(\frac{\bar p_{30}^2-\bar p_{31}^2}{m_3^2-\bar p_{31}^2}\biggr)
-2\cLi(x_2,x_1^{-1})
\nn\\[.3em]
&& \quad {}
\hspace{-15em}
+\sum_{k=1}^2 \sum_{l=3}^4 (-1)^k \cLi(x_k,x_l) \Biggr\},
\label{eq:DOc1s0massreg2}
%\\[1em]\alt
%%
%\Green{\DOp{0}{p_{21}^2}{p_{32}^2}{p_{30}^2}%
%{p_{20}^2}{p_{31}^2}% 
%\DOm{\la^2}{\la^2}{m_2^2}{m_3^2}{(n_{\soft}=0)} 
%&=&
%\frac{1}{(p_{20}^2-m_2^2)(p_{31}^2-m_3^2)-(p_{21}^2-m_2^2)(p_{30}^2-m_3^2)}
%\nn\\[-3em]
%&& \quad {}
%\times\Biggl\{ 
%2\ln\biggl(\frac{\la m_2}{m_2^2-\bar p_{21}^2}\biggr)[\ln(x_1)-\ln(x_2)]
%\nn\\[.3em]
%&& \quad {}
%\hspace{-15em}
%-2\Li\biggl(\frac{\bar p_{21}^2-\bar p_{20}^2}{m_2^2-\bar p_{20}^2}\biggr)
%+2\Li\biggl(\frac{\bar p_{31}^2-\bar p_{30}^2}{m_3^2-\bar p_{30}^2}\biggr)
%-2\cLi(x_2,x_1^{-1})
%\nn\\[.3em]
%&& \quad {}
%\hspace{-15em}
%-\sum_{k=1}^2 \sum_{l=3}^4 (-1)^k \cLi(x_k^{-1},x_l^{-1}) \Biggr\},
%\label{eq:DOc1s0massreg2}
%}%
\eeqar
with the quantities
\beqar
x_1 &=& \frac{m_3^2-\bar p_{30}^2}{m_2^2-\bar p_{20}^2}, \qquad
x_2 = \frac{m_3^2-\bar p_{31}^2}{m_2^2-\bar p_{21}^2}, \\[1ex]
x_{3,4} &=& \frac{m_2^2+m_3^2-\bar p_{32}^2 
\pm \sqrt{\la(\bar p_{32}^2,m_2^2,m_3^2)}}{2m_3^2}
=\frac{2m_2^2}{m_2^2+m_3^2-\bar p_{32}^2 
\mp \sqrt{\la(\bar p_{32}^2,m_2^2,m_3^2)}}.\nn
\eeqar
We remark that 
$x_{3} = ({m_2}/{m_3})x_{32}^{-1}=1/r_{23,1}=r_{32,2}$, 
$x_{4} = ({m_2}/{m_3})x_{32}=1/r_{23,2}=r_{32,1}$,
with $x_{32}$ and $r_{23,k}$ defined in \refeq{eq:xij} and
\refeq{rijdef}, respectively.

%The special cases $m_2=0$ or $m_3=0$ do not lead to singularities
%and can be easily obtained from the above formulas, if necessary upon
%interchanging the respective propagators.
The results for $m_2=m_3=0$ are obtained by the substitutions
\beqar
&& \hspace{-2em}
\sum_{k=1}^2 \sum_{l=3}^4 (-1)^k \cLi(x_k,x_l)
+ 2\ln\biggl(\frac{\nu m_3}{m_3^2-\bar p_{30}^2}\biggr)[\ln(x_1)-\ln(x_2)]
\nl
&\to\;&
\frac{1}{2}\ln^2(x_1)-\frac{1}{2}\ln^2(x_2)
+ \biggl[\ln\biggl(\frac{\bar p_{32}^2}{\bar p_{30}^2}\biggr)
+ \ln\biggl(\frac{\nu^2}{-\bar p_{30}^2}\biggr)\biggr]
[\ln(x_1)-\ln(x_2)]
\eeqar
with $\nu=\la$ or $\nu=\mu$.

The dimensionally regularized integral \refeq{eq:DOc1s0dimreg}
corresponds to  ``Box~13'' of
\citere{Ellis:2007qk}.

The following four cases with no or one soft singularity have
the common finite part
\beq
\DOfin = D_0 - \frac{C_0(2)}{p_{20}^2-m_2^2}
- \frac{(p_{20}^2-p_{21}^2)C_0(3)}{(p_{20}^2-m_2^2)(p_{31}^2-m_3^2)}.
\eeq
Here the mass $m_3$ is real, but $m_2$ can be complex,
\beqar
%\DOp{0}{p_{21}^2}{p_{32}^2}{m_3^2}%
%{p_{20}^2}{p_{31}^2}% 
%\DOm{0}{0}{m_2^2}{m_3^2}{(n_{\soft}=1)} 
%&=&
%\frac{1}{(p_{20}^2-m_2^2)(p_{31}^2-m_3^2)}\Biggl\{
%\frac{c_\eps}{2\eps^2}
%+ \frac{c_\eps}{\eps} \biggl[
%\ln\biggl(\frac{\mu m_3}{m_3^2-\bar p_{31}^2}\biggr)
%\nn\\[-3em]
%&& \quad {}
%+\ln\biggl(\frac{m_2^2-\bar p_{21}^2}{m_2^2-\bar p_{20}^2}\biggr)
%\biggr]
%+2\ln\biggl(\frac{\mu m_2}{m_2^2-\bar p_{20}^2}\biggr)
%\ln\biggl(\frac{\mu m_3}{m_3^2-\bar p_{31}^2}\biggr)
%\nn\\[.3em]
%&& \quad {}
%\hspace{-15em}
%-\ln^2\biggl(\frac{\mu m_2}{m_2^2-\bar p_{21}^2}\biggr)
%-\ln^2(x_{32})
%-2\Li\biggl(\frac{\bar p_{21}^2-\bar p_{20}^2}{m_2^2-\bar p_{20}^2}\biggr)
%-\sum_{l=\pm1} \cLi\biggl(\frac{m_3^2}{m_3^2-\bar p_{31}^2},
%\frac{m_2^2-\bar p_{21}^2}{m_2 m_3},x_{32}^l\biggr)
%\nn\\[.3em]
%&& \quad {}
%\hspace{-15em}
%-\frac{\pi^2}{6}
%\Biggr\},
%\label{eq:DOc1s1dimregonsh3}
%%
%\\[1em]\alt
\DOp{0}{p_{21}^2}{p_{32}^2}{m_3^2}%
{p_{20}^2}{p_{31}^2}% 
\DOm{0}{0}{m_2^2}{m_3^2}{(n_{\soft}=1)} 
&=&
\frac{1}{(p_{20}^2-m_2^2)(p_{31}^2-m_3^2)}\Biggl\{
\frac{c_\eps}{2\eps^2}
+ \frac{c_\eps}{\eps} \biggl[
\ln\biggl(\frac{\mu m_3}{m_3^2-\bar p_{31}^2}\biggr)
\nn\\[-3em]
&& \quad {}
+\ln\biggl(\frac{m_2^2-\bar p_{21}^2}{m_2^2-\bar p_{20}^2}\biggr)
\biggr]
+2\ln\biggl(\frac{m_2^2-\bar p_{21}^2}{m_2^2-\bar p_{20}^2}\biggr)
\ln\biggl(\frac{\mu m_3}{m_3^2-\bar p_{31}^2}\biggr)
\nn\\[.3em]
&& \quad {}
\hspace{-15em}
+\ln^2\biggl(\frac{\mu m_3}{m_3^2-\bar p_{31}^2}\biggr)
-2\Li\biggl(\frac{\bar p_{21}^2-\bar p_{20}^2}{m_2^2-\bar p_{20}^2}\biggr)
+\sum_{l=\pm1} \cLi\biggl(\frac{m_3^2-\bar p_{31}^2}{m_2^2-\bar
  p_{21}^2},\frac{m_2}{ m_3}x_{32}^l\biggr)
-\frac{\pi^2}{6}
\Biggr\},
\label{eq:DOc1s1dimregonsh3}
\\[1em]
%
%\DOp{\la^2}{p_{21}^2}{p_{32}^2}{m_3^2}%
%{p_{20}^2}{p_{31}^2}% 
%\DOm{0}{\la^2}{m_2^2}{m_3^2}{(n_{\soft}=1)} 
%&=&
%\frac{1}{(p_{20}^2-m_2^2)(p_{31}^2-m_3^2)}\Biggl\{
%\frac{c_\eps}{\eps}
%\ln\biggl(\frac{\la m_3}{m_3^2-\bar p_{31}^2}\biggr)
%\nn\\[-3em]
%&& \quad {}
%+2\ln\biggl(\frac{\mu m_2}{m_2^2-\bar p_{20}^2}\biggr)
%\ln\biggl(\frac{\la m_3}{m_3^2-\bar p_{31}^2}\biggr)
%\nn\\[.3em]
%&& \quad {}
%\hspace{-15em}
%-\ln^2\biggl(\frac{\la m_2}{m_2^2-\bar p_{21}^2}\biggr)
%-\ln^2(x_{32})
%-\sum_{l=\pm1} \cLi\biggl(\frac{m_3^2}{m_3^2-\bar p_{31}^2},
%\frac{m_2^2-\bar p_{21}^2}{m_2 m_3},x_{32}^l\biggr)
%-\frac{\pi^2}{6}
%\Biggr\},
%\label{eq:DOc1s1dimmassreg1}
%%
%\\[1em]\alt
%
\DOp{\la^2}{p_{21}^2}{p_{32}^2}{m_3^2}%
{p_{20}^2}{p_{31}^2}% 
\DOm{0}{\la^2}{m_2^2}{m_3^2}{(n_{\soft}=1)} 
&=&
\frac{1}{(p_{20}^2-m_2^2)(p_{31}^2-m_3^2)}\Biggl\{
\nn\\[-3em]
&& \quad {}
\biggl[ 
\frac{c_\eps}{\eps}
+2\ln\biggl(\frac{m_2^2-\bar p_{21}^2}{m_2^2-\bar p_{20}^2}\biggr)
+\ln\biggl(\frac{\mu^2}{\la^2}\biggr)
\biggr]
\ln\biggl(\frac{\la m_3}{m_3^2-\bar p_{31}^2}\biggr)
\nn\\[.3em]
&& \quad {}
\hspace{-15em}
+\ln^2\biggl(\frac{\la m_3}{m_3^2-\bar p_{31}^2}\biggr)
+\sum_{l=\pm1} 
\cLi\biggl(\frac{m_3^2-\bar p_{31}^2}{m_2^2-\bar p_{21}^2},
\frac{m_2}{ m_3}x_{32}^l\biggr)
-\frac{\pi^2}{6}
\Biggr\},
\label{eq:DOc1s1dimmassreg1}
\\[1em]
%
%\DOp{\la^2}{p_{21}^2}{p_{32}^2}{m_3^2}%
%{p_{20}^2}{p_{31}^2}% 
%\DOm{\la^2}{0}{m_2^2}{m_3^2}{(n_{\soft}=0)} 
%&=&
%\frac{1}{(p_{20}^2-m_2^2)(p_{31}^2-m_3^2)}\Biggl\{
%-\ln^2\biggl(\frac{\la m_2}{m_2^2-\bar p_{20}^2}\biggr)
%\nn\\[-3em]
%&& \quad {}
%+2\ln\biggl(\frac{\la m_2}{m_2^2-\bar p_{20}^2}\biggr)
%\biggl[
%\ln\biggl(\frac{\la m_3}{m_3^2-\bar p_{31}^2}\biggr)
%+\ln\biggl(\frac{m_2^2-\bar p_{21}^2}{m_2^2-\bar p_{20}^2}\biggr)
%\biggr]
%\nn\\[.3em]
%&& \quad {}
%\hspace{-15em}
%-\ln^2(x_{32})
%-\sum_{l=\pm1} \cLi\biggl(\frac{m_3^2}{m_3^2-\bar p_{31}^2},
%\frac{m_2^2-\bar p_{21}^2}{m_2 m_3},x_{32}^l\biggr)
%+\frac{\pi^2}{4}
%\Biggr\},
%\label{eq:DOc1s1dimmassreg2}
%%
%\\[1em]\alt
%
\DOp{\la^2}{p_{21}^2}{p_{32}^2}{m_3^2}%
{p_{20}^2}{p_{31}^2}% 
\DOm{\la^2}{0}{m_2^2}{m_3^2}{(n_{\soft}=0)} 
&=&
\frac{1}{(p_{20}^2-m_2^2)(p_{31}^2-m_3^2)}\Biggl\{
\nn\\[-3em]&& \quad {}
\biggl[\ln\biggl(\frac{\la m_3}{m_3^2-\bar p_{31}^2}\biggr)
+\ln\biggl(\frac{m_2^2-\bar p_{21}^2}{m_2^2-\bar p_{20}^2}\biggr)
\biggr]^2
\nn\\[.3em]
&& \quad {}
\hspace{-15em}
+\sum_{l=\pm1}\cLi\biggl(\frac{m_3^2-\bar p_{31}^2}{m_2^2-\bar p_{21}^2},\frac{m_2}{m_3}x_{32}^l\biggr)
+\frac{\pi^2}{4}
\Biggr\},
\label{eq:DOc1s1dimmassreg2}
\\[1em]
%
%\DOp{0}{p_{21}^2}{p_{32}^2}{m_3^2}%
%{p_{20}^2}{p_{31}^2}% 
%\DOm{\la^2}{\la^2}{m_2^2}{m_3^2}{(n_{\soft}=0)} 
%&=&
%\frac{1}{(p_{20}^2-m_2^2)(p_{31}^2-m_3^2)}\Biggl\{
%2\ln\biggl(\frac{\la m_2}{m_2^2-\bar p_{20}^2}\biggr)
%\ln\biggl(\frac{\la m_3}{m_3^2-\bar p_{31}^2}\biggr)
%\nn\\[-3em]
%&& \quad {}
%-\ln^2\biggl(\frac{\la m_2}{m_2^2-\bar p_{21}^2}\biggr)
%-\ln^2(x_{32})
%-2\Li\biggl(\frac{\bar p_{21}^2-\bar p_{20}^2}{m_2^2-\bar p_{20}^2}\biggr)
%\nn\\[.3em]
%&& \quad {}
%\hspace{-15em}
%-\sum_{l=\pm1} \cLi\biggl(\frac{m_3^2}{m_3^2-\bar p_{31}^2},
%\frac{m_2^2-\bar p_{21}^2}{m_2 m_3},x_{32}^l\biggr)
%-\frac{\pi^2}{12}
%\Biggr\},
%\label{eq:DOc1s1massreg}
%\\[1em]\alt
%
\DOp{0}{p_{21}^2}{p_{32}^2}{m_3^2}%
{p_{20}^2}{p_{31}^2}% 
\DOm{\la^2}{\la^2}{m_2^2}{m_3^2}{(n_{\soft}=0)} 
&=&
\frac{1}{(p_{20}^2-m_2^2)(p_{31}^2-m_3^2)}\Biggl\{
\nn\\[-3em]&& \quad {}
\biggl[\ln\biggl(\frac{\la m_3}{m_3^2-\bar p_{31}^2}\biggr)
+\ln\biggl(\frac{m_2^2-\bar p_{21}^2}{m_2^2-\bar p_{20}^2}\biggr)\biggr]^2
\nn\\[.3em]
&& \quad {}
\hspace{-15em}
+2\Li\biggl(\frac{\bar p_{20}^2-\bar p_{21}^2}{m_2^2-\bar p_{21}^2}\biggr)
+\sum_{l=\pm1} \cLi\biggl(\frac{m_3^2-\bar p_{31}^2}{m_2^2-\bar
  p_{21}^2},\frac{m_2}{m_3}x_{32}^l\biggr)
-\frac{\pi^2}{12}
\Biggr\},
%}
\label{eq:DOc1s1massreg}
\eeqar
with $x_{32}$ defined as in \refeq{eq:xij}.
We note that $({m_2}/{m_3})x_{32}^{-1}=1/r_{23,1}=r_{32,2}$,
$({m_2}/{m_3})x_{32}=1/r_{23,2}=r_{32,1}$, with $r_{ij,k}$ defined in
\refeq{rijdef}.

The above results are also valid for $m_2=0$, even if $p_{32}^2=m_3^2$.
For $m_2\to0$ the variable $x_{32}$ behaves as
\beq
\frac{m_2}{m_3}x_{32}^{+1} = {\cal O}(m_2^2), \qquad
\frac{m_2}{m_3}x_{32}^{-1} = \frac{m_3^2-\bar p_{32}^2}{m_3^2}+{\cal O}(m_2^2).
\eeq

The dimensionally regularized integral \refeq{eq:DOc1s1dimregonsh3} can be
found in (A.4) of \citere{Beenakker:2002nc} and corresponds to ``Box
12'' of \citere{Ellis:2007qk}. 
Integral \refeq{eq:DOc1s1dimmassreg1} is case (ia) of
\citere{Beenakker:1990jr} (modulo the translation from mass
to dimensional regularization which is trivial for the purely 
soft-singular case) and corresponds to 
\refeq{eq:DOc0s1dimreg} for small $m_1$.

The following three cases with up to two soft singularities have
the common finite part
\beq
\DOfin = D_0 - \frac{C_0(2)}{p_{20}^2-m_2^2}
- \frac{C_0(3)}{p_{31}^2-m_3^2}.
\eeq
Here $m_2$ and $m_3$ have to be real quantities,
\beqar
\DOp{0}{m_2^2}{p_{32}^2}{m_3^2}%
{p_{20}^2}{p_{31}^2}% 
\DOm{0}{0}{m_2^2}{m_3^2}{(n_{\soft}=2)} 
&=&
\frac{1}{(p_{20}^2-m_2^2)(p_{31}^2-m_3^2)}\Biggl\{
\frac{c_\eps}{\eps^2}
+\frac{c_\eps}{\eps}
\ln\biggl(\frac{\mu m_2}{m_2^2-\bar p_{20}^2}\biggr)
\nn\\*[-3em]
&& \quad {}
+\frac{c_\eps}{\eps}
\ln\biggl(\frac{\mu m_3}{m_3^2-\bar p_{31}^2}\biggr)
+2\ln\biggl(\frac{\mu m_2}{m_2^2-\bar p_{20}^2}\biggr)
\ln\biggl(\frac{\mu m_3}{m_3^2-\bar p_{31}^2}\biggr)
\nn\\*[.3em]
&& \quad {}
\hspace{-15em}
-\ln^2(x_{32})
-\frac{2\pi^2}{3}
\Biggr\},
\label{eq:DOc1s2dimreg}
\\[1em]
\DOp{\la^2}{m_2^2}{p_{32}^2}{m_3^2}%
{p_{20}^2}{p_{31}^2}% 
\DOm{0}{\la^2}{m_2^2}{m_3^2}{(n_{\soft}=1)} 
&=&
\frac{1}{(p_{20}^2-m_2^2)(p_{31}^2-m_3^2)}\Biggl\{
\frac{c_\eps}{\eps} 
\ln\biggl(\frac{\la m_3}{m_3^2-\bar p_{31}^2}\biggr)
\nn\\[-3em]
&& \quad {}
+2\ln\biggl(\frac{\mu m_2}{m_2^2-\bar p_{20}^2}\biggr)
\ln\biggl(\frac{\la m_3}{m_3^2-\bar p_{31}^2}\biggr)
-\ln^2(x_{32})
-\frac{\pi^2}{4}
\Biggr\},
\nn\\
\label{eq:DOc1s2dimmassreg}
\\[1em]
\DOp{0}{m_2^2}{p_{32}^2}{m_3^2}%
{p_{20}^2}{p_{31}^2}% 
\DOm{\la^2}{\la^2}{m_2^2}{m_3^2}{(n_{\soft}=0)} 
&=&
\frac{1}{(p_{20}^2-m_2^2)(p_{31}^2-m_3^2)}\Biggl\{
2\ln\biggl(\frac{\la m_2}{m_2^2-\bar p_{20}^2}\biggr)
\ln\biggl(\frac{\la m_3}{m_3^2-\bar p_{31}^2}\biggr)
\nn\\[-3em]
&& \quad {}
-\ln^2(x_{32})
-\frac{\pi^2}{2}
\Biggr\},
\label{eq:DOc1s2massregintern}
\eeqar
with again $x_{32}$ defined as in \refeq{eq:xij}.

The dimensionally regularized integral \refeq{eq:DOc1s2dimreg} can be
found in (A.4) of \citere{Beenakker:2002nc} and corresponds to ``Box
11'' of \citere{Ellis:2007qk}.
Integral \refeq{eq:DOc1s2dimmassreg} can be calculated from case (i)
of \citere{Beenakker:1990jr} for $m_0^2=m_2^2$ and $m_1$ small 
(with the trivial translation to dimensional regularization) or from
\refeq{eq:DOc0s1dimreg} for $p_{21}^2=m_2^2$ and small $m_1$.

\paragraph{Cases with two collinear singularities on adjacent legs
$(n_{\coll}=2)$}

Here we distinguish six different cases with no or one soft
singularity which have the following common finite part
\beq
\DOfin = D_0 
- \frac{(p_{31}^2-p_{32}^2)C_0(0)+(p_{31}^2-p_{30}^2)C_0(2)}
{p_{20}^2(p_{31}^2-m_3^2)}
- \frac{C_0(3)}{p_{31}^2-m_3^2}.
\eeq
The mass $m_3$ can be complex,
\beqar
\DOp{0}{0}{p_{32}^2}{p_{30}^2}%
{p_{20}^2}{p_{31}^2}% 
\DOm{0}{0}{0}{m_3^2}{(n_{\soft}=1)} 
&=&
\frac{1}{p_{20}^2(p_{31}^2-m_3^2)}\Biggl\{
\frac{c_\eps}{\eps^2}
+\frac{c_\eps}{\eps} 
\biggl[ 
\ln\biggl(\frac{\mu^2}{-\bar p_{20}^2}\biggr)
+\ln\biggl(\frac{m_3^2-\bar p_{32}^2}{m_3^2-\bar p_{31}^2}\biggr)
\nn\\[-3em]
&& \quad {}
+\ln\biggl(\frac{m_3^2-\bar p_{30}^2}{m_3^2-\bar p_{31}^2}\biggr) \biggr]
+\frac{1}{2}\ln^2\biggl(\frac{\mu^2}{-\bar p_{20}^2}\biggr)
\nn\\[.3em]
&& \quad {}
\hspace{-15em}
+\ln\biggl(\frac{\mu^2}{-\bar p_{20}^2}\biggr)
\biggl[ \ln\biggl(\frac{m_3^2-\bar p_{32}^2}{m_3^2-\bar p_{31}^2}\biggr)
+\ln\biggl(\frac{m_3^2-\bar p_{30}^2}{m_3^2-\bar p_{31}^2}\biggr) \biggr]
-\frac{1}{2} \ln^2\biggl(\frac{m_3^2-\bar p_{32}^2}{m_3^2-\bar p_{30}^2}\biggr) 
\nn\\[.3em]
&& \quad {}
\hspace{-15em}
-2\Li\biggl(\frac{\bar p_{32}^2-\bar p_{31}^2}{m_3^2-\bar p_{31}^2}\biggr)
-2\Li\biggl(\frac{\bar p_{30}^2-\bar p_{31}^2}{m_3^2-\bar p_{31}^2}\biggr)
+\cLi\biggl(\frac{\bar p_{20}^2}{\bar p_{32}^2-m_3^2},
\frac{m_3^2}{m_3^2-\bar p_{30}^2}\biggr)
-\frac{\pi^2}{3}
\Biggr\},
\label{eq:DOc2s1dimreg}
\\[1em]
\DOp{\la^2}{0}{p_{32}^2}{p_{30}^2}%
{p_{20}^2}{p_{31}^2}% 
\DOm{\la^2}{0}{0}{m_3^2}{(n_{\soft}=1)} 
&=&
\frac{1}{p_{20}^2(p_{31}^2-m_3^2)}\Biggl\{
\frac{c_\eps}{2\eps^2}
+\frac{c_\eps}{\eps} \biggl[
\ln\biggl(\frac{\mu\la}{-\bar p_{20}^2}\biggr)
+\ln\biggl(\frac{m_3^2-\bar p_{32}^2}{m_3^2-\bar p_{31}^2}\biggr)
\biggr]
\nn\\[-3em]
&& \quad {}
-\ln^2\biggl(\frac{\mu}{\la}\biggr)
+\frac{1}{2}\ln^2\biggl(\frac{\mu^2}{-\bar p_{20}^2}\biggr)
\nn\\[.3em]
&& \quad {}
\hspace{-15em}
+\ln\biggl(\frac{\mu^2}{-\bar p_{20}^2}\biggr)
\ln\biggl(\frac{m_3^2-\bar p_{32}^2}{m_3^2-\bar p_{31}^2}\biggr)
+\ln\biggl(\frac{\la^2}{-\bar p_{20}^2}\biggr)
\ln\biggl(\frac{m_3^2-\bar p_{30}^2}{m_3^2-\bar p_{31}^2}\biggr) 
-\frac{1}{2} \ln^2\biggl(\frac{m_3^2-\bar p_{32}^2}{m_3^2-\bar p_{30}^2}\biggr) 
\nn\\[.3em]
&& \quad {}
\hspace{-15em}
-2\Li\biggl(\frac{\bar p_{32}^2-\bar p_{31}^2}{m_3^2-\bar p_{31}^2}\biggr)
+\cLi\biggl(\frac{\bar p_{20}^2}{\bar p_{32}^2-m_3^2},
\frac{m_3^2}{m_3^2-\bar p_{30}^2}\biggr)
-\frac{\pi^2}{3}
\Biggr\},
\label{eq:DOc2s1dimmassreg1}
\\[1em]
\DOp{\la_1^2}{\la_2^2}{p_{32}^2}{p_{30}^2}%
{p_{20}^2}{p_{31}^2}% 
\DOm{\la_1^2}{0}{\la_2^2}{m_3^2}{(n_{\soft}=1)} 
&=&
\frac{1}{p_{20}^2(p_{31}^2-m_3^2)}\Biggl\{
\frac{c_\eps}{\eps} 
\ln\biggl(\frac{\la_1\la_2}{-\bar p_{20}^2}\biggr)
+\ln\biggl(\frac{\mu^2}{-\bar p_{20}^2}\biggr)
\ln\biggl(\frac{\la_1\la_2}{-\bar p_{20}^2}\biggr)
\nn\\[-3em]
&& \quad {}
-\frac{1}{4}\ln^2\biggl(\frac{\la_1^2}{-\bar p_{20}^2}\biggr)
-\frac{1}{4}\ln^2\biggl(\frac{\la_2^2}{-\bar p_{20}^2}\biggr)
\nn\\[.3em]
&& \quad {}
\hspace{-15em}
+\ln\biggl(\frac{\la_1^2}{-\bar p_{20}^2}\biggr)
\ln\biggl(\frac{m_3^2-\bar p_{30}^2}{m_3^2-\bar p_{31}^2}\biggr)
+\ln\biggl(\frac{\la_2^2}{-\bar p_{20}^2}\biggr)
\ln\biggl(\frac{m_3^2-\bar p_{32}^2}{m_3^2-\bar p_{31}^2}\biggr)
-\frac{1}{2} \ln^2\biggl(\frac{m_3^2-\bar p_{32}^2}{m_3^2-\bar p_{30}^2}\biggr) 
\nn\\[.3em]
&& \quad {}
\hspace{-15em}
+\cLi\biggl(\frac{\bar p_{20}^2}{\bar p_{32}^2-m_3^2},
\frac{m_3^2}{m_3^2-\bar p_{30}^2}\biggr)
-\frac{\pi^2}{3}
\Biggr\},
\label{eq:DOc2s1dimmassreg2}
\\[1em]
\DOp{\la^2}{\la^2}{p_{32}^2}{p_{30}^2}%
{p_{20}^2}{p_{31}^2}% 
\DOm{0}{\la^2}{0}{m_3^2}{(n_{\soft}=0)} 
&=&
\frac{1}{p_{20}^2(p_{31}^2-m_3^2)}\Biggl\{
\frac{1}{2} \biggl[
\ln\biggl(\frac{\la^2}{-\bar p_{20}^2}\biggr)
+\ln\biggl(\frac{m_3^2-\bar p_{32}^2}{m_3^2-\bar p_{31}^2}\biggr)
\nn\\[-3em]
&& \quad {}
+\ln\biggl(\frac{m_3^2-\bar p_{30}^2}{m_3^2-\bar p_{31}^2}\biggr) \biggr]^2
\nn\\[.3em]
&& \quad {}
\hspace{-15em}
+\cLi\biggl(\frac{\bar p_{20}^2}{\bar p_{32}^2-m_3^2},
\frac{m_3^2}{m_3^2-\bar p_{30}^2}\biggr)
+\frac{\pi^2}{2}
\Biggr\},
\label{eq:DOc2s1massreg2}
\\[1em]
\DOp{0}{\la^2}{p_{32}^2}{p_{30}^2}%
{p_{20}^2}{p_{31}^2}% 
\DOm{\la^2}{\la^2}{0}{m_3^2}{(n_{\soft}=0)} 
&=&
\frac{1}{p_{20}^2(p_{31}^2-m_3^2)}\Biggl\{
\frac{1}{2} \biggl[
\ln\biggl(\frac{\la^2}{-\bar p_{20}^2}\biggr)
+\ln\biggl(\frac{m_3^2-\bar p_{32}^2}{m_3^2-\bar p_{31}^2}\biggr)
\nn\\[-3em]
&& \quad {}
+\ln\biggl(\frac{m_3^2-\bar p_{30}^2}{m_3^2-\bar p_{31}^2}\biggr) \biggr]^2
+2\Li\biggl(\frac{\bar p_{31}^2-\bar p_{30}^2}{m_3^2-\bar p_{30}^2}\biggr) 
\nn\\[.3em]
&& \quad {}
\hspace{-15em}
+\cLi\biggl(\frac{\bar p_{20}^2}{\bar p_{32}^2-m_3^2},
\frac{m_3^2}{m_3^2-\bar p_{30}^2}\biggr)
+\frac{\pi^2}{6}
\Biggr\},
\label{eq:DOc2s1massreg1}
\\[1em]
\DOp{0}{0}{p_{32}^2}{p_{30}^2}%
{p_{20}^2}{p_{31}^2}% 
\DOm{\la^2}{\la^2}{\la^2}{m_3^2}{(n_{\soft}=0)} 
&=&
\frac{1}{p_{20}^2(p_{31}^2-m_3^2)}\Biggl\{
\frac{1}{2}\ln^2\biggl(\frac{\la^2}{-\bar p_{20}^2}\biggr)
\nn\\[-3em]
&& \quad {}
+\ln\biggl(\frac{\la^2}{-\bar p_{20}^2}\biggr)
\biggl[ \ln\biggl(\frac{m_3^2-\bar p_{32}^2}{m_3^2-\bar p_{31}^2}\biggr)
+\ln\biggl(\frac{m_3^2-\bar p_{30}^2}{m_3^2-\bar p_{31}^2}\biggr) \biggr]
\nn\\[.3em]
&& \quad {}
\hspace{-15em}
-\frac{1}{2} \ln^2\biggl(\frac{m_3^2-\bar p_{32}^2}{m_3^2-\bar p_{30}^2}\biggr) 
-2\Li\biggl(\frac{\bar p_{32}^2-\bar p_{31}^2}{m_3^2-\bar p_{31}^2}\biggr)
-2\Li\biggl(\frac{\bar p_{30}^2-\bar p_{31}^2}{m_3^2-\bar p_{31}^2}\biggr)
\nn\\[.3em]
&& \quad {}
\hspace{-15em}
+\cLi\biggl(\frac{\bar p_{20}^2}{\bar p_{32}^2-m_3^2},
\frac{m_3^2}{m_3^2-\bar p_{30}^2}\biggr)
-\frac{\pi^2}{6}
\Biggr\}.
\label{eq:DOc2s1massreg3}
\eeqar
The dimensionally regularized integral \refeq{eq:DOc2s1dimreg} was
calculated in \citere{Berger:2000iu} and  corresponds to ``Box 8'' of
\citere{Ellis:2007qk}.
Integral \refeq{eq:DOc2s1dimmassreg1} can be obtained from
\refeq{eq:DOc1s1dimregonsh3} for $m_3$ small.
Integral \refeq{eq:DOc2s1dimmassreg2} is case (ib) of
\citere{Beenakker:1990jr} (modulo the trivial translation to dimensional
regularization) and corresponds to 
\refeq{eq:DOc1s1dimmassreg1} for small $m_3$. 

The following eight different cases with up to two soft
singularities have the common finite part
\beq
\DOfin = D_0 
- \frac{(p_{31}^2-p_{32}^2)C_0(0)}{p_{20}^2(p_{31}^2-m_3^2)}
- \frac{C_0(2)}{p_{20}^2} - \frac{C_0(3)}{p_{31}^2-m_3^2},
\eeq
with the real mass $m_3$,
\beqar
\DOp{0}{0}{p_{32}^2}{m_3^2}%
{p_{20}^2}{p_{31}^2}% 
\DOm{0}{0}{0}{m_3^2}{(n_{\soft}=2)} 
&=&
\frac{1}{p_{20}^2(p_{31}^2-m_3^2)}\Biggl\{
\frac{3c_\eps}{2\eps^2}
+\frac{c_\eps}{\eps} 
\biggl[ 
\ln\biggl(\frac{\mu^2}{-\bar p_{20}^2}\biggr) 
-\ln\biggl(\frac{\mu m_3}{m_3^2-\bar p_{32}^2}\biggr) 
\nn\\[-3em]
&& \quad {}
+2\ln\biggl(\frac{\mu m_3}{m_3^2-\bar p_{31}^2}\biggr)
\biggr]
-\ln^2\biggl(\frac{\mu m_3}{m_3^2-\bar p_{32}^2}\biggr) 
\nn\\[.3em]
&& \quad {}
\hspace{-15em}
+2\ln\biggl(\frac{\mu^2}{-\bar p_{20}^2}\biggr) 
\ln\biggl(\frac{\mu m_3}{m_3^2-\bar p_{31}^2}\biggr) 
-2\Li\biggl(\frac{\bar p_{32}^2-\bar p_{31}^2}{m_3^2-\bar p_{31}^2}\biggr)
-\frac{2\pi^2}{3}
\Biggr\},
\label{eq:DOc2s2dimreg}
\\[1em]
\DOp{0}{\la^2}{p_{32}^2}{m_3^2}%
{p_{20}^2}{p_{31}^2}% 
\DOm{0}{0}{\la^2}{m_3^2}{(n_{\soft}=2)} 
&=&
\frac{1}{p_{20}^2(p_{31}^2-m_3^2)}\Biggl\{
\frac{c_\eps}{\eps^2} 
+\frac{c_\eps}{\eps} \biggl[
\ln\biggl(\frac{\mu\la}{-\bar p_{20}^2}\biggr) 
+\ln\biggl(\frac{\mu m_3}{m_3^2-\bar p_{31}^2}\biggr) 
\biggr]
\nn\\[-3em]
&& \quad {}
+2\ln\biggl(\frac{\mu\la}{-\bar p_{20}^2}\biggr)
\ln\biggl(\frac{\mu m_3}{m_3^2-\bar p_{31}^2}\biggr) 
-\ln^2\biggl(\frac{\la m_3}{m_3^2-\bar p_{32}^2}\biggr) 
\nn\\[.3em]
&& \quad {}
-\frac{2\pi^2}{3}
\Biggr\},
\label{eq:DOc2s2dimmassreg2}
\\[1em]
\DOp{\la^2}{0}{p_{32}^2}{m_3^2}%
{p_{20}^2}{p_{31}^2}% 
\DOm{\la^2}{0}{0}{m_3^2}{(n_{\soft}=1)} 
&=&
\frac{1}{p_{20}^2(p_{31}^2-m_3^2)}\Biggl\{
\frac{c_\eps}{2\eps^2}
+\frac{c_\eps}{\eps} 
\biggl[ 
\ln\biggl(\frac{\mu\la}{-\bar p_{20}^2}\biggr)
+\ln\biggl(\frac{m_3^2-\bar p_{32}^2}{m_3^2-\bar p_{31}^2}\biggr) 
\biggr]
\nn\\[-3em]
&& \quad {}
-\ln^2\biggl(\frac{\mu m_3}{m_3^2-\bar p_{32}^2}\biggr) 
+2\ln\biggl(\frac{\mu\la}{-\bar p_{20}^2}\biggr) 
\ln\biggl(\frac{\mu m_3}{m_3^2-\bar p_{31}^2}\biggr) 
\nn\\[.3em]
&& \quad {}
-2\Li\biggl(\frac{\bar p_{32}^2-\bar p_{31}^2}{m_3^2-\bar p_{31}^2}\biggr)
-\frac{\pi^2}{4}
\Biggr\},
\label{eq:DOc2s2dimmassreg1}
\\[1em]
%
%\DOp{\la^2}{\la^2}{p_{32}^2}{m_3^2}%
%{p_{20}^2}{p_{31}^2}% 
%\DOm{0}{\la^2}{0}{m_3^2}{(n_{\soft}=1)} 
%&=&
%\frac{1}{p_{20}^2(p_{31}^2-m_3^2)}\Biggl\{
%\biggl[ \frac{c_\eps}{\eps}
%+2\ln\biggl(\frac{\mu\la}{-\bar p_{20}^2}\biggr) 
%\biggr]
%\ln\biggl(\frac{\la m_3}{m_3^2-\bar p_{31}^2}\biggr) 
%\nn\\[-3em]
%&& \quad {}
%-\ln^2\biggl(\frac{\la m_3}{m_3^2-\bar p_{32}^2}\biggr) 
%+\ln^2\biggl(\frac{m_3^2-\bar p_{32}^2}{m_3^2-\bar p_{31}^2}\biggr) 
%+\frac{\pi^2}{6}
%\Biggr\},
%\label{eq:DOc2s2dimmassreg4}
%%
%\\[1em]\alt
%
\DOp{\la^2}{\la^2}{p_{32}^2}{m_3^2}%
{p_{20}^2}{p_{31}^2}% 
\DOm{0}{\la^2}{0}{m_3^2}{(n_{\soft}=1)} 
&=&
\frac{1}{p_{20}^2(p_{31}^2-m_3^2)}\Biggl\{
\biggl[ \frac{c_\eps}{\eps}
+2\ln\biggl(\frac{\bar p_{32}^2-m_3^2}{\bar p_{20}^2}\biggr) 
+\ln\biggl(\frac{\mu^2}{m_3^2}\biggr)
\biggr]
\nn\\[-3em]
&& \quad {}
\times 
\ln\biggl(\frac{\la m_3}{m_3^2-\bar p_{31}^2}\biggr) 
+\ln^2\biggl(\frac{\la m_3}{m_3^2-\bar p_{31}^2}\biggr) 
+\frac{\pi^2}{6}
\Biggr\},
\label{eq:DOc2s2dimmassreg4}
\\[1em]
\DOp{\la^2}{0}{p_{32}^2}{m_3^2}%
{p_{20}^2}{p_{31}^2}% 
\DOm{0}{\la^2}{\la^2}{m_3^2}{(n_{\soft}=1)} 
&=&
\frac{1}{p_{20}^2(p_{31}^2-m_3^2)}\Biggl\{
\biggl[ \frac{c_\eps}{\eps}
+2\ln\biggl(\frac{\mu\la}{-\bar p_{20}^2}\biggr) 
\biggr]
\ln\biggl(\frac{\la m_3}{m_3^2-\bar p_{31}^2}\biggr) 
\nn\\[-3em]
&& \quad {}
-\ln^2\biggl(\frac{\la m_3}{m_3^2-\bar p_{32}^2}\biggr) 
-2\Li\biggl(\frac{\bar p_{32}^2-\bar p_{31}^2}{m_3^2-\bar p_{31}^2}\biggr)
-\frac{\pi^2}{6}
\Biggr\},
\label{eq:DOc2s2dimmassreg3}
\\[1em]
\DOp{\la_1^2}{\la_2^2}{p_{32}^2}{m_3^2}%
{p_{20}^2}{p_{31}^2}%5
\DOm{\la_1^2}{0}{\la_2^2}{m_3^2}{(n_{\soft}=1)} 
&=&
\frac{1}{p_{20}^2(p_{31}^2-m_3^2)}\Biggl\{
%\frac{c_\eps}{\eps} 
%\ln\biggl(\frac{\la_1\la_2}{-\bar p_{20}^2}\biggr)
%+2\ln\biggl(\frac{\la_1 \la_2}{-\bar p_{20}^2}\biggr) 
%\ln\biggl(\frac{\mu m_3}{m_3^2-\bar p_{31}^2}\biggr) 
\biggl[\frac{c_\eps}{\eps} 
+2
\ln\biggl(\frac{\mu m_3}{m_3^2-\bar p_{31}^2}\biggr) \biggr]
\ln\biggl(\frac{\la_1\la_2}{-\bar p_{20}^2}\biggr)
\nn\\[-3em]
&& \quad {}
-\ln^2\biggl(\frac{\la_2 m_3}{m_3^2-\bar p_{32}^2}\biggr) 
-\frac{\pi^2}{4}
\Biggr\},
\label{eq:DOc2s2dimmassreg5}
\\[1em]
\DOp{0}{\la^2}{p_{32}^2}{m_3^2}%
{p_{20}^2}{p_{31}^2}% 
\DOm{\la^2}{\la^2}{0}{m_3^2}{(n_{\soft}=0)} 
&=&
\frac{1}{p_{20}^2(p_{31}^2-m_3^2)}\Biggl\{
\ln\biggl(\frac{\la m_3}{m_3^2-\bar p_{31}^2}\biggr) 
\biggl[
\ln\biggl(\frac{\la m_3}{m_3^2-\bar p_{31}^2}\biggr) 
\nn\\[-3em]
&& \quad {}
+2 \ln\biggl(\frac{\la^2}{-\bar p_{20}^2}\biggr) 
-2 \ln\biggl(\frac{\la m_3}{m_3^2-\bar p_{32}^2}\biggr) 
\biggr]
-\frac{\pi^2}{12}
\Biggr\},
\label{eq:DOc2s2massreg1}
\\[1em]
\DOp{0}{0}{p_{32}^2}{m_3^2}%
{p_{20}^2}{p_{31}^2}% 
\DOm{\la^2}{\la^2}{\la^2}{m_3^2}{(n_{\soft}=0)} 
&=&
\frac{1}{p_{20}^2(p_{31}^2-m_3^2)}\Biggl\{
2\ln\biggl(\frac{\la^2}{-\bar p_{20}^2}\biggr) 
\ln\biggl(\frac{\la m_3}{m_3^2-\bar p_{31}^2}\biggr) 
\nn\\[-3em]
&& \quad {}
-\ln^2\biggl(\frac{\la m_3}{m_3^2-\bar p_{32}^2}\biggr) 
-2\Li\biggl(\frac{\bar p_{32}^2-\bar p_{31}^2}{m_3^2-\bar p_{31}^2}\biggr)
-\frac{5\pi^2}{12}
\Biggr\}.
\hspace*{2em}
\nn\\
\label{eq:DOc2s2massreg2}
\eeqar
The dimensionally regularized integral \refeq{eq:DOc2s2dimreg}, e.g., 
appeared in the NLO QCD
correction to $\Pp\Pp\to\Pt\bar\Pt\PH$~\cite{Beenakker:2002nc} [see
Eq.~(A.4) there] and corresponds to ``Box 7'' of
\citere{Ellis:2007qk}.
Integral \refeq{eq:DOc2s2dimmassreg2} can be obtained
from \refeq{eq:DOc1s2dimreg} for $m_2$ small.
Integral \refeq{eq:DOc2s2dimmassreg1} can be obtained
from \refeq{eq:DOc1s1dimregonsh3}
for $p_{32}^2=m_2^2$ and $m_3$ small. 
Integral \refeq{eq:DOc2s2dimmassreg4} is case (iic) of
\citere{Beenakker:1990jr} (modulo translation to dimensional 
regularization)
and corresponds to \refeq{eq:DOc0s1dimreg} with $p_{21}^2=m_1^2$ small
and $m_2=0$. Similar comments apply to the next two cases.
Integral \refeq{eq:DOc2s2dimmassreg3} results from case (ic) of
\citere{Beenakker:1990jr}  and corresponds to 
\refeq{eq:DOc0s1dimreg} with $p_{21}^2=0$ and $m_2=m_1$ both small.
Integral \refeq{eq:DOc2s2dimmassreg5} coincides with case (i) of
\citere{Beenakker:1990jr} for $m_0^2=m_2^2$ and both $m_1$ and $m_4$
small and corresponds to \refeq{eq:DOc0s1dimreg} with $p_{32}^2=m_2^2$
and both $m_1$ and $m_3$ small.

The following six different cases with up to three soft
singularities have the common finite part
\beq
\DOfin = D_0 
- \frac{C_0(0)+C_0(2)}{p_{20}^2} - \frac{C_0(3)}{p_{31}^2-m_3^2}.
\eeq
Again the mass $m_3$ has to be a real quantity here,
\beqar
\DOp{0}{0}{m_3^2}{m_3^2}%
{p_{20}^2}{p_{31}^2}% 
\DOm{0}{0}{0}{m_3^2}{(n_{\soft}=3)} 
&=&
\frac{1}{p_{20}^2(p_{31}^2-m_3^2)}\Biggl\{
\frac{2c_\eps}{\eps^2}
+\frac{c_\eps}{\eps} \biggl[ 
\ln\biggl(\frac{\mu^2}{-\bar p_{20}^2}\biggr)
+2\ln\biggl(\frac{\mu m_3}{m_3^2-\bar p_{31}^2}\biggr) \biggr]
\nn\\[-3em]
&& \quad {}
+2\ln\biggl(\frac{\mu^2}{-\bar p_{20}^2}\biggr)
\ln\biggl(\frac{\mu m_3}{m_3^2-\bar p_{31}^2}\biggr) 
-\frac{5\pi^2}{6} \Biggr\},
\hspace{2em}
\label{eq:DOc2s3dimreg}
\\[1em]
\DOp{\la^2}{0}{m_3^2}{m_3^2}%
{p_{20}^2}{p_{31}^2}% 
\DOm{\la^2}{0}{0}{m_3^2}{(n_{\soft}=2)} 
&=&
\frac{1}{p_{20}^2(p_{31}^2-m_3^2)}\Biggl\{
\frac{c_\eps}{\eps^2} 
+\frac{c_\eps}{\eps} \biggl[
\ln\biggl(\frac{\mu\la}{-\bar p_{20}^2}\biggr)
+ \ln\biggl(\frac{\mu m_3}{m_3^2-\bar p_{31}^2}\biggr)
\biggr]
\nn\\[-3em]
&& \quad {}
+2\ln\biggl(\frac{\mu\la}{-\bar p_{20}^2}\biggr)
\ln\biggl(\frac{\mu m_3}{m_3^2-\bar p_{31}^2}\biggr)
-\frac{5\pi^2}{12} \Biggr\},
\label{eq:DOc2s3dimmassreg1}
\\[1em]
\DOp{\la^2}{\la^2}{m_3^2}{m_3^2}%
{p_{20}^2}{p_{31}^2}% 
\DOm{0}{\la^2}{0}{m_3^2}{(n_{\soft}=2)} 
&=&
\frac{2}{p_{20}^2(p_{31}^2-m_3^2)}
\biggl[
\frac{c_\eps}{\eps} 
+\ln\biggl(\frac{\mu^2}{-\bar p_{20}^2}\biggr)
\biggr]
\ln\biggl(\frac{\la m_3}{m_3^2-\bar p_{31}^2}\biggr),
\label{eq:DOc2s3dimmassreg3}
\\[1em]
\DOp{0}{\la^2}{m_3^2}{m_3^2}%
{p_{20}^2}{p_{31}^2}% 
\DOm{\la^2}{\la^2}{0}{m_3^2}{(n_{\soft}=1)} 
&=&
\frac{1}{p_{20}^2(p_{31}^2-m_3^2)}
\Biggl\{
\biggl[
\frac{c_\eps}{\eps}
+2\ln\biggl(\frac{\mu\la}{-\bar p_{20}^2}\biggr)
\biggr]
\ln\biggl(\frac{\la m_3}{m_3^2-\bar p_{31}^2}\biggr)
\nn\\[-3em]
&& \quad {}
-\frac{\pi^2}{4}
\biggr\},
\label{eq:DOc2s3dimmassreg2}
\\[1em]
\DOp{\la_1^2}{\la_2^2}{m_3^2}{m_3^2}%
{p_{20}^2}{p_{31}^2}% 
\DOm{\la_1^2}{0}{\la_2^2}{m_3^2}{(n_{\soft}=1)} 
&=&
\frac{1}{p_{20}^2(p_{31}^2-m_3^2)}
\biggl[
\frac{c_\eps}{\eps}
+2\ln\biggl(\frac{\mu m_3}{m_3^2-\bar p_{31}^2}\biggr)
\biggr]
\ln\biggl(\frac{\la_1\la_2}{-\bar p_{20}^2}\biggr),
\label{eq:DOc2s3dimmassreg4}
\\[1em]
\DOp{0}{0}{m_3^2}{m_3^2}%
{p_{20}^2}{p_{31}^2}% 
\DOm{\la^2}{\la^2}{\la^2}{m_3^2}{(n_{\soft}=0)} 
&=&
\frac{1}{p_{20}^2(p_{31}^2-m_3^2)}\Biggl\{
2\ln\biggl(\frac{\la^2}{-\bar p_{20}^2}\biggr)
\ln\biggl(\frac{\la m_3}{m_3^2-\bar p_{31}^2}\biggr)
-\frac{\pi^2}{2} \Biggr\}.
\nn\\[-3em]
\phantom{\biggl(}
\label{eq:DOc2s3massreg}
\eeqar
The integral \refeq{eq:DOc2s3dimreg} was given in Eq.~(A.4) of
\citere{Beenakker:1988bq} and corresponds to ``Box 6'' of
\citere{Ellis:2007qk}.
Integral \refeq{eq:DOc2s3dimmassreg1} can be obtained
from \refeq{eq:DOc1s2dimreg} for $p_{32}^2=m_2^2$ and $m_3$
small.
Integral \refeq{eq:DOc2s3dimmassreg3} is case (iiia) of
\citere{Beenakker:1990jr} (again to be translated to dimensional
regularization, also in the next two cases) and corresponds to 
\refeq{eq:DOc0s2dimreg} for small $m_1^2$.
Integral \refeq{eq:DOc2s3dimmassreg2} results from case (i) of
\citere{Beenakker:1990jr} for $m_2^2=0$, $m_3^2=m_4^2$ and small
$m_1=m_0$, or analogously from \refeq{eq:DOc0s1dimreg} with
$p_{21}^2=0$, $p_{32}^2=m_3^2$ and $m_1^2=m_2^2$ small.
Integral \refeq{eq:DOc2s3dimmassreg4} results from case (i) of
\citere{Beenakker:1990jr} for $m_0^2=m_3^2=m_2^2$ and both $m_1$ and
$m_4$ small and corresponds to \refeq{eq:DOc0s1dimreg} with
$p_{21}^2=p_{32}^2=m_2^2$ and both $m_1^2$ and $m_3^2$ small.

\paragraph{Cases with two collinear singularities on opposite legs
$(n_{\coll}=2)$}

We distinguish seven different cases which do not contain a soft
singularity and have the following common finite part
\beq
\DOfin = D_0 
- \frac{(p_{31}^2-p_{21}^2)C_0(0)
+(p_{20}^2-p_{30}^2)C_0(1)
+(p_{31}^2-p_{30}^2)C_0(2)
+(p_{20}^2-p_{21}^2)C_0(3)}
{p_{20}^2 p_{31}^2-p_{21}^2 p_{30}^2}.
\eeq
The explicit results read
\beqar
%\DOp{0}{p_{21}^2}{0}{p_{30}^2}%
%{p_{20}^2}{p_{31}^2}% 
%\DOm{0}{0}{0}{0}{(n_{\soft}=0)} 
%&=&
%\frac{1}{p_{20}^2 p_{31}^2 - p_{21}^2 p_{30}^2} 
%\Biggl\{
%\ln\biggl(\frac{\bar p_{21}^2}{\bar p_{31}^2}\biggr)
%\biggl[\frac{c_\eps}{\eps}+\ln\biggl(\frac{\mu^2}{-\bar p_{31}^2}\biggr)\biggr]
%\nn\\[-3em]
%&& \quad {}
%+\ln\biggl(\frac{\bar p_{30}^2}{\bar p_{20}^2}\biggr)
%\biggl[\frac{c_\eps}{\eps}+\ln\biggl(\frac{\mu^2}{-\bar p_{20}^2}\biggr)\biggr]
%\nn\\[.3em]
%&& \quad {}
%\hspace{-15em}
%+\ln\biggl(\frac{\bar p_{30}^2}{\bar p_{31}^2}\biggr)
%\biggl[\frac{c_\eps}{\eps}+\ln\biggl(\frac{\mu^2}{-\bar p_{21}^2}\biggr)\biggr]
%+\ln\biggl(\frac{\bar p_{21}^2}{\bar p_{20}^2}\biggr)
%\biggl[\frac{c_\eps}{\eps}+\ln\biggl(\frac{\mu^2}{-\bar p_{30}^2}\biggr)\biggr]
%+2\Li\biggl(1-\frac{\bar p_{20}^2}{\bar p_{21}^2}\biggr)
%\nn\\[.3em]
%&& \quad {}
%\hspace{-15em}
%+2\Li\biggl(1-\frac{\bar p_{20}^2}{\bar p_{30}^2}\biggr)
%+2\Li\biggl(1-\frac{\bar p_{31}^2}{\bar p_{21}^2}\biggr)
%+2\Li\biggl(1-\frac{\bar p_{31}^2}{\bar p_{30}^2}\biggr) 
%-2\cLi\biggl(\frac{\bar p_{20}^2}{\bar p_{21}^2},
%\frac{\bar p_{31}^2}{\bar p_{30}^2}\biggr)
%\Biggr\},
%\label{eq:DOc2opps0dimreg}
%%
%\\[1em]\alt
%
\DOp{0}{p_{21}^2}{0}{p_{30}^2}%
{p_{20}^2}{p_{31}^2}% 
\DOm{0}{0}{0}{0}{(n_{\soft}=0)} 
&=&
\frac{2}{p_{20}^2 p_{31}^2 - p_{21}^2 p_{30}^2} 
\Biggl\{\cLi\biggl(\frac{\bar p_{21}^2}{\bar p_{20}^2},
\frac{\bar p_{30}^2}{\bar p_{31}^2}\biggr)
\nn\\[-3em]
&& \quad {}
+
\biggl[\frac{c_\eps}{\eps}+\ln\biggl(\frac{\mu^2}{-\bar p_{20}^2}\biggr)\biggr]
\biggl[\ln\biggl(\frac{\bar p_{30}^2}{\bar p_{31}^2}\biggr)
-\ln\biggl(\frac{\bar p_{20}^2}{\bar p_{21}^2}\biggr)\biggr]
\nn\\[.3em]
&& \quad {}
\hspace{-15em}
+\Li\biggl(1-\frac{\bar p_{20}^2}{\bar p_{21}^2}\biggr)
+\Li\biggl(1-\frac{\bar p_{20}^2}{\bar p_{30}^2}\biggr)
-\Li\biggl(1-\frac{\bar p_{21}^2}{\bar p_{31}^2}\biggr)
-\Li\biggl(1-\frac{\bar p_{30}^2}{\bar p_{31}^2}\biggr) 
\Biggr\},
\label{eq:DOc2opps0dimreg}
\\[1em]
%
%\DOp{\la^2}{p_{21}^2}{0}{p_{30}^2}%
%{p_{20}^2}{p_{31}^2}% 
%\DOm{\la^2}{0}{0}{0}{(n_{\soft}=0)} 
%&=&
%\frac{1}{p_{20}^2 p_{31}^2 - p_{21}^2 p_{30}^2} 
%\Biggl\{
%\ln\biggl(\frac{\bar p_{21}^2}{\bar p_{31}^2}\biggr)
%\biggl[ \frac{c_\eps}{\eps} 
%+\ln\biggl(\frac{\mu^2}{-\bar p_{31}^2}\biggr) \biggr]
%\nn\\[-3em]
%&& \quad {}
%+\ln\biggl(\frac{\bar p_{30}^2}{\bar p_{20}^2}\biggr)
%\biggl[ \frac{c_\eps}{\eps}
%+\ln\biggl(\frac{\mu^2}{-\bar p_{20}^2}\biggr) \biggr]
%+\ln\biggl(\frac{\bar p_{30}^2}{\bar p_{31}^2}\biggr)
%\ln\biggl(\frac{\la^2}{-\bar p_{21}^2}\biggr)
%\nn\\[.3em]
%&& \quad {}
%\hspace{-15em}
%+ \ln\biggl(\frac{\bar p_{21}^2}{\bar p_{20}^2}\biggr)
%\ln\biggl(\frac{\la^2}{-\bar p_{30}^2}\biggr)
%-\ln^2\biggl(\frac{\bar p_{31}^2}{\bar p_{30}^2}\biggr) 
%+2\Li\biggl(1-\frac{\bar p_{31}^2}{\bar p_{21}^2}\biggr)
%+2\Li\biggl(1-\frac{\bar p_{20}^2}{\bar p_{30}^2}\biggr)
%\nn\\[.3em]
%&& \quad {}
%\hspace{-15em}
%-2\cLi\biggl(\frac{\bar p_{20}^2}{\bar p_{21}^2},
%\frac{\bar p_{31}^2}{\bar p_{30}^2}\biggr)
%\Biggr\},
%\hspace{2em}
%\label{eq:DOc2opps0dimmassreg1}
%%
%\\[1em]\alt
%
\DOp{\la^2}{p_{21}^2}{0}{p_{30}^2}%
{p_{20}^2}{p_{31}^2}% 
\DOm{\la^2}{0}{0}{0}{(n_{\soft}=0)} 
&=&
\frac{2}{p_{20}^2 p_{31}^2 - p_{21}^2 p_{30}^2} 
\Biggl\{\cLi\biggl(\frac{\bar p_{21}^2}{\bar p_{20}^2},
\frac{\bar p_{30}^2}{\bar p_{31}^2}\biggr)
\nn\\[-3em]
&& \quad {}
+
\biggl[\frac{c_\eps}{2\eps}+\ln\biggl(\frac{\mu\la}{-\bar p_{20}^2}\biggr)\biggr]
\biggl[\ln\biggl(\frac{\bar p_{30}^2}{\bar p_{31}^2}\biggr)
-\ln\biggl(\frac{\bar p_{20}^2}{\bar p_{21}^2}\biggr)\biggr]
\nn\\[.3em]
&& \quad {}
\hspace{-15em}
%+\Li\biggl(1-\frac{\bar p_{20}^2}{\bar p_{21}^2}\biggr)
+\Li\biggl(1-\frac{\bar p_{20}^2}{\bar p_{30}^2}\biggr)
-\Li\biggl(1-\frac{\bar p_{21}^2}{\bar p_{31}^2}\biggr)
%-\Li\biggl(1-\frac{\bar p_{30}^2}{\bar p_{31}^2}\biggr) 
\Biggr\},
\label{eq:DOc2opps0dimmassreg1}
\\[1em]
%
%\DOp{0}{p_{21}^2}{0}{p_{30}^2}%
%{p_{20}^2}{p_{31}^2}% 
%\DOm{\la^2}{\la^2}{0}{0}{(n_{\soft}=0)} 
%&=&
%\frac{1}{p_{20}^2 p_{31}^2 - p_{21}^2 p_{30}^2} 
%\Biggl\{
%\ln\biggl(\frac{\bar p_{21}^2}{\bar p_{31}^2}\biggr)
%\biggl[\frac{c_\eps}{\eps}+\ln\biggl(\frac{\mu^2}{-\bar p_{31}^2}\biggr)\biggr]
%\nn\\[-3em]
%&& \quad {}
%+\ln\biggl(\frac{\bar p_{30}^2}{\bar p_{20}^2}\biggr)
%\biggl[\frac{c_\eps}{\eps}+\ln\biggl(\frac{\mu^2}{-\bar p_{20}^2}\biggr)\biggr]
%\nn\\[.3em]
%&& \quad {}
%\hspace{-15em}
%+\ln\biggl(\frac{\bar p_{30}^2}{\bar p_{31}^2}\biggr)
%\ln\biggl(\frac{\la^2}{-\bar p_{21}^2}\biggr)
%+\ln\biggl(\frac{\bar p_{21}^2}{\bar p_{20}^2}\biggr)
%\ln\biggl(\frac{\la^2}{-\bar p_{30}^2}\biggr)
%+2\Li\biggl(1-\frac{\bar p_{20}^2}{\bar p_{21}^2}\biggr)
%+2\Li\biggl(1-\frac{\bar p_{20}^2}{\bar p_{30}^2}\biggr)
%\nn\\[.3em]
%&& \quad {}
%\hspace{-15em}
%+2\Li\biggl(1-\frac{\bar p_{31}^2}{\bar p_{21}^2}\biggr)
%+2\Li\biggl(1-\frac{\bar p_{31}^2}{\bar p_{30}^2}\biggr) 
%-2\cLi\biggl(\frac{\bar p_{20}^2}{\bar p_{21}^2},
%\frac{\bar p_{31}^2}{\bar p_{30}^2}\biggr)
%\Biggr\},
%\label{eq:DOc2opps0dimmassreg2}
%%
%\\[1em]\alt
%
\DOp{0}{p_{21}^2}{0}{p_{30}^2}%
{p_{20}^2}{p_{31}^2}% 
\DOm{\la^2}{\la^2}{0}{0}{(n_{\soft}=0)} 
&=&
\frac{2}{p_{20}^2 p_{31}^2 - p_{21}^2 p_{30}^2} 
\Biggl\{\cLi\biggl(\frac{\bar p_{21}^2}{\bar p_{20}^2},
\frac{\bar p_{30}^2}{\bar p_{31}^2}\biggr)
\nn\\[-3em]
&& \quad {}
+
\biggl[\frac{c_\eps}{2\eps}+\ln\biggl(\frac{\mu\la}{-\bar p_{20}^2}\biggr)\biggr]
\biggl[\ln\biggl(\frac{\bar p_{30}^2}{\bar p_{31}^2}\biggr)
-\ln\biggl(\frac{\bar p_{20}^2}{\bar p_{21}^2}\biggr)\biggr]
\nn\\[.3em]
&& \quad {}
\hspace{-15em}
+\Li\biggl(1-\frac{\bar p_{20}^2}{\bar p_{21}^2}\biggr)
+\Li\biggl(1-\frac{\bar p_{20}^2}{\bar p_{30}^2}\biggr)
-\Li\biggl(1-\frac{\bar p_{21}^2}{\bar p_{31}^2}\biggr)
-\Li\biggl(1-\frac{\bar p_{30}^2}{\bar p_{31}^2}\biggr) 
\Biggr\},
\label{eq:DOc2opps0dimmassreg2}
\\[1em]
%
%\DOp{\la_1^2}{p_{21}^2}{\la_2^2}{p_{30}^2}%
%{p_{20}^2}{p_{31}^2}% 
%\DOm{\la_1^2}{0}{\la_2^2}{0}{(n_{\soft}=0)} 
%&=&
%\frac{1}{p_{20}^2 p_{31}^2 - p_{21}^2 p_{30}^2} 
%\Biggl\{
%\ln\biggl(\frac{\bar p_{21}^2}{\bar p_{31}^2}\biggr)
%\ln\biggl(\frac{\la_2^2}{-\bar p_{21}^2}\biggr) 
%\nn\\[-3em]
%&& \quad {}
%+\ln\biggl(\frac{\bar p_{30}^2}{\bar p_{20}^2}\biggr)
%\ln\biggl(\frac{\la_2^2}{-\bar p_{20}^2}\biggr)
%+\ln\biggl(\frac{\bar p_{30}^2}{\bar p_{31}^2}\biggr)
%\ln\biggl(\frac{\la_1^2}{-\bar p_{21}^2}\biggr)
%\nn\\[.3em]
%&& \quad {}
%\hspace{-15em}
%+ \ln\biggl(\frac{\bar p_{21}^2}{\bar p_{20}^2}\biggr)
%\ln\biggl(\frac{\la_1^2}{-\bar p_{30}^2}\biggr)
%-\ln^2\biggl(\frac{\bar p_{31}^2}{\bar p_{30}^2}\biggr) 
%-2\cLi\biggl(\frac{\bar p_{20}^2}{\bar p_{21}^2},
%\frac{\bar p_{31}^2}{\bar p_{30}^2}\biggr)
%\Biggr\},
%\label{eq:DOc2opps0massreg1}
%%
%\\[1em]\alt
%
\DOp{\la_1^2}{p_{21}^2}{\la_2^2}{p_{30}^2}%
{p_{20}^2}{p_{31}^2}% 
\DOm{\la_1^2}{0}{\la_2^2}{0}{(n_{\soft}=0)} 
&=&
\frac{2}{p_{20}^2 p_{31}^2 - p_{21}^2 p_{30}^2} 
\Biggl\{\cLi\biggl(\frac{\bar p_{21}^2}{\bar p_{20}^2},
\frac{\bar p_{30}^2}{\bar p_{31}^2}\biggr)
\nn\\[-3em]
&& \quad {}
+
\ln\biggl(\frac{\la_1\la_2}{-\bar p_{20}^2}\biggr)
\biggl[\ln\biggl(\frac{\bar p_{30}^2}{\bar p_{31}^2}\biggr)
-\ln\biggl(\frac{\bar p_{20}^2}{\bar p_{21}^2}\biggr)\biggr]
%\nn\\[.3em]
%&& \quad {}
%\hspace{-15em}
%+\Li\biggl(1-\frac{\bar p_{20}^2}{\bar p_{21}^2}\biggr)
%+\Li\biggl(1-\frac{\bar p_{20}^2}{\bar p_{30}^2}\biggr)
%-\Li\biggl(1-\frac{\bar p_{21}^2}{\bar p_{31}^2}\biggr)
%-\Li\biggl(1-\frac{\bar p_{30}^2}{\bar p_{31}^2}\biggr) 
\Biggr\},
\label{eq:DOc2opps0massreg1}
\\[1em]
%
%\DOp{\la_1^2}{p_{21}^2}{\la_2^2}{p_{30}^2}%
%{p_{20}^2}{p_{31}^2}% 
%\DOm{\la_1^2}{0}{0}{\la_2^2}{(n_{\soft}=0)} 
%&=&
%\frac{1}{p_{20}^2 p_{31}^2 - p_{21}^2 p_{30}^2} 
%\Biggl\{
%\ln\biggl(\frac{\bar p_{21}^2}{\bar p_{31}^2}\biggr)
%\ln\biggl(\frac{\la_2^2}{-\bar p_{31}^2}\biggr)
%\nn\\[-3em]
%&& \quad {}
%+\ln\biggl(\frac{\bar p_{30}^2}{\bar p_{20}^2}\biggr)
%\ln\biggl(\frac{\la_2^2}{-\bar p_{30}^2}\biggr)
%+\ln\biggl(\frac{\bar p_{30}^2}{\bar p_{31}^2}\biggr)
%\ln\biggl(\frac{\la_1^2}{-\bar p_{21}^2}\biggr)
%\nn\\[.3em]
%&& \quad {}
%\hspace{-15em}
%+ \ln\biggl(\frac{\bar p_{21}^2}{\bar p_{20}^2}\biggr)
%\ln\biggl(\frac{\la_1^2}{-\bar p_{30}^2}\biggr)
%-\ln^2\biggl(\frac{\bar p_{31}^2}{\bar p_{30}^2}\biggr) 
%-2\cLi\biggl(\frac{\bar p_{20}^2}{\bar p_{21}^2},
%\frac{\bar p_{31}^2}{\bar p_{30}^2}\biggr)
%\Biggr\},
%\label{eq:DOc2opps0massreg2}
%%
%\\[1em]\alt
%
\DOp{\la_1^2}{p_{21}^2}{\la_2^2}{p_{30}^2}%
{p_{20}^2}{p_{31}^2}% 
\DOm{\la_1^2}{0}{0}{\la_2^2}{(n_{\soft}=0)} 
&=&\hspace{4em}\left.
\DOp{\la_1^2}{p_{21}^2}{\la_2^2}{p_{30}^2}%
{p_{20}^2}{p_{31}^2}% 
\DOm{\la_1^2}{0}{\la_2^2}{0}{} 
\right|_{ p_{21}^2\leftrightarrow  p_{31}^2,
 p_{20}^2\leftrightarrow  p_{30}^2}
\nn\\
&& \hspace{-14em}
= \frac{2}{p_{20}^2 p_{31}^2 - p_{21}^2 p_{30}^2} 
\Biggl\{-\cLi\biggl(\frac{\bar p_{31}^2}{\bar p_{30}^2},
\frac{\bar p_{20}^2}{\bar p_{21}^2}\biggr) 
-\ln\biggl(\frac{\la_1\la_2}{- {\bar p_{30}^2}}\biggr)
\biggl[\ln\biggl(\frac{\bar p_{20}^2}{\bar p_{21}^2}\biggr)
-\ln\biggl(\frac{\bar p_{30}^2}{\bar p_{31}^2}\biggr)\biggr]
\Biggr\}, 
\label{eq:DOc2opps0massreg2}
\\[1em]
%
%\DOp{\la_1^2}{p_{21}^2}{0}{p_{30}^2}%
%{p_{20}^2}{p_{31}^2}% 
%\DOm{\la_1^2}{0}{\la_2^2}{\la_2^2}{(n_{\soft}=0)} 
%&=&
%\frac{1}{p_{20}^2 p_{31}^2 - p_{21}^2 p_{30}^2} 
%\Biggl\{
%\ln\biggl(\frac{\bar p_{21}^2}{\bar p_{31}^2}\biggr)
%\ln\biggl(\frac{\la_2^2}{-\bar p_{31}^2}\biggr)
%\nn\\[-3em]
%&& \quad {}
%+\ln\biggl(\frac{\bar p_{30}^2}{\bar p_{20}^2}\biggr)
%\ln\biggl(\frac{\la_2^2}{-\bar p_{20}^2}\biggr)
%+\ln\biggl(\frac{\bar p_{30}^2}{\bar p_{31}^2}\biggr)
%\ln\biggl(\frac{\la_1^2}{-\bar p_{21}^2}\biggr)
%\nn\\[.3em]
%&& \quad {}
%\hspace{-15em}
%+ \ln\biggl(\frac{\bar p_{21}^2}{\bar p_{20}^2}\biggr)
%\ln\biggl(\frac{\la_1^2}{-\bar p_{30}^2}\biggr)
%-\ln^2\biggl(\frac{\bar p_{31}^2}{\bar p_{30}^2}\biggr) 
%+2\Li\biggl(1-\frac{\bar p_{31}^2}{\bar p_{21}^2}\biggr)
%+2\Li\biggl(1-\frac{\bar p_{20}^2}{\bar p_{30}^2}\biggr)
%\nn\\[.3em]
%&& \quad {}
%\hspace{-15em}
%-2\cLi\biggl(\frac{\bar p_{20}^2}{\bar p_{21}^2},
%\frac{\bar p_{31}^2}{\bar p_{30}^2}\biggr)
%\Biggr\},
%\label{eq:DOc2opps0massreg3}
%%
%\\[1em]\alt
%
\DOp{\la_1^2}{p_{21}^2}{0}{p_{30}^2}%
{p_{20}^2}{p_{31}^2}% 
\DOm{\la_1^2}{0}{\la_2^2}{\la_2^2}{(n_{\soft}=0)} 
&=&
\frac{2}{p_{20}^2 p_{31}^2 - p_{21}^2 p_{30}^2} 
\Biggl\{\cLi\biggl(\frac{\bar p_{21}^2}{\bar p_{20}^2},
\frac{\bar p_{30}^2}{\bar p_{31}^2}\biggr)
\nn\\[-3em]
&& \quad {}
+
\ln\biggl(\frac{\la_1\la_2}{-\bar p_{20}^2}\biggr)
\biggl[\ln\biggl(\frac{\bar p_{30}^2}{\bar p_{31}^2}\biggr)
-\ln\biggl(\frac{\bar p_{20}^2}{\bar p_{21}^2}\biggr)\biggr]
\nn\\[.3em]
&& \quad {}
\hspace{-15em}
%+\Li\biggl(1-\frac{\bar p_{20}^2}{\bar p_{21}^2}\biggr)
+\Li\biggl(1-\frac{\bar p_{20}^2}{\bar p_{30}^2}\biggr)
-\Li\biggl(1-\frac{\bar p_{21}^2}{\bar p_{31}^2}\biggr)
%-\Li\biggl(1-\frac{\bar p_{30}^2}{\bar p_{31}^2}\biggr) 
\Biggr\},
\label{eq:DOc2opps0massreg3}
\\[1em]
%
%\DOp{0}{p_{21}^2}{0}{p_{30}^2}%
%{p_{20}^2}{p_{31}^2}% 
%\DOm{\la_1^2}{\la_1^2}{\la_2^2}{\la_2^2}{(n_{\soft}=0)} 
%&=&
%\frac{1}{p_{20}^2 p_{31}^2 - p_{21}^2 p_{30}^2} 
%\Biggl\{
%\ln\biggl(\frac{\bar p_{21}^2}{\bar p_{31}^2}\biggr)
%\ln\biggl(\frac{\la_2^2}{-\bar p_{31}^2}\biggr)
%\nn\\[-3em]
%&& \quad {}
%+\ln\biggl(\frac{\bar p_{30}^2}{\bar p_{20}^2}\biggr)
%\ln\biggl(\frac{\la_2^2}{-\bar p_{20}^2}\biggr)
%+\ln\biggl(\frac{\bar p_{30}^2}{\bar p_{31}^2}\biggr)
%\ln\biggl(\frac{\la_1^2}{-\bar p_{21}^2}\biggr)
%\nn\\[.3em]
%&& \quad {}
%\hspace{-15em}
%+\ln\biggl(\frac{\bar p_{21}^2}{\bar p_{20}^2}\biggr)
%\ln\biggl(\frac{\la_1^2}{-\bar p_{30}^2}\biggr)
%+2\Li\biggl(1-\frac{\bar p_{20}^2}{\bar p_{21}^2}\biggr)
%+2\Li\biggl(1-\frac{\bar p_{20}^2}{\bar p_{30}^2}\biggr)
%+2\Li\biggl(1-\frac{\bar p_{31}^2}{\bar p_{21}^2}\biggr)
%\nn\\[.3em]
%&& \quad {}
%\hspace{-15em}
%+2\Li\biggl(1-\frac{\bar p_{31}^2}{\bar p_{30}^2}\biggr) 
%-2\cLi\biggl(\frac{\bar p_{20}^2}{\bar p_{21}^2},
%\frac{\bar p_{31}^2}{\bar p_{30}^2}\biggr)
%\Biggr\},
%\label{eq:DOc2opps0massreg4}
%%
%\\[1em]\alt
%
\DOp{0}{p_{21}^2}{0}{p_{30}^2}%
{p_{20}^2}{p_{31}^2}% 
\DOm{\la_1^2}{\la_1^2}{\la_2^2}{\la_2^2}{(n_{\soft}=0)} 
&=&
\frac{2}{p_{20}^2 p_{31}^2 - p_{21}^2 p_{30}^2} 
\Biggl\{\cLi\biggl(\frac{\bar p_{21}^2}{\bar p_{20}^2},
\frac{\bar p_{30}^2}{\bar p_{31}^2}\biggr)
\nn\\[-3em]
&& \quad {}
+
\ln\biggl(\frac{\la_1\la_2}{-\bar p_{20}^2}\biggr)
\biggl[\ln\biggl(\frac{\bar p_{30}^2}{\bar p_{31}^2}\biggr)
-\ln\biggl(\frac{\bar p_{20}^2}{\bar p_{21}^2}\biggr)\biggr]
\nn\\[.3em]
&& \quad {}
\hspace{-15em}
+\Li\biggl(1-\frac{\bar p_{20}^2}{\bar p_{21}^2}\biggr)
+\Li\biggl(1-\frac{\bar p_{20}^2}{\bar p_{30}^2}\biggr)
-\Li\biggl(1-\frac{\bar p_{21}^2}{\bar p_{31}^2}\biggr)
-\Li\biggl(1-\frac{\bar p_{30}^2}{\bar p_{31}^2}\biggr) 
\Biggr\}.
\label{eq:DOc2opps0massreg4}
\eeqar
The integral \refeq{eq:DOc2opps0dimreg} can be found in
\citeres{Bern:1993kr,Duplancic:2000sk,Ellis:1980wv} and corresponds to
``Box 3'' of \citere{Ellis:2007qk}.
Integrals 
\refeq{eq:DOc2opps0dimmassreg1}--\refeq{eq:DOc2opps0massreg4} can be obtained
from \refeq{eq:DOc1s0dimreg}--\refeq{eq:DOc1s0massreg2}
with appropriate substitutions.

\paragraph{Cases with three collinear singularities $(n_{\coll}=3)$}

There are ten different cases with up to two soft
singularities and the following common finite part
\beq
\DOfin = D_0 
-\frac{C_0(0)}{p_{20}^2}
-\frac{(p_{20}^2-p_{30}^2)C_0(1)+(p_{31}^2-p_{30}^2)C_0(2)}{p_{20}^2 p_{31}^2}
-\frac{C_0(3)}{p_{31}^2}.
\eeq

\beqar
\DOp{0}{0}{0}{p_{30}^2}%
{p_{20}^2}{p_{31}^2}% 
\DOm{0}{0}{0}{0}{(n_{\soft}=2)} 
&=& \frac{1}{p_{20}^2 p_{31}^2}\Biggl\{
\frac{2c_\eps}{\eps^2}
+\frac{2c_\eps}{\eps}
\biggl[
\ln\biggl(\frac{\mu^2}{-\bar p_{20}^2}\biggr)
+\ln\biggl(\frac{\bar p_{30}^2}{\bar p_{31}^2}\biggr)
\biggr]
\nn\\[-3em]
&& \quad {}
+2\ln\biggl(\frac{\mu^2}{-\bar p_{31}^2}\biggr)
\ln\biggl(\frac{\mu^2}{-\bar p_{20}^2}\biggr)
-\ln^2\biggl(\frac{\mu^2}{-\bar p_{30}^2}\biggr)
\nn\\[.3em]
&& \quad {}
-2\Li\biggl(1-\frac{\bar p_{30}^2}{\bar p_{20}^2}\biggr)
-2\Li\biggl(1-\frac{\bar p_{30}^2}{\bar p_{31}^2}\biggr)
-\frac{2\pi^2}{3}
\Biggl\},
\label{D0ms3ir2zm4} 
\\[1.5em]
\DOp{\la^2}{0}{0}{p_{30}^2}%
{p_{20}^2}{p_{31}^2}% 
\DOm{\la^2}{0}{0}{0}{(n_{\soft}=2)} 
&=& \frac{1}{p_{20}^2 p_{31}^2}\Biggl\{
\frac{3c_\eps}{2\eps^2}
+\frac{c_\eps}{\eps}
\biggl[
2\ln\biggl(\frac{\mu\la}{-\bar p_{20}^2}\biggr)
-\ln\biggl(\frac{\mu\la}{-\bar p_{30}^2}\biggr)
+\ln\biggl(\frac{\mu^2}{-\bar p_{31}^2}\biggr)
\biggr]
\nn\\[-3em]
&& \quad {}
+2\ln\biggl(\frac{\mu^2}{-\bar p_{31}^2}\biggr)
\ln\biggl(\frac{\mu\la}{-\bar p_{20}^2}\biggr)
-\ln^2\biggl(\frac{\mu\la}{-\bar p_{30}^2}\biggr)
\nn\\[.3em]
&& \quad {}
-2\Li\biggl(1-\frac{\bar p_{30}^2}{\bar p_{20}^2}\biggr)
-\frac{2\pi^2}{3}
\Biggl\},
\label{D0ms3ir2zm3} 
\\[1em]
\DOp{\la_1^2}{0}{\la_2^2}{p_{30}^2}%
{p_{20}^2}{p_{31}^2}% 
\DOm{\la_1^2}{0}{0}{\la_2^2}{(n_{\soft}=2)} 
&=& \frac{1}{p_{20}^2 p_{31}^2}\Biggl\{
\frac{c_\eps}{\eps^2}
+\frac{c_\eps}{\eps}\biggl[
\ln\biggl(\frac{\mu\la_1}{-\bar p_{20}^2}\biggr)
+\ln\biggl(\frac{\mu\la_2}{-\bar p_{31}^2}\biggr)
\biggr]
\nn\\[-3em]
&& \quad {}
+2\ln\biggl(\frac{\mu\la_1}{-\bar p_{20}^2}\biggr)
\ln\biggl(\frac{\mu\la_2}{-\bar p_{31}^2}\biggr)
-\ln^2\biggl(\frac{\la_1\la_2}{-\bar p_{30}^2}\biggr)
-\frac{2\pi^2}{3}
\Biggl\},
\nn\\
\label{D0ms3ir2zm2} 
\\[1em]
%
%\DOp{\la^2}{\la^2}{0}{p_{30}^2}%
%{p_{20}^2}{p_{31}^2}% 
%\DOm{0}{\la^2}{0}{0}{(n_{\soft}=1)} 
%&=& \frac{1}{p_{20}^2 p_{31}^2}\Biggl\{
%\frac{c_\eps}{2\eps^2}
%+\frac{c_\eps}{\eps}\biggl[
%\ln\biggl(\frac{\mu\la}{-\bar p_{20}^2}\biggr)
%+\ln\biggl(\frac{\bar p_{30}^2}{\bar p_{31}^2}\biggr)
%\biggr]
%\nn\\[-3em]
%&& \quad {}
%+\ln\biggl(\frac{\mu\la}{-\bar p_{20}^2}\biggr)
%\biggl[
%\ln\biggl(\frac{\mu\la}{-\bar p_{20}^2}\biggr)
%+2\ln\biggl(\frac{\bar p_{30}^2}{\bar p_{31}^2}\biggr)
%\biggr]
%+\ln^2\biggl(\frac{\bar p_{30}^2}{\bar p_{31}^2}\biggr)
%\nn\\[.3em]
%&& \quad {}
%+2\Li\biggl(1-\frac{\bar p_{20}^2}{\bar p_{30}^2}\biggr)
%+\frac{\pi^2}{6}
%\Biggl\},
%%
%\label{D0ms3ir1zm3} 
%\\[1em] \alt
%
\DOp{\la^2}{\la^2}{0}{p_{30}^2}%
{p_{20}^2}{p_{31}^2}% 
\DOm{0}{\la^2}{0}{0}{(n_{\soft}=1)} 
&=& \frac{1}{p_{20}^2 p_{31}^2}\Biggl\{
\frac{c_\eps}{2\eps^2}
+\frac{c_\eps}{\eps}\biggl[
\ln\biggl(\frac{\mu\la}{-\bar p_{31}^2}\biggr)
+\ln\biggl(\frac{\bar p_{30}^2}{\bar p_{20}^2}\biggr)
\biggr]
\nn\\[-3em]
&& \quad {}
+\biggl[
\ln\biggl(\frac{\mu\la}{-\bar p_{31}^2}\biggr)
+\ln\biggl(\frac{\bar p_{30}^2}{\bar p_{20}^2}\biggr)
\biggr]^2
+2\Li\biggl(1-\frac{\bar p_{20}^2}{\bar p_{30}^2}\biggr)
+\frac{\pi^2}{6}
\Biggl\},\nn\\
\label{D0ms3ir1zm3} 
\\[1em]
%
%\DOp{0}{\la^2}{0}{p_{30}^2}%
%{p_{20}^2}{p_{31}^2}% 
%\DOm{\la^2}{\la^2}{0}{0}{(n_{\soft}=1)} 
%&=& \frac{1}{p_{20}^2 p_{31}^2}\Biggl\{
%\frac{c_\eps}{2\eps^2}
%+\frac{c_\eps}{\eps}
%\biggl[ 
%\ln\biggl(\frac{\mu\la}{-\bar p_{31}^2}\biggr)
%+\ln\biggl(\frac{\bar p_{30}^2}{\bar p_{20}^2}\biggr) \biggr]
%\nn\\[-3em]
%&& \quad {}
%+\ln\biggl(\frac{\mu\la}{-\bar p_{20}^2}\biggr)
%\biggl[
%\ln\biggl(\frac{\mu\la}{-\bar p_{20}^2}\biggr)
%+2\ln\biggl(\frac{\bar p_{30}^2}{\bar p_{31}^2}\biggr)
%\biggr]
%\nn\\
%&& \quad {}
%+2\Li\biggl(1-\frac{\bar p_{20}^2}{\bar p_{30}^2}\biggr)
%-2\Li\biggl(1-\frac{\bar p_{30}^2}{\bar p_{31}^2}\biggr)
%-\frac{\pi^2}{6}
%\Biggl\},
%%
%\label{D0ms3ir1zm2a} 
%\\[1em]\alt
%
\DOp{0}{\la^2}{0}{p_{30}^2}%
{p_{20}^2}{p_{31}^2}% 
\DOm{\la^2}{\la^2}{0}{0}{(n_{\soft}=1)} 
&=& \frac{1}{p_{20}^2 p_{31}^2}\Biggl\{
\frac{c_\eps}{2\eps^2}
+\frac{c_\eps}{\eps}
\biggl[ 
\ln\biggl(\frac{\mu\la}{-\bar p_{20}^2}\biggr)
+\ln\biggl(\frac{\bar p_{30}^2}{\bar p_{31}^2}\biggr) \biggr]
\nn\\[-3em]
&& \quad {}
+2\ln\biggl(\frac{\mu\la}{-\bar p_{20}^2}\biggr)
\ln\biggl(\frac{\mu\la}{-\bar p_{31}^2}\biggr)
-\ln^2\biggl(\frac{\mu\la}{\bar p_{30}^2}\biggr)
\nn\\
&& \quad {}
-2\Li\biggl(1-\frac{\bar p_{30}^2}{\bar p_{20}^2}\biggr)
-2\Li\biggl(1-\frac{\bar p_{30}^2}{\bar p_{31}^2}\biggr)
-\frac{\pi^2}{6}
\Biggl\},
\label{D0ms3ir1zm2a} 
\\[1em]
\DOp{\la_1^2}{\la_1^2}{\la_2^2}{p_{30}^2}%
{p_{20}^2}{p_{31}^2}% 
\DOm{0}{\la_1^2}{0}{\la_2^2}{(n_{\soft}=1)} 
&=& \frac{1}{p_{20}^2 p_{31}^2}\Biggl\{
\ln\biggl(\frac{\la_1\la_2}{-\bar p_{31}^2}\biggr) \biggl[
\frac{c_\eps}{\eps}
+\ln\biggl(\frac{\la_1\la_2}{-\bar p_{31}^2}\biggr)
-2\ln\biggl(\frac{\la_2\mu}{-\bar p_{30}^2}\biggr) 
\nn\\[-3em]
&& \quad {}
+2\ln\biggl(\frac{\mu^2}{-\bar p_{20}^2}\biggr) 
\biggr]
+\frac{\pi^2}{6}
\Biggl\},
\label{D0ms3ir1zm2o} 
\\[1em]
\DOp{\la_1^2}{\la_2^2}{0}{p_{30}^2}%
{p_{20}^2}{p_{31}^2}% 
\DOm{\la_1^2}{0}{\la_2^2}{\la_2^2}{(n_{\soft}=1)} 
&=& \frac{1}{p_{20}^2 p_{31}^2}\Biggl\{
\ln\biggl(\frac{\la_1\la_2}{-\bar p_{20}^2}\biggr)
\biggl[ 
\frac{c_\eps}{\eps}
+\ln\biggl(\frac{\la_1\la_2}{-\bar p_{20}^2}\biggr)
-2\ln\biggl(\frac{\la_1\la_2}{-\bar p_{30}^2}\biggr) 
\nn\\[-3em]
&& \quad {}
+2\ln\biggl(\frac{\la_2\mu}{-\bar p_{31}^2}\biggr) 
\biggr]
+2\Li\biggl(1-\frac{\bar p_{20}^2}{\bar p_{30}^2}\biggr)
-\frac{\pi^2}{6}
\Biggl\},
\label{D0ms3ir1zm1} 
\\[1em]
\DOp{\la^2}{0}{\la^2}{p_{30}^2}%
{p_{20}^2}{p_{31}^2}% 
\DOm{0}{\la^2}{\la^2}{0}{(n_{\soft}=0)} 
&=& \frac{1}{p_{20}^2 p_{31}^2}\Biggl\{
\biggl[
\ln\biggl(\frac{\la^2}{-\bar p_{20}^2}\biggr)
+\ln\biggl(\frac{\bar p_{30}^2}{\bar p_{31}^2}\biggr)
\biggr]^2
+\frac{\pi^2}{3}
\Biggl\},
\label{D0ms3ir0zm2} 
\\[1em]
%
%\DOp{0}{0}{\la^2}{p_{30}^2}%
%{p_{20}^2}{p_{31}^2}% 
%\DOm{\la^2}{\la^2}{\la^2}{0}{(n_{\soft}=0)} 
%&=& \frac{1}{p_{20}^2 p_{31}^2}\Biggl\{
%\ln\biggl(\frac{\la^2}{-\bar p_{20}^2}\biggr)
%\biggl[
%\ln\biggl(\frac{\la^2}{-\bar p_{20}^2}\biggr)
%+2\ln\biggl(\frac{\bar p_{30}^2}{\bar p_{31}^2}\biggr)
%\biggr]
%\nn\\[-3em]
%&& \quad {}
%-2\Li\biggl(1-\frac{\bar p_{30}^2}{\bar p_{31}^2}\biggr)
%\Biggl\},
%%
%\label{D0ms3ir0zm1} 
%\\[1em]\alt
%
\DOp{0}{0}{\la^2}{p_{30}^2}%
{p_{20}^2}{p_{31}^2}% 
\DOm{\la^2}{\la^2}{\la^2}{0}{(n_{\soft}=0)} 
&=& \frac{1}{p_{20}^2 p_{31}^2}\Biggl\{
\biggl[
\ln\biggl(\frac{\la^2}{-\bar p_{20}^2}\biggr)
+\ln\biggl(\frac{\bar p_{30}^2}{\bar p_{31}^2}\biggr)
\biggr]^2
+2\Li\biggl(1-\frac{\bar p_{31}^2}{\bar p_{30}^2}\biggr)
\Biggl\},
\nl[-6ex]
\label{D0ms3ir0zm1} 
\\[1em]
%
%
%\DOp{0}{0}{0}{p_{30}^2}%
%{p_{20}^2}{p_{31}^2}% 
%\DOm{\la^2}{\la^2}{\la^2}{\la^2}{(n_{\soft}=0)} 
%&=& \frac{1}{p_{20}^2 p_{31}^2}\Biggl\{
%\ln\biggl(\frac{\la^2}{-\bar p_{20}^2}\biggr)
%\biggl[
%\ln\biggl(\frac{\la^2}{-\bar p_{20}^2}\biggr)
%+2\ln\biggl(\frac{\bar p_{30}^2}{\bar p_{31}^2}\biggr)
%\biggr]
%\nn\\[-3em]
%&& \quad {}
%+2\Li\biggl(1-\frac{\bar p_{20}^2}{\bar p_{30}^2}\biggr)
%-2\Li\biggl(1-\frac{\bar p_{30}^2}{\bar p_{31}^2}\biggr)
%-\frac{\pi^2}{3}
%\Biggl\},
%%
%\label{D0ms3ir0zm0} 
%\\[1em]\alt
%
\DOp{0}{0}{0}{p_{30}^2}%
{p_{20}^2}{p_{31}^2}% 
\DOm{\la^2}{\la^2}{\la^2}{\la^2}{(n_{\soft}=0)} 
&=& \frac{1}{p_{20}^2 p_{31}^2}\Biggl\{
\biggl[
\ln\biggl(\frac{\la^2}{-\bar p_{20}^2}\biggr)
+\ln\biggl(\frac{\bar p_{30}^2}{\bar p_{31}^2}\biggr)
\biggr]^2
\nn\\[-3em]
&& \quad {}
+2\Li\biggl(1-\frac{\bar p_{20}^2}{\bar p_{30}^2}\biggr)
+2\Li\biggl(1-\frac{\bar p_{31}^2}{\bar p_{30}^2}\biggr)
-\frac{\pi^2}{3}
\Biggl\}.
\label{D0ms3ir0zm0} 
\eeqar
The integral \refeq{D0ms3ir2zm4} can be found in
\citeres{Bern:1993kr,Duplancic:2000sk,Ellis:1980wv}, corresponds to
``Box 2'' of \citere{Ellis:2007qk}, and
had already been calculated in \citeres{Fabricius:1979tb,Papadopoulos:1981ju}.
Integral \refeq{D0ms3ir2zm3} can be obtained
from (A.4) of \citere{Beenakker:2002nc} and corresponds to 
\refeq{eq:DOc2s2dimreg} with small $m_3$.
Integral \refeq{D0ms3ir2zm2} corresponds to 
\refeq{eq:DOc1s2dimreg} with both $m_2$ and $m_3$ small.
Integral \refeq{D0ms3ir1zm3} 
% is a special case of ``Box 9'' of \citere{Ellis:2007qk} and 
can be obtained from \refeq{eq:DOc1s1dimregonsh3} for $m_2^2=0$,
$p^2_{32}=m_3^2$, and $m_3^2$ small.
Integral \refeq{D0ms3ir1zm2a} is a particular case of
\refeq{eq:DOc1s1dimregonsh3}.
Integrals \refeq{D0ms3ir1zm2o} and \refeq{D0ms3ir1zm1}
are cases (iid) and (id) of \citere{Beenakker:1990jr}
(translated to dimensional regularization), respectively,
and can also be derived from
\refeq{eq:DOc2s2dimmassreg4} and \refeq{eq:DOc2s2dimmassreg3}
for small $m_3$, respectively.

\paragraph{Cases with four collinear singularities $(n_{\coll}=4)$}

There are six different cases with up to four soft
singularities and the following common finite part
\beq
\DOfin = D_0 
-\frac{C_0(0)+C_0(2)}{p_{20}^2}
-\frac{C_0(1)+C_0(3)}{p_{31}^2}.
\eeq

\beqar
\DOp{0}{0}{0}{0}%
{p_{20}^2}{p_{31}^2}% 
\DOm{0}{0}{0}{0}{(n_{\soft}=4)} 
&=& \frac{1}{p_{20}^2 p_{31}^2}\Biggl\{
\frac{4c_\eps}{\eps^2}
+\frac{2c_\eps}{\eps}
\biggl[
\ln\biggl(\frac{\mu^2}{-\bar p_{20}^2}\biggr)
+\ln\biggl(\frac{\mu^2}{-\bar p_{31}^2}\biggr)
\biggr]
\nn\\[-3em]
&& \quad {}
+2\ln\biggl(\frac{\mu^2}{-\bar p_{31}^2}\biggr)
\ln\biggl(\frac{\mu^2}{-\bar p_{20}^2}\biggr)
-\frac{5\pi^2}{3}\Biggl\},
\label{D0ms4ir4zm4}
\\[1em]
\DOp{\la^2}{\la^2}{0}{0}%
{p_{20}^2}{p_{31}^2}% 
\DOm{0}{\la^2}{0}{0}{(n_{\soft}=3)} 
&=& \frac{1}{p_{20}^2 p_{31}^2}\Biggl\{
\frac{2c_\eps}{\eps^2}
+\frac{c_\eps}{\eps}
\biggl[
\ln\biggl(\frac{\mu^2}{-\bar p_{20}^2}\biggr)
+2\ln\biggl(\frac{\mu\la}{-\bar p_{31}^2}\biggr) 
\biggr]
\nn\\[-3em]
&& \quad {}
+2\ln\biggl(\frac{\mu\la}{-\bar p_{31}^2}\biggr)
\ln\biggl(\frac{\mu^2}{-\bar p_{20}^2}\biggr) 
-\frac{5\pi^2}{6}\Biggl\},
\label{D0ms4ir3zm3}
\\[1em]
\DOp{\la^2}{0}{\la^2}{0}%
{p_{20}^2}{p_{31}^2}% 
\DOm{0}{\la^2}{\la^2}{0}{(n_{\soft}=2)} 
&=& \frac{1}{p_{20}^2 p_{31}^2}\Biggl\{
\frac{c_\eps}{\eps^2}
+\frac{c_\eps}{\eps}
\biggl[
 \ln\biggl(\frac{\mu\la}{-\bar p_{20}^2}\biggr)
+\ln\biggl(\frac{\mu\la}{-\bar p_{31}^2}\biggr) 
\biggr]
\nn\\*[-3em]
&& \quad {}
+2\ln\biggl(\frac{\mu\la}{-\bar p_{20}^2}\biggr)
\ln\biggl(\frac{\mu\la}{-\bar p_{31}^2}\biggr)
-\frac{2\pi^2}{3}\Biggl\},
\label{D0ms4ir2zm2a}
\\[1em]
\DOp{\la_1^2}{\la_1^2}{\la_2^2}{\la_2^2}%
{p_{20}^2}{p_{31}^2}% 
\DOm{0}{\la_1^2}{0}{\la_2^2}{(n_{\soft}=2)} 
&=& \frac{2}{p_{20}^2 p_{31}^2} 
\biggl[ \frac{c_\eps}{\eps}
+\ln\biggl(\frac{\mu^2}{-\bar p_{20}^2}\biggr) \biggr]
\ln\biggl(\frac{\la_1\la_2}{-\bar p_{31}^2}\biggr),
\label{D0ms4ir2zm2o}
\\[1em]
\DOp{\la^2}{0}{0}{\la^2}%
{p_{20}^2}{p_{31}^2}% 
\DOm{0}{\la^2}{\la^2}{\la^2}{(n_{\soft}=1)} 
&=& \frac{1}{p_{20}^2 p_{31}^2} \Biggl\{
\biggl[ \frac{c_\eps}{\eps}
+2\ln\biggl(\frac{\mu\la}{-\bar p_{20}^2}\biggr) \biggr]
\ln\biggl(\frac{\la^2}{-\bar p_{31}^2}\biggr)
-\frac{\pi^2}{2} \Biggr\},
\hspace{2em}
\label{D0ms4ir2zm1}
\\[1em]
\DOp{0}{0}{0}{0}%
{p_{20}^2}{p_{31}^2}% 
\DOm{\la^2}{\la^2}{\la^2}{\la^2}{(n_{\soft}=0)} 
&=& \frac{2}{p_{20}^2 p_{31}^2} \Biggl\{
\ln\biggl(\frac{\la^2}{-\bar p_{20}^2}\biggr)
\ln\biggl(\frac{\la^2}{-\bar p_{31}^2}\biggr)
-\frac{\pi^2}{2} \Biggr\}.
 \label{D0ms4ir0zm0}
\eeqar
The integral \refeq{D0ms4ir4zm4} is given in (4.25) of
\citere{Bern:1993kr}, corresponds to ``Box 1'' of
\citere{Ellis:2007qk} and had already been calculated in
\citeres{Fabricius:1979tb,Papadopoulos:1981ju}.
Integral \refeq{D0ms4ir3zm3} can be obtained
from \citere{Beenakker:1988bq} and corresponds to 
\refeq{eq:DOc2s3dimreg} for small $m_3$. 
Integral \refeq{D0ms4ir2zm2a} is a special case of 
\refeq{eq:DOc1s2dimreg} for $p_{32}^2=0$ and small masses.
Integral \refeq{D0ms4ir2zm2o} is case (iii b)
of \citere{Beenakker:1990jr} (in dimensional regularization)
and corresponds to \refeq{eq:DOc2s3dimmassreg3} for small $m_3$. 
Integral \refeq{D0ms4ir2zm1} is case (ie) of \citere{Beenakker:1990jr}
(in dimensional regularization) and corresponds to a special case of
\refeq{eq:DOc0s1dimreg}, for vanishing $p_{21}^2$ and $p_{32}^2$ and
if again all masses are small.

\section{Summary}
\label{se:sum}

The calculation of scalar one-loop integrals is at the heart of any
perturbative calculation of radiative corrections at the one-loop level,
independent of the method employed. In diagrammatic calculations, which
do not rely on numerical integration of loop integrals,
tensor integrals as well as any one-loop integral with more than
four external particles (legs) are algebraically reduced to standard
scalar integrals with at most four propagators. Recently so-called
unitarity-based methods became more and more popular at the one-loop
level, i.e.\ methods that abandon the direct use of Feynman diagrams,
but also those methods need the standard scalar integrals with up to
four propagators as input. The purpose of this paper is to close some
gaps in the literature and to deliver a complete set of 4-point (``box'')
scalar integrals that appear in any kind of one-loop calculation.
For the lower-point integrals exhaustive lists of results already exist.

The considered scalar box integrals are naturally categorised into
finite, regular cases and singular cases that involve soft and/or
collinear divergences. For the most general regular case with complex
internal masses, which appear in applications that involve unstable
intermediate particles, we have presented two new analytical expressions
that involve less (72 and 32)
dilogarithmic functions than a recently published independent result
of other authors. 
In addition we have listed a complete set of soft- and/or 
collinear-singular cases which appear in the calculation of
strong or electroweak radiative corrections, i.e.\ we support
both dimensional regularization and regularizations schemes that employ
small mass parameters. Mixed regularizations of these types are supported
as well, and all results are valid for complex masses of internal
particles. Many of those results have not yet been published elsewhere.

In combination with available results on scalar 1-, 2-, and 3-point
integrals now all relevant standard scalar integrals for one-loop
calculations in QCD, QED, the electroweak Standard Model and typical
extensions thereof are known for all not too exotic (mass or
dimensional) regularization schemes, covering also internal unstable
particles appearing as resonances.

\section*{Acknowledgements}
This work is supported in part by the European Community's Marie-Curie
Research Training Network under contract MRTN-CT-2006-035505 ``Tools
and Precision Calculations for Physics Discoveries at Colliders''.

\end{document}